\documentclass[12pt]{iopart}

\usepackage{axodraw}

\usepackage{graphicx}
\usepackage{rotating}
\usepackage{color}

\begin{document}
\begin{flushright}
hep-ph/0605236   \\
{D\O\ Note CONF 5116 (2006)}
\end{flushright}

\title[New Physics Searches at the Tevatron and the LHC]
{New~Physics~Searches~at~the~Tevatron~and~the~LHC\footnote[7]
{Presented at PHENO 05: World Year of Phenomenology, 
University of Wisconsin-Madison, May 2-4, 2005; 
and Collider Physics Workshop: From the Tevatron to the LHC to 
the Linear Collider, Aspen Center for Physics, Summer 2005. }}

\author{Andr\'e Sopczak}

\address{Lancaster University}
\ead{andre.sopczak@cern.ch}
\begin{abstract}
The Tevatron Run-II started data-taking in spring 2001 and several 
searches for new particles have been performed. The preliminary 2005 
results are concisely reviewed for the experiments CDF and D\O.
Model-independent and model-dependent limits on Higgs boson and 
Supersymmetric particle production are set and interpretations are given.
Several limits from the LEP era have been extended.
The outlook for the Tevatron and the prospects for the ATLAS and CMS 
experiments at the LHC for selected searches are briefly addressed.
\end{abstract}

\maketitle

\tableofcontents

%\documentclass[slac_one]{revtex4}
%
%\usepackage{axodraw}
%\usepackage{pstricks}
%\usepackage{color}
%
%\usepackage{graphicx}
%\usepackage{rotating}
%\usepackage{fancyhdr}
%\pagestyle{fancy}
%\fancyhead{} % clear all fields
%\fancyfoot{} % clear all fields
%\renewcommand{\headrulewidth}{0pt}
%\renewcommand{\footrulewidth}{0pt}
%\renewcommand{\sfdefault}{phv}
%\renewcommand{\baselinestretch}{1.2}

%AS
%\setlength{\textheight}{255mm}
%\setlength{\textwidth}{178mm}%{170mm}
%\setlength{\topmargin}{-15mm}

\newcommand{\bb}{b\bar b}
\newcommand{\nn}{\nu\bar \nu}
\newcommand{\tautau}{\tau^+\tau^-}
\newcommand{\ee}{\mbox{$\mathrm{e}^{+}\mathrm{e}^{-}$}}

\newcommand{\Zo} {{\mathrm {Z}}}
\newcommand{\db}    {{d_{\rm B}}}%
\newcommand{\dgz}  {{\Delta g_1^{\Zo}}}%
\newcommand{\dkg}   {{\Delta \kappa_\gamma}}%

\newcommand{\pb}   {\mbox{$\rm pb^{-1}$}}
\newcommand{\fb}   {\mbox{$\rm fb^{-1}$}}

\clearpage
\section{Introduction} 
\begin{figure}[h!]
%\vspace*{-0.4cm}
\includegraphics[width=0.49\textwidth]{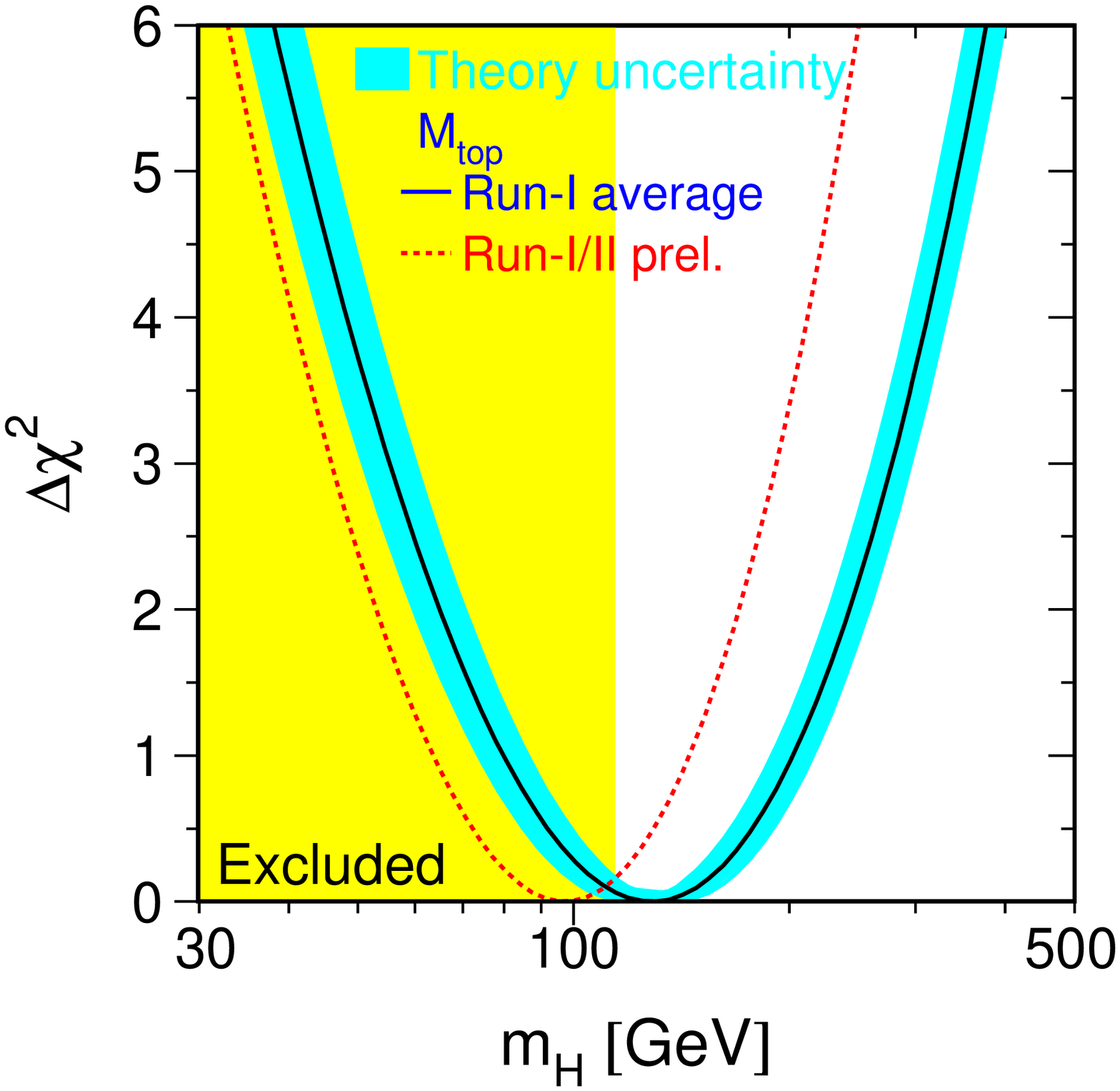}\hfill
\includegraphics[width=0.49\textwidth]{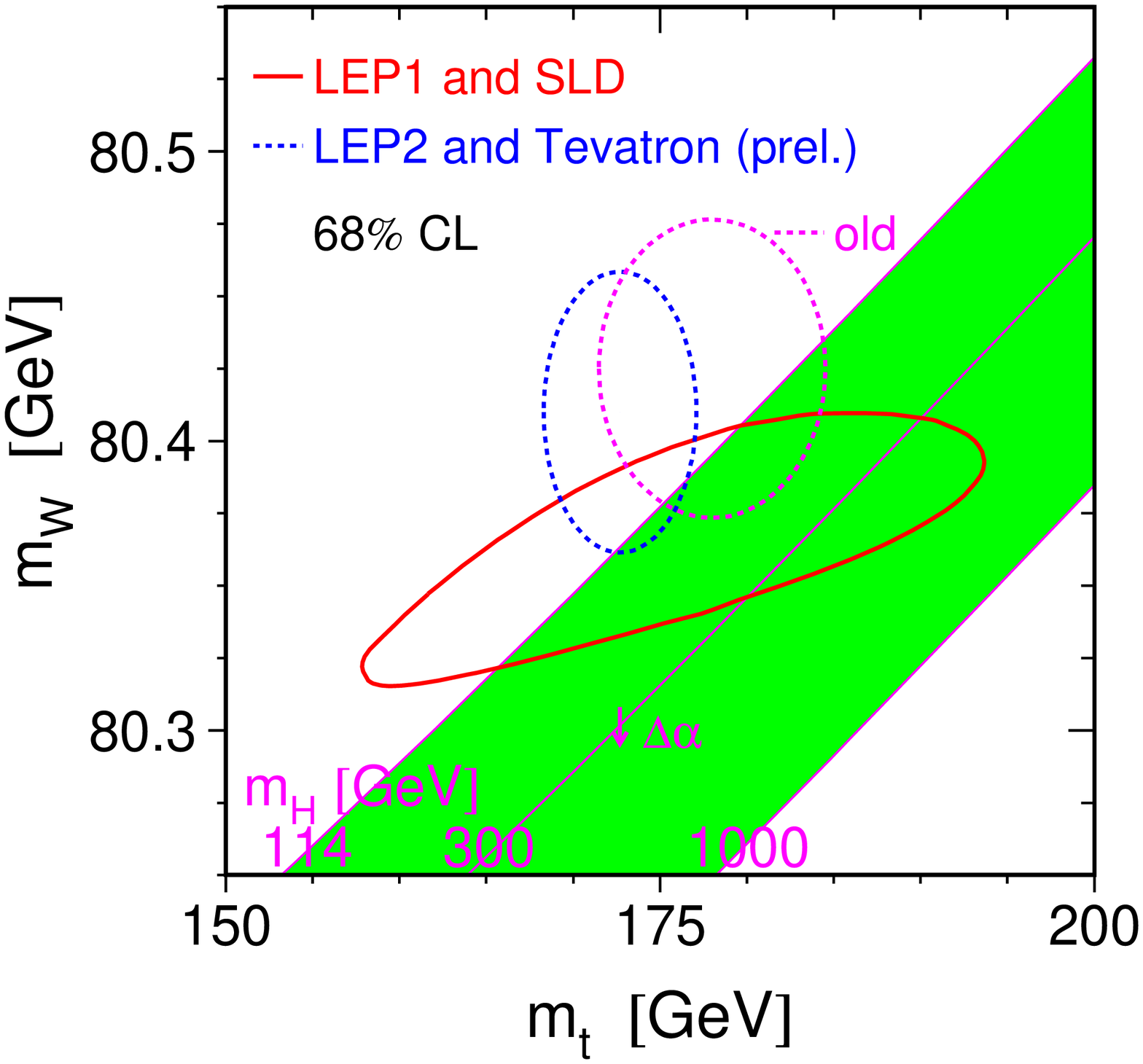}
\vspace*{-0.4cm}
\caption{
Left: Higgs boson mass prediction in the SM framework. 
      The shift to a lower value (dotted line) from the new top quark mass measurement at the Tevatron Run-II 
      is clearly visible.
Right: old (pre-summer'05) results and the new smaller ellipse (dashed lines) prefer a region 
       outside the SM Higgs boson mass band ($m_{\rm H} = 114$ to 1000 GeV).
       The combined results from LEP-1 and SLD are shown separately (solid line).
}
\label{fig:ew}
%\vspace*{-0.6cm}
\end{figure}

\begin{figure}[h!]
\begin{center}
\vspace*{0.5cm}
\includegraphics[width=0.9\textwidth,height=8cm]{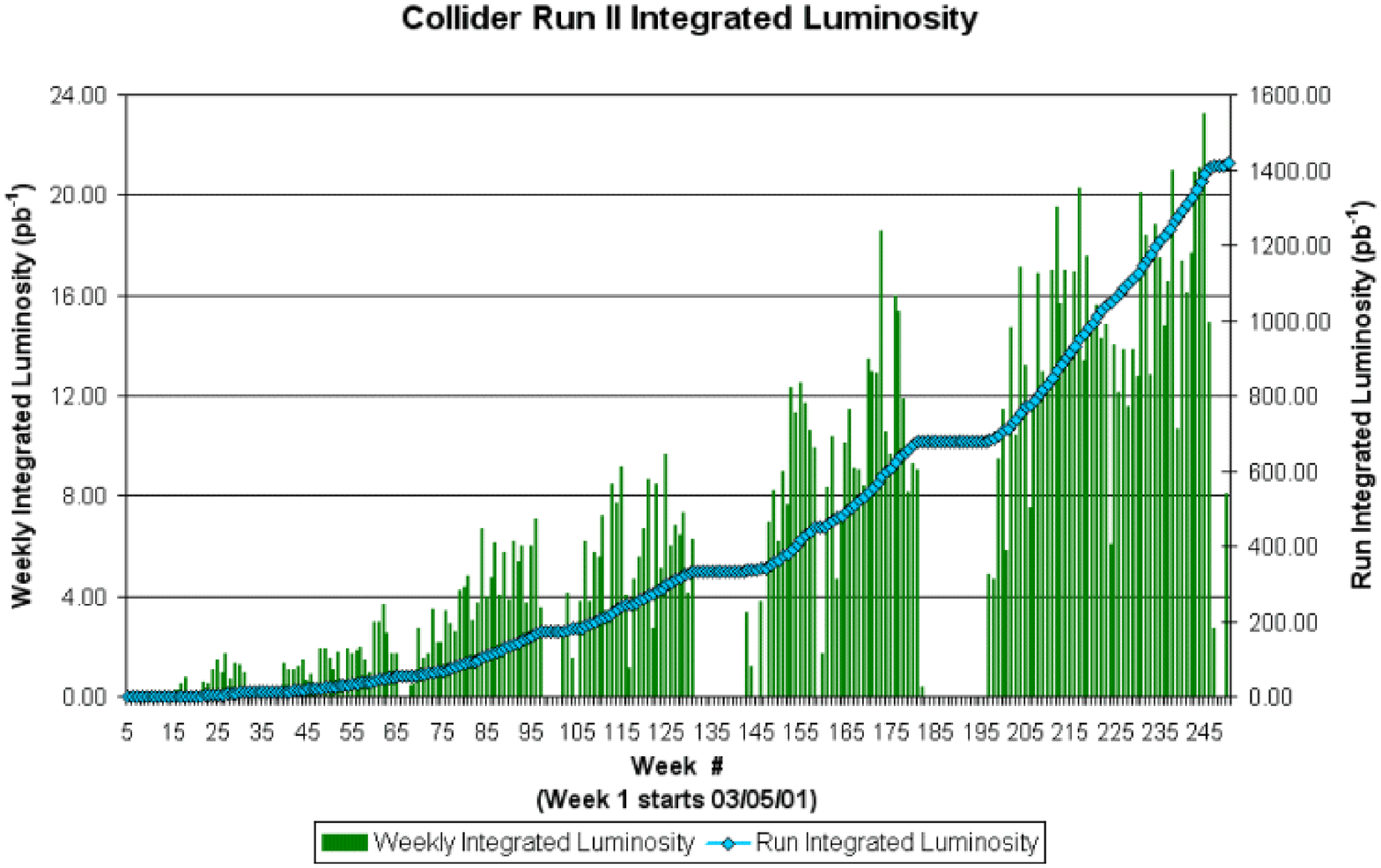} \\
\vspace*{1cm}
\includegraphics[width=0.8\textwidth,height=7cm]{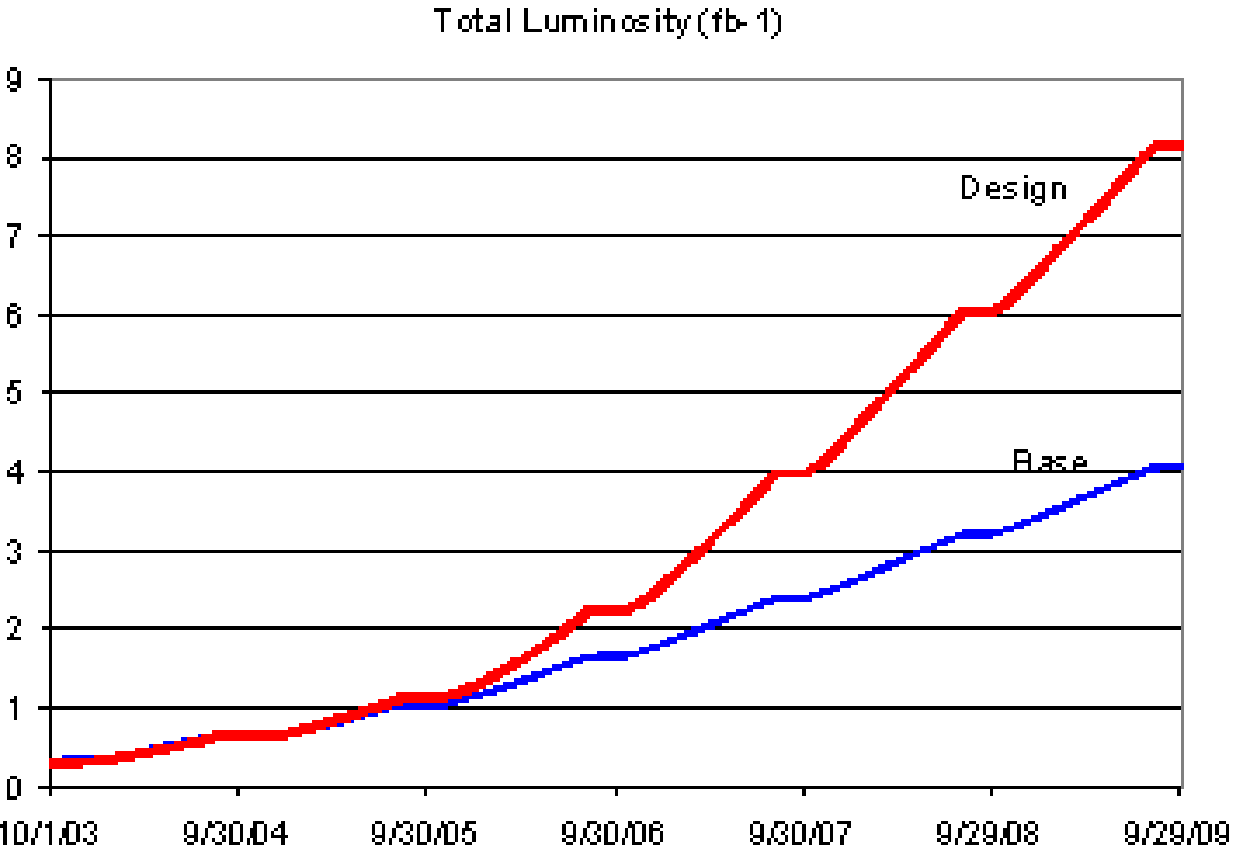}
%\vspace*{-0.3cm}
\caption{Upper: delivered Tevatron luminosity.
         Lower: expected integrated delivered luminosity.
}
\label{fig:lumi}
\vspace*{-0.8cm}
\end{center}
\end{figure}

The search for new particles is at the forefront of High Energy Physics. The discovery of a Higgs 
boson would shed light on electroweak symmetry breaking and the generation of mass in the Universe.
Experimental evidence for Supersymmetric particles would be of equal importance and extend the 
Standard Model (SM) of particle physics.
Many searches for new particles were performed at LEP and stringent limits on Higgs bosons 
in the SM and beyond were set. These limits are summarized in Table~\ref{tab:lep} (from~\cite{lep05}) and
new model-independent limits and benchmark results in the 
Minimal Supersymmetric extension of the SM were recently presented~\cite{susy05}.
The interpretations of the MSSM results depend significantly on the top quark mass. Results are presented 
for various top quark masses and the current (summer'05) value is $m_{\rm t} = 172.7 \pm 2.9$~GeV~\cite{topmass}.
In addition to the limits from direct searches, some indication on the Higgs boson mass exist from
precision electro-weak measurements, as shown in Fig.~\ref{fig:ew} (from~\cite{lep-ew}).
About 300~pb$^{-1}$ of data have been analyzed so far by each Tevatron experiment,
while above 1~fb$^{-1}$ data have been recorded. 
Figure~\ref{fig:lumi} (from~\cite{lumi}) shows the delivered luminosity and its expectations.

%\clearpage
The Higgs boson searches at the Tevatron are discussed in Sec.~2.
In Sec.~2.1 Higgs boson production and decay is addressed.
Aspects of b-quark tagging are summarized in Sec. 2.2.
The gluon fusion process $\rm gg\rightarrow H$ ($\rm H\to WW$) is presented in Sec.~2.3.
Subsequently, the associated production processes WH ($\rm H\to\bb,WW$) and $\rm ZH\to \nn\bb$
are discussed in Sec.~2.4. A summary of the cross section limits is given in Sec.~2.5.
The following Higgs boson searches beyond the SM are presented in Sec.~2.6:
$\rm \bb A$, $\rm A\to \tautau$, $\rm H\to\gamma\gamma$, $\rm t\to H^+b$, and $\rm H^{++}$.
Supersymmetric particle searches at the Tevatron are reviewed in Sec.~3.
First, the di-photon process in the GMSB interpretation is addressed in Sec.~3.1, then
tri-lepton signatures (Sec.~3.2), scalar quarks and gluinos (Sec.~3.3),
scalar tops and scalar bottoms (Sec.~3.4) and charged massive particles (Sec.~3.5).
For the LHC, the prospects are briefly discussed in Sec.~4, first, for the SM Higgs boson (Sec.~4.1),
then for Higgs boson searches beyond the SM (Sec.~4.2) and for Supersymmetric particles (Sec.~4.3).

\clearpage
\begin{table}[htb]
\renewcommand{\arraystretch}{1.2} % enlarge line spacing
\caption{Summary of Higgs boson mass limits at 95\% CL.
`LEP' indicates a combination of the results from ALEPH, DELPHI, L3 and OPAL.
If results from the experiments are not (yet) combined, examples
which represent the different search areas from individual experiments
are given. Details of the limit parameters are given in Ref.~\cite{lep05}.
\label{tab:lep} }
\begin{center}
{
\begin{tabular}{c|c|r}
Search                     & experiment & limit \\\hline 
Standard Model              &   LEP  
   & $m^{\rm SM}_{\rm H} > 114.4$ GeV \\ 
Reduced rate and SM decay &       
  & $\xi^2>0.05:$ $ m_{\rm H} > 85$ GeV \\
& & $\xi^2>0.3:$ $ m_{\rm H} > 110$ GeV \\
Reduced rate and $\rm b\bar b$ decay  &
  & $\xi^2>0.04:$ $ m_{\rm H} > 80$ GeV \\
& & $\xi^2>0.25:$ $ m_{\rm H} >110$ GeV \\ 
Reduced rate and $\tau^+\tau^-$ decay & 
  & $\xi^2>0.2:$ $ m_{\rm H} > 113$ GeV \\ 
\hspace*{-4mm} Reduced rate and hadronic decay\hspace*{-2mm} &
  & $\xi^2=1:$   $m_{\rm H} >112.9$ GeV\\ 
& & $\xi^2>0.3:$ $ m_{\rm H} > 97$ GeV \\ 
&ALEPH& $\xi^2>0.04:$ $m_{\rm H} \approx 90$ GeV \\ 
Anomalous couplings & L3 & $d,~\db,~\dgz,~\dkg$ exclusions \\ \hline
MSSM (no scalar top mixing) & LEP 
  & almost entirely excluded\\ 
General MSSM scan & DELPHI &  $m_{\rm h} > 87$ GeV, $m_{\rm A} >90$ GeV\\ 
Larger top-quark mass     & LEP & strongly reduced $\tan\beta$ limits \\ \hline
MSSM with CP-violating phases  & LEP    &  strongly reduced mass limits  \\ \hline
Visible/invisible Higgs decays & DELPHI & $m_{\rm H} >111.8$ GeV\\ 
Majoron model (max. mixing) &  & $m_{\rm H,S} >112.1$ GeV\\ \hline
Two-doublet Higgs model   & DELPHI
  & $\rm hA\to b\bar b b\bar b:$
    $m_{\rm h}+m_{\rm A} >  150$ GeV\\
(for $\sigma_{\rm max}$) & 
  & $\tau^+\tau^-\tau^+\tau^-:$
    $m_{\rm h}+m_{\rm A} >  160$ GeV\\
& & $\rm (AA)A\to 6b:$ $m_{\rm h}+m_{\rm A} >  150$ GeV\\
& & $\rm (AA)Z\to 4b~Z:$ $m_{\rm h} >  90$ GeV\\
& & $\rm hA\to q\bar q q\bar q:$ 
      $m_{\rm h}+m_{\rm A} >  110$ GeV\\
Two-doublet model scan & OPAL
  & $\tan\beta > 1:$ $ m_{\rm h} \approx m_{\rm A} > 85$ GeV \\\hline 
Yukawa process & DELPHI & $C > 40:$ $m_{\rm h,A} > 40$ GeV \\\hline 
Singly-charged Higgs bosons & LEP 
  & $m_{\rm H^\pm} > 78.6$ GeV \\
$\rm W^\pm A$ decay mode & DELPHI& $m_{\rm H^\pm} > 76.7$ GeV \\ \hline
Doubly-charged Higgs bosons &  DELPHI/OPAL 
  & 
$m_{\rm H^{++}} > 99$ GeV \\
$\ee\to\ee$ &L3 &$h_{\rm ee} > 0.5:$ $m_{\rm H^{++}} > 700$ GeV \\ \hline
Fermiophobic $\rm H\to WW, ZZ, \gamma\gamma$ & L3 
  &  $m_{\rm H} > 108.3$ GeV \\
$\rm H\to \gamma\gamma$ &LEP &  $ m_{\rm H} > 109.7$ GeV \\ \hline
Uniform and stealthy scenarios & OPAL & depending on model parameters
\end{tabular}
}
\end{center}
\vspace*{-3.9cm}
\end{table}

\clearpage
\section{Higgs Boson Searches at the Tevatron}

\begin{figure}[hp]
\begin{minipage}{0.8\textwidth}
\includegraphics[width=\textwidth]{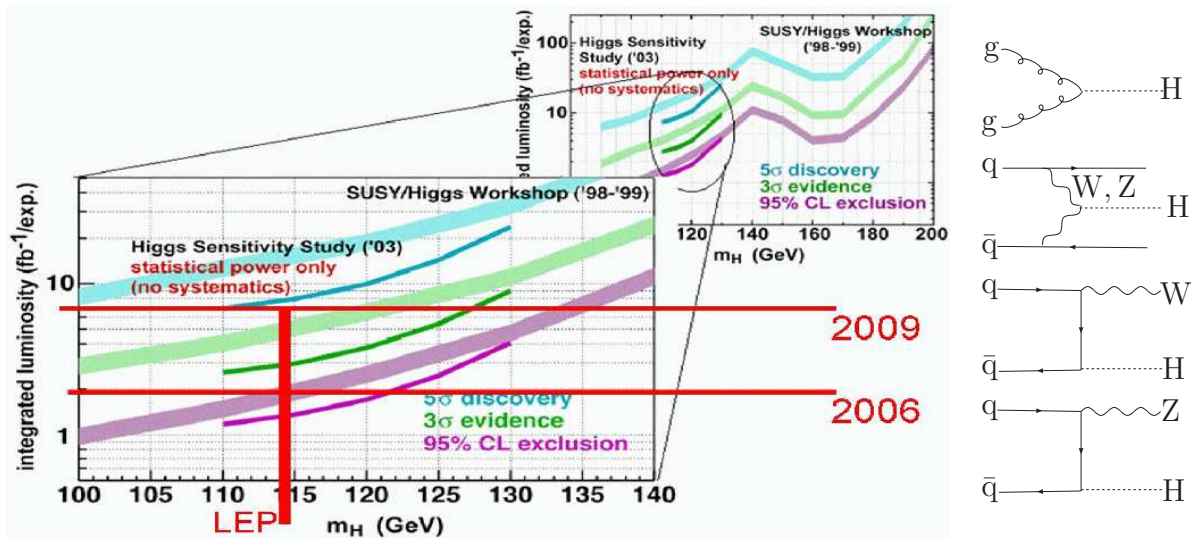}
\end{minipage}\hfill
\begin{minipage}{0.19\textwidth}
\SetScale{0.5}      % picture size control
\unitlength=0.5 pt  % picture size control
\fcolorbox{white}{white}{
  \begin{picture}(165,72) (31,-25)
    \SetWidth{0.5}
    \SetColor{Black}
    \Gluon(107,10)(46,-17){3}{3.34}
    \DashLine(109,10)(168,10){2}
    \Gluon(46,41)(106,10){3}{3.4}
    \Text(166,12)[l]{\normalsize{\Black{$\rm H$}}}
    \Text(31,-17)[l]{\normalsize{\Black{$\rm g$}}}
    \Text(33,39)[l]{\normalsize{\Black{$\rm g$}}}
  \end{picture}
}
\\
\fcolorbox{white}{white}{
  \begin{picture}(170,76) (43,-22)
    \SetWidth{0.5}
    \SetColor{Black}
    \Text(43,-14)[l]{\normalsize{\Black{$\rm \bar{q}$}}}
    \Text(111,30)[l]{\normalsize{\Black{$\rm W,Z$}}}
    \SetWidth{0.5}
    \Photon(89,-12)(118,18){3}{2}
    \Photon(89,46)(117,18){3}{2}
    \Text(43,46)[l]{\normalsize{\Black{$\rm q$}}}
    \DashLine(117,16)(178,16){2}
    \ArrowLine(168,-14)(55,-13)
    \ArrowLine(59,46)(167,46)
    \Text(183,17)[l]{\normalsize{\Black{$\rm H$}}}
  \end{picture}
}

\fcolorbox{white}{white}{
  \begin{picture}(167,76) (75,-36)
    \SetWidth{0.5}
    \SetColor{Black}
    \Photon(150,31)(210,31){3}{3}
    \ArrowLine(150,31)(150,-29)
    \ArrowLine(87,32)(150,31)
    \ArrowLine(150,-29)(88,-30)
    \DashLine(211,-28)(150,-28){2}
    \Text(75,32)[l]{\normalsize{\Black{$\rm q$}}}
    \Text(75,-28)[l]{\normalsize{\Black{$\rm \bar{q}$}}}
    \Text(210,29)[l]{\normalsize{\Black{$\rm W$}}}
    \Text(212,-28)[l]{\normalsize{\Black{$\rm H$}}}
  \end{picture}
}
\\
\fcolorbox{white}{white}{
  \begin{picture}(167,76) (75,-36)
    \SetWidth{0.5}
    \SetColor{Black}
    \Photon(150,31)(210,31){3}{3}
    \ArrowLine(150,31)(150,-29)
    \ArrowLine(87,32)(150,31)
    \ArrowLine(150,-29)(88,-30)
    \DashLine(211,-28)(150,-28){2}
    \Text(75,32)[l]{\normalsize{\Black{$\rm q$}}}
    \Text(75,-28)[l]{\normalsize{\Black{$\rm \bar{q}$}}}
    \Text(212,-28)[l]{\normalsize{\Black{$\rm H$}}}
    \Text(211,32)[l]{\normalsize{\Black{$\rm Z$}}}
  \end{picture}
}
\end{minipage}
\caption{
Left: expected Tevatron SM Higgs boson sensitivities from 1999 and 2003 studies.
Right: Higgs boson production graphs: gluon fusion, vector boson fusion (VBF), associated 
       production WH and WZ.
}
\label{fig:prospects}
\end{figure}

Figure~\ref{fig:prospects} (from~\cite{prospects99,prospects03}) 
shows the estimated discovery and exclusion potential for the SM Higgs boson at the Tevatron,
and the dominant Higgs boson production diagrams.
Currently about 1~\fb\ luminosity has been recorded per experiment. 
It is expected that by the end of 2006 about 2~\fb\ will have been recorded, and about 8~\fb\
by the end of 2009, corresponding to a 95\% CL exclusion sensitivity up to 120~GeV and 180~GeV, 
respectively.
With the anticipated delivered luminosity, it is particularly important to increase the expected 
sensitivity in the vicinity of 140 GeV.

\subsection{Production and Decay}

\begin{figure}
\begin{minipage}{0.58\textwidth}
\begin{turn}{270}
\includegraphics[height=\textwidth,width=6cm]{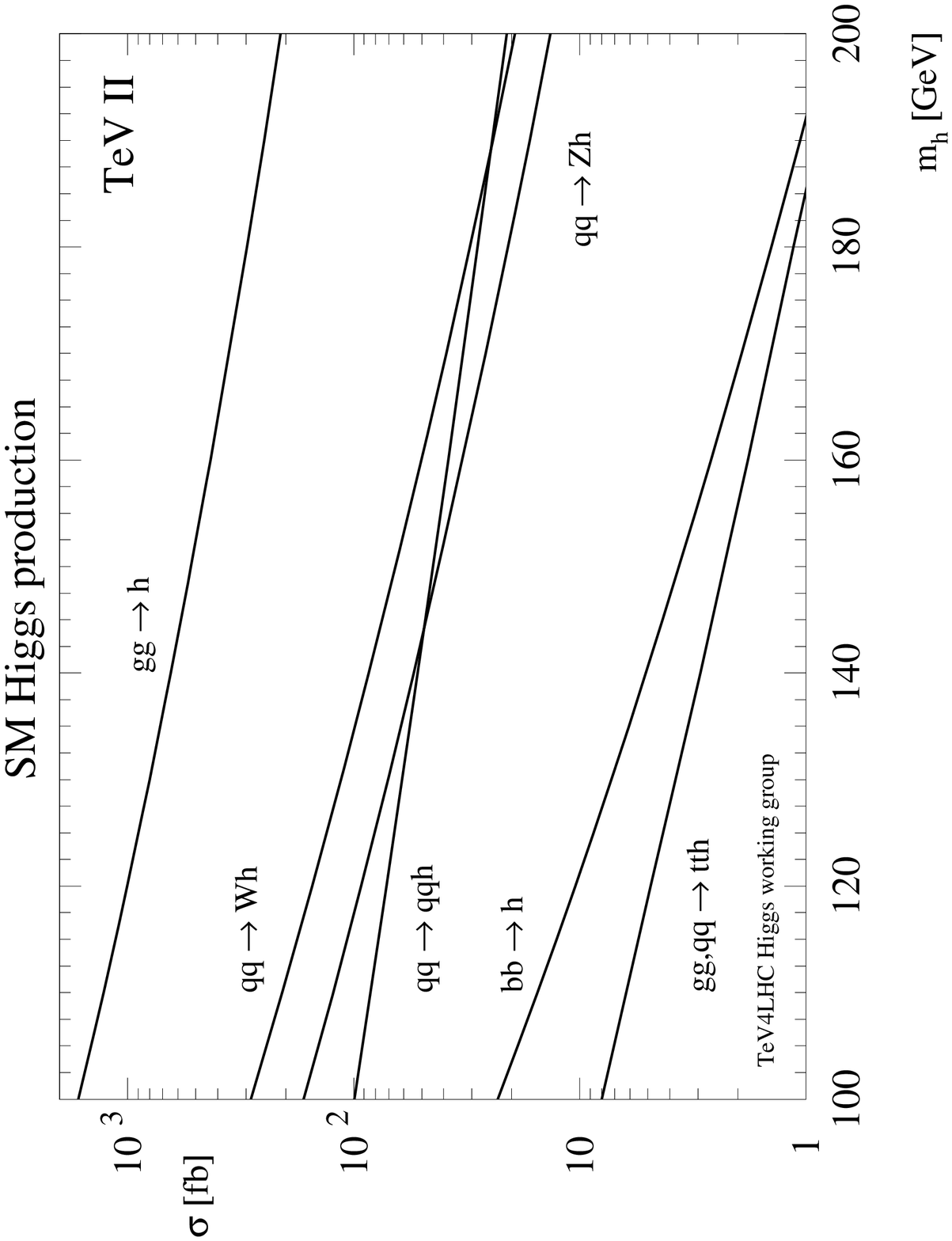} \hfill
\end{turn}\hfill
\end{minipage}
\begin{minipage}{0.40\textwidth}
\includegraphics[width=\textwidth,height=6cm]{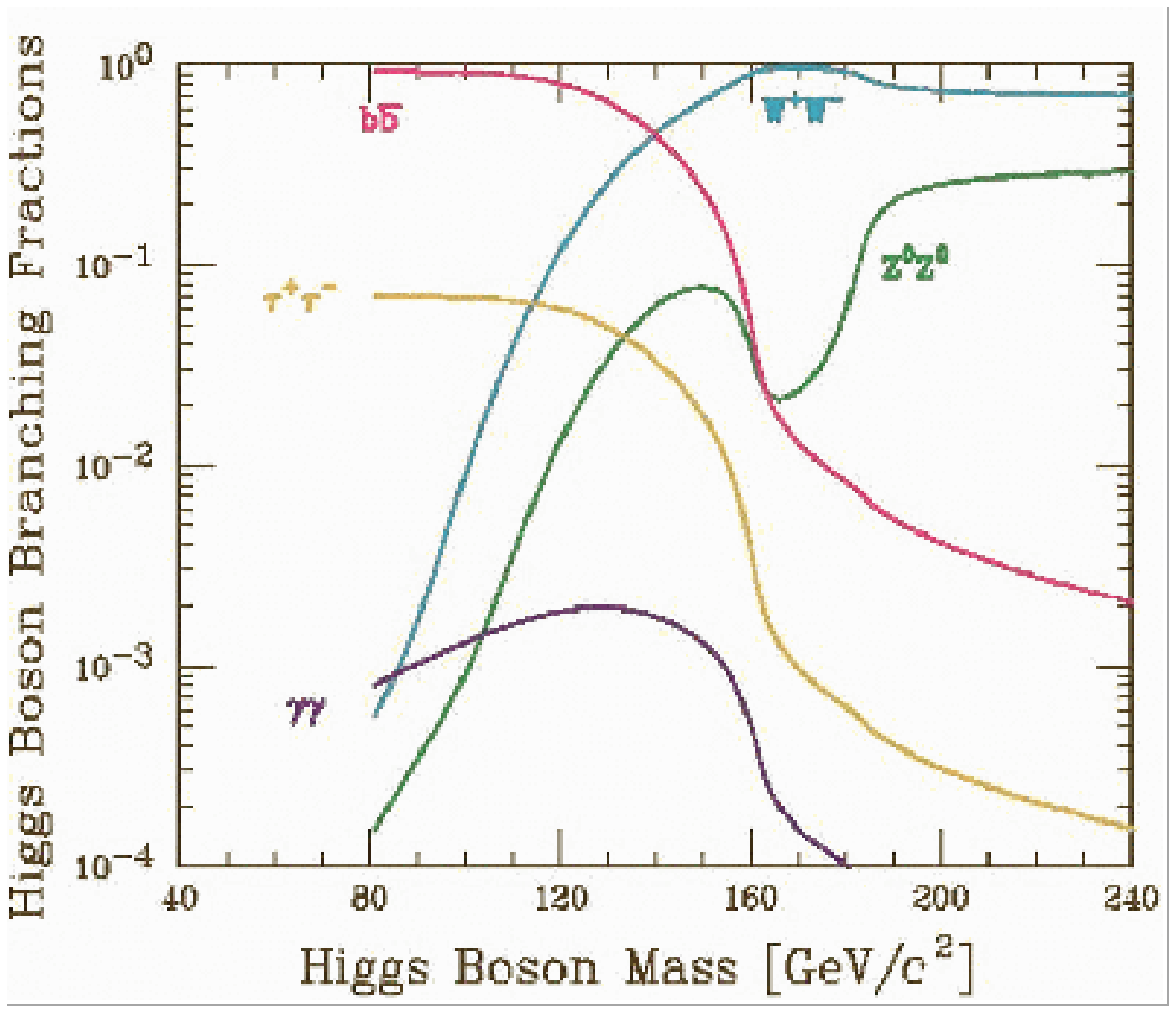}
\end{minipage}
\caption{
Left: expected SM Higgs boson production cross sections at the Tevatron (1.96~TeV).
Right: expected Higgs boson decay branching ratios for a SM Higgs boson masses up to 240~GeV. 
       At the Tevatron, $\rm \bb$ and WW decays are dominant, in addition the $\tautau$ decay
       mode has a significant contribution.
}
\label{fig:tev-xsec}
\end{figure}

The expected cross section and branching ratios are shown 
in Fig.~\ref{fig:tev-xsec} (from~\cite{xsec} and~\cite{br}) as a function of the Higgs boson
mass.
It is interesting to note that corresponding to the current collected data sample
of about 1~\fb\ about 1000 SM Higgs bosons of 120 GeV could have already 
been produced at each experiment.
For a SM  Higgs boson mass below about 200~GeV the decay width is below 1~GeV which is much
below the detector resolution. 

\subsection{b-Quark Tagging}

\begin{figure}[hp]
\vspace*{1cm}
\includegraphics[width=0.32\textwidth,height=5cm]{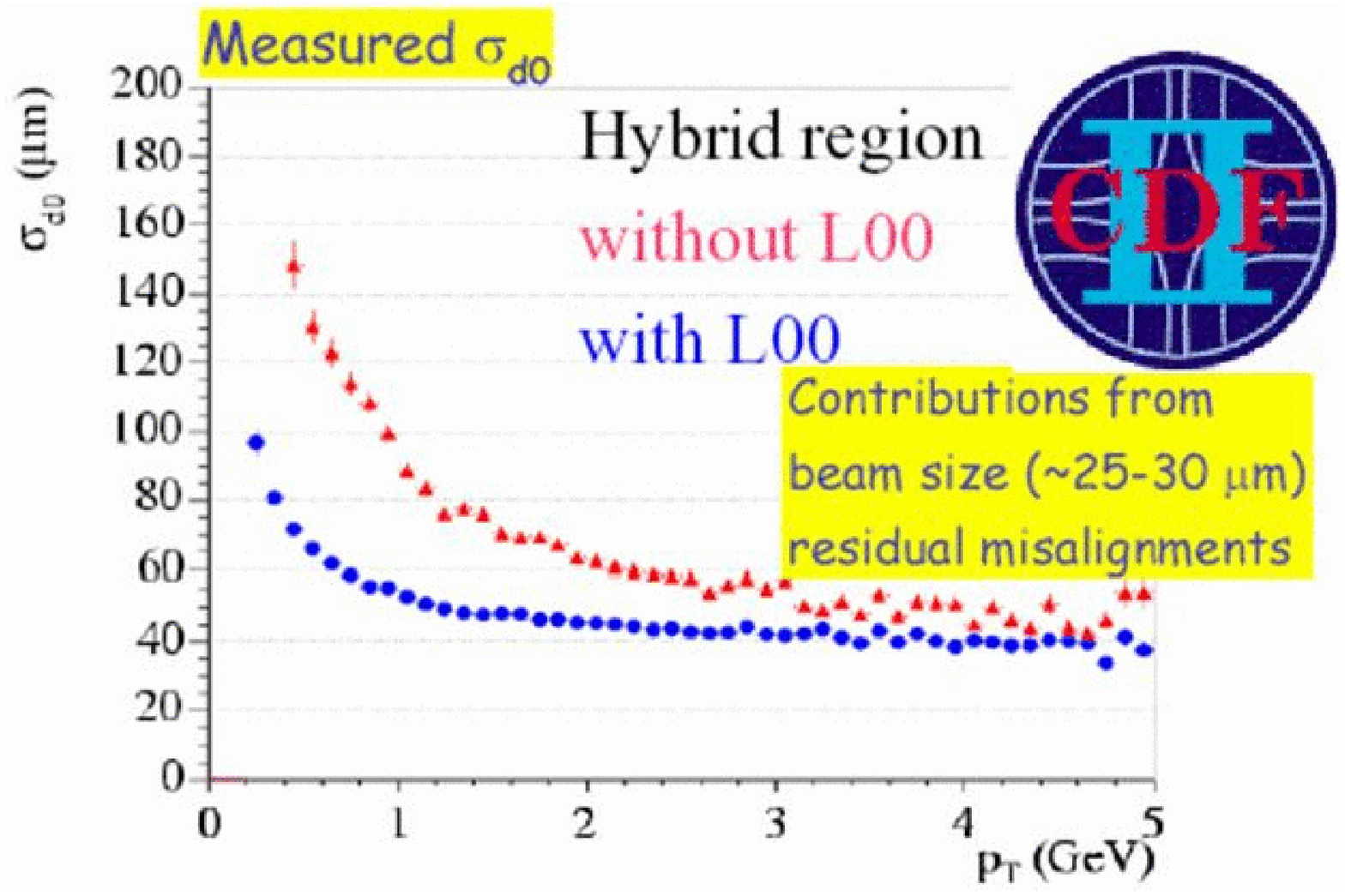}\hfill
\includegraphics[width=0.32\textwidth,height=5cm]{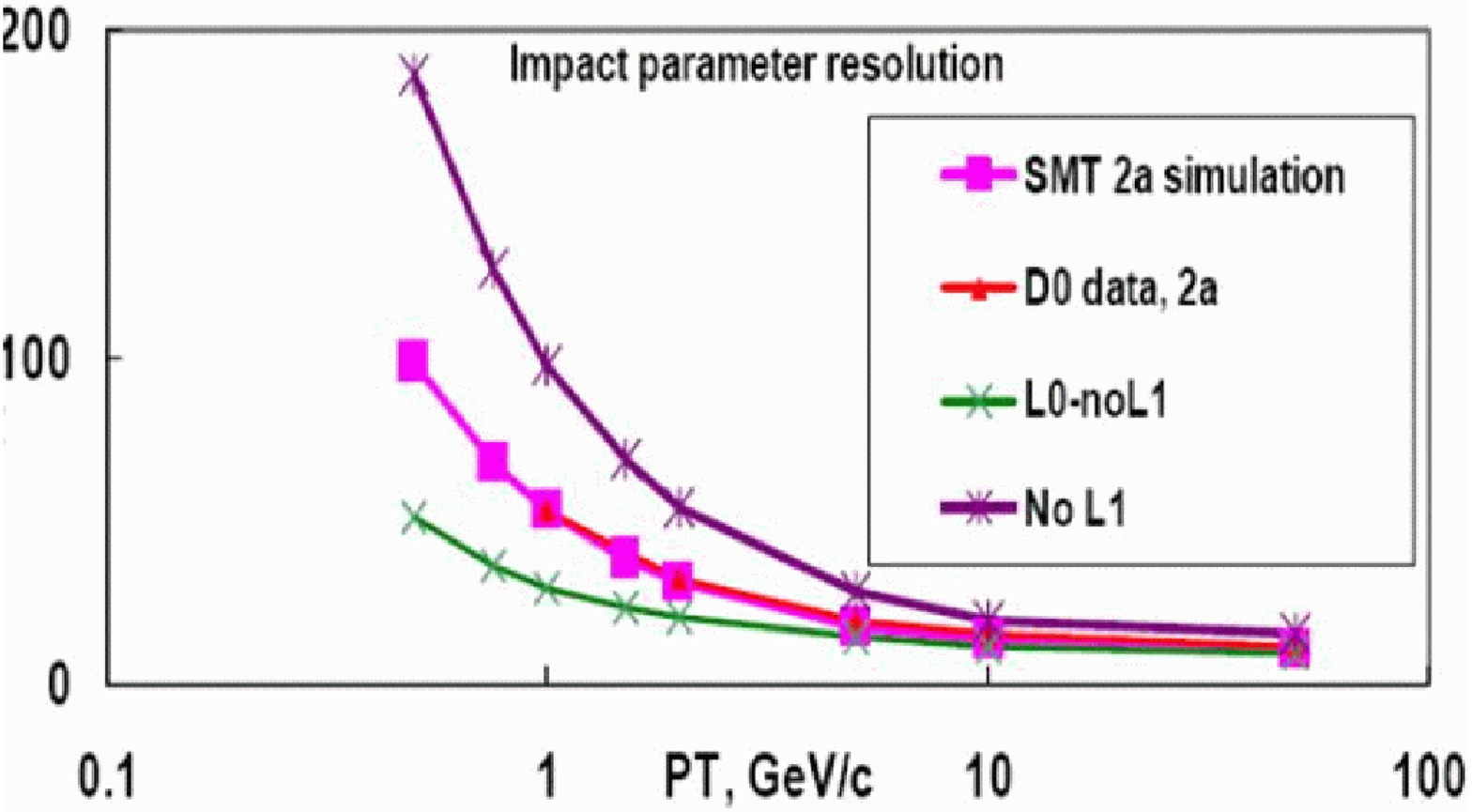} \hfill
\includegraphics[width=0.32\textwidth,height=5cm]{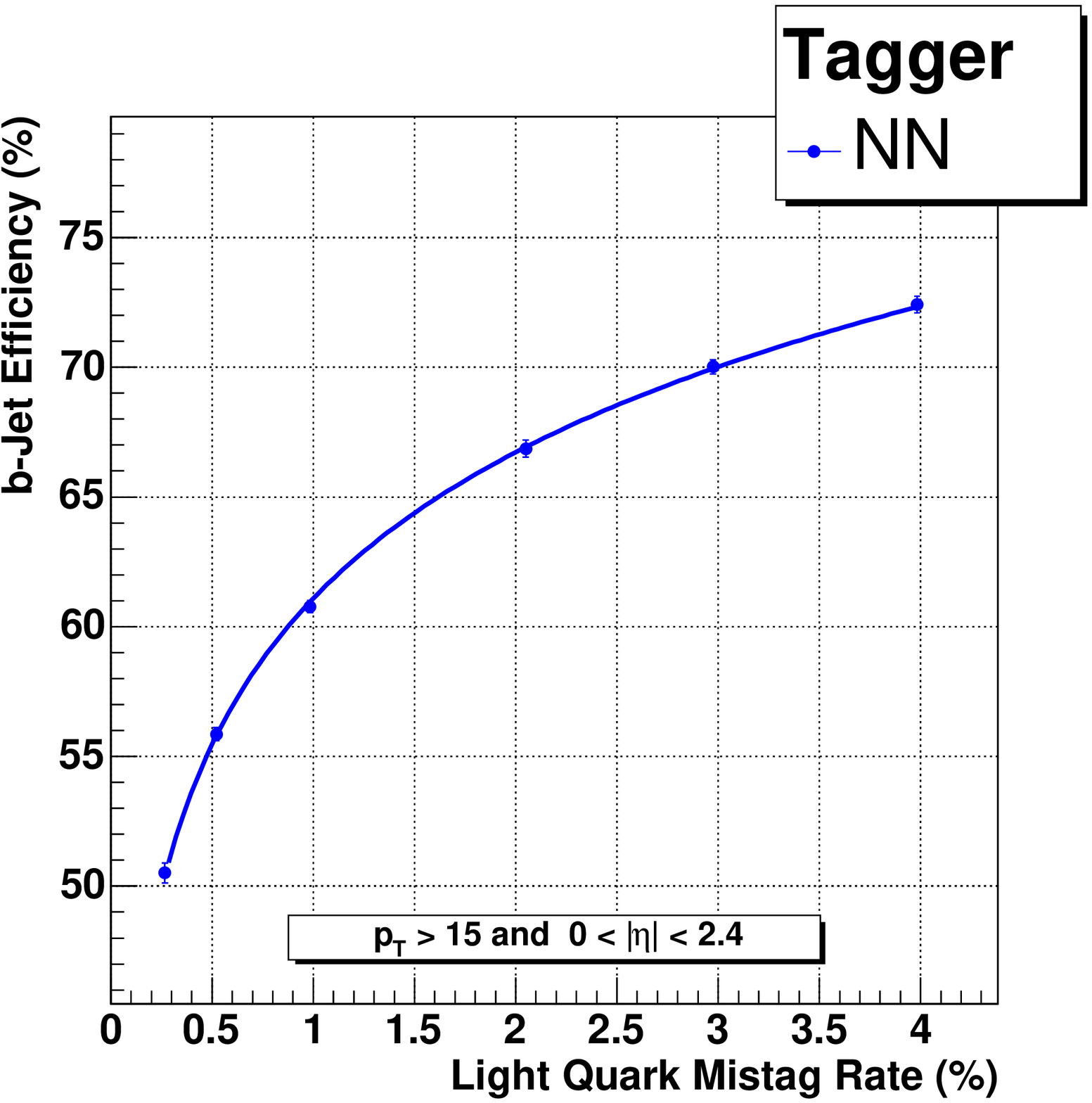}
\put(-285,120){\tiny $\sigma$(IP)}
\put(-285,110){\tiny ($\mu$m)}
%\vspace*{-0.5cm}
\caption{
Left: CDF impact parameter resolution as a function of $p_T$ for 
tracks traversing passive material in vertex detector, with (blue dots) and 
without (red triangles) use of L00 hits.
Center: D\O\ expected improved impact parameter resolution with the installation of a new
vertex detector layer (L0).
Right: D\O\ b-quark tagging performance.
}
\label{fig:vertex}
\end{figure}

\begin{figure}[htbp]
\vspace*{1cm}
\begin{minipage}{0.49\textwidth}
\includegraphics[width=\textwidth]{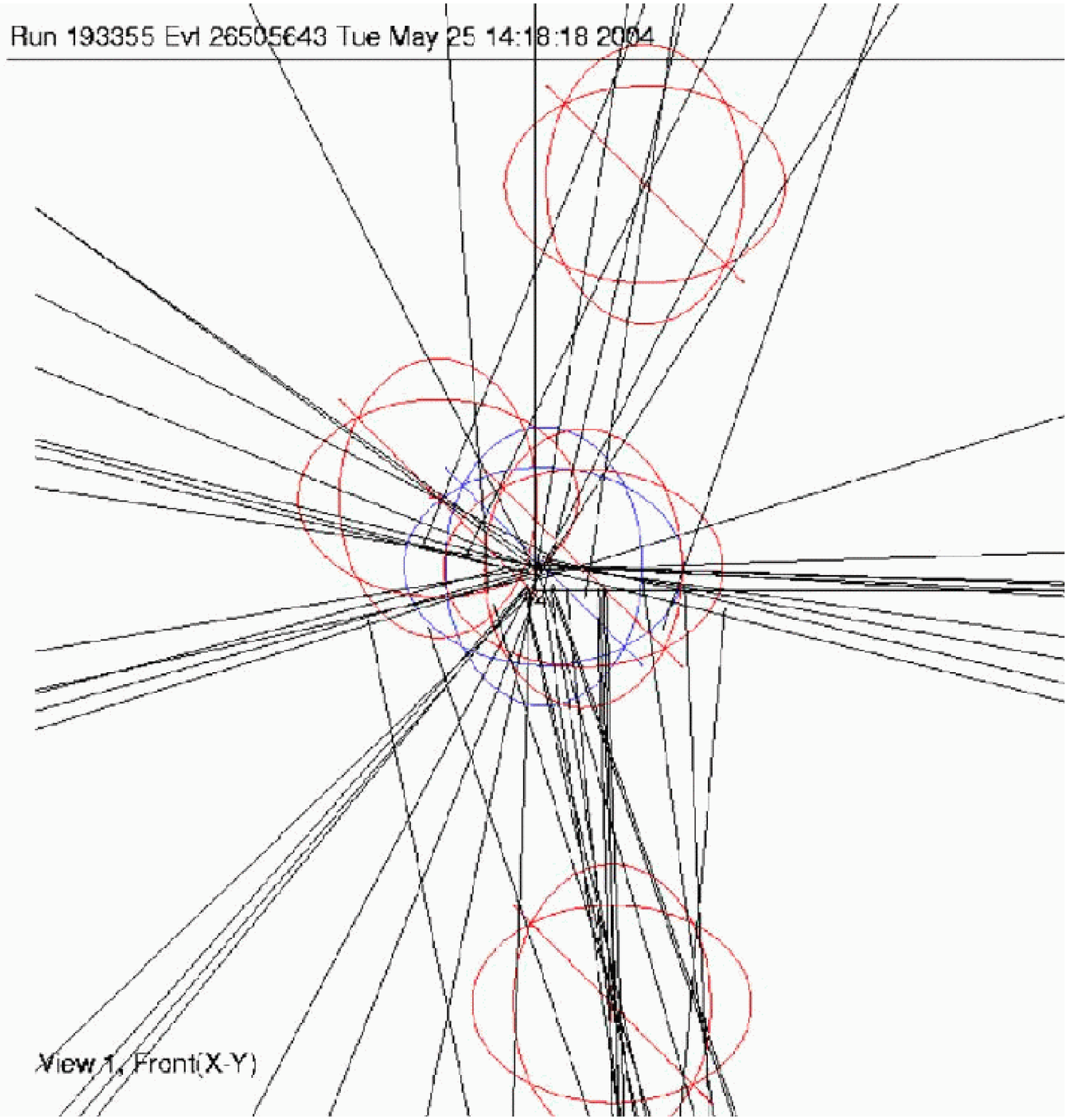} 
\end{minipage}\hfill
\begin{minipage}{0.49\textwidth}
\includegraphics[width=\textwidth,height=6cm]{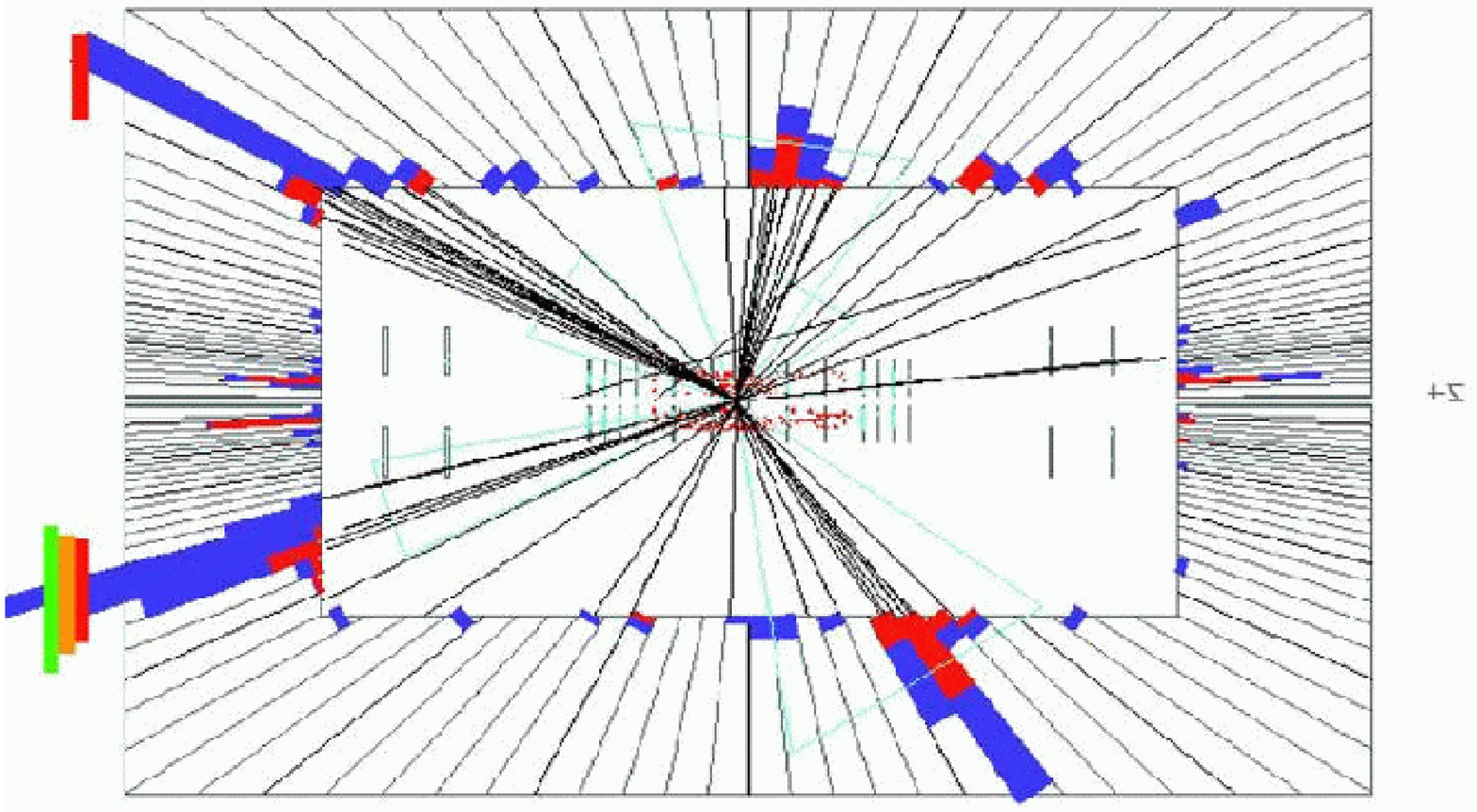} \\
\includegraphics[width=\textwidth,height=6cm]{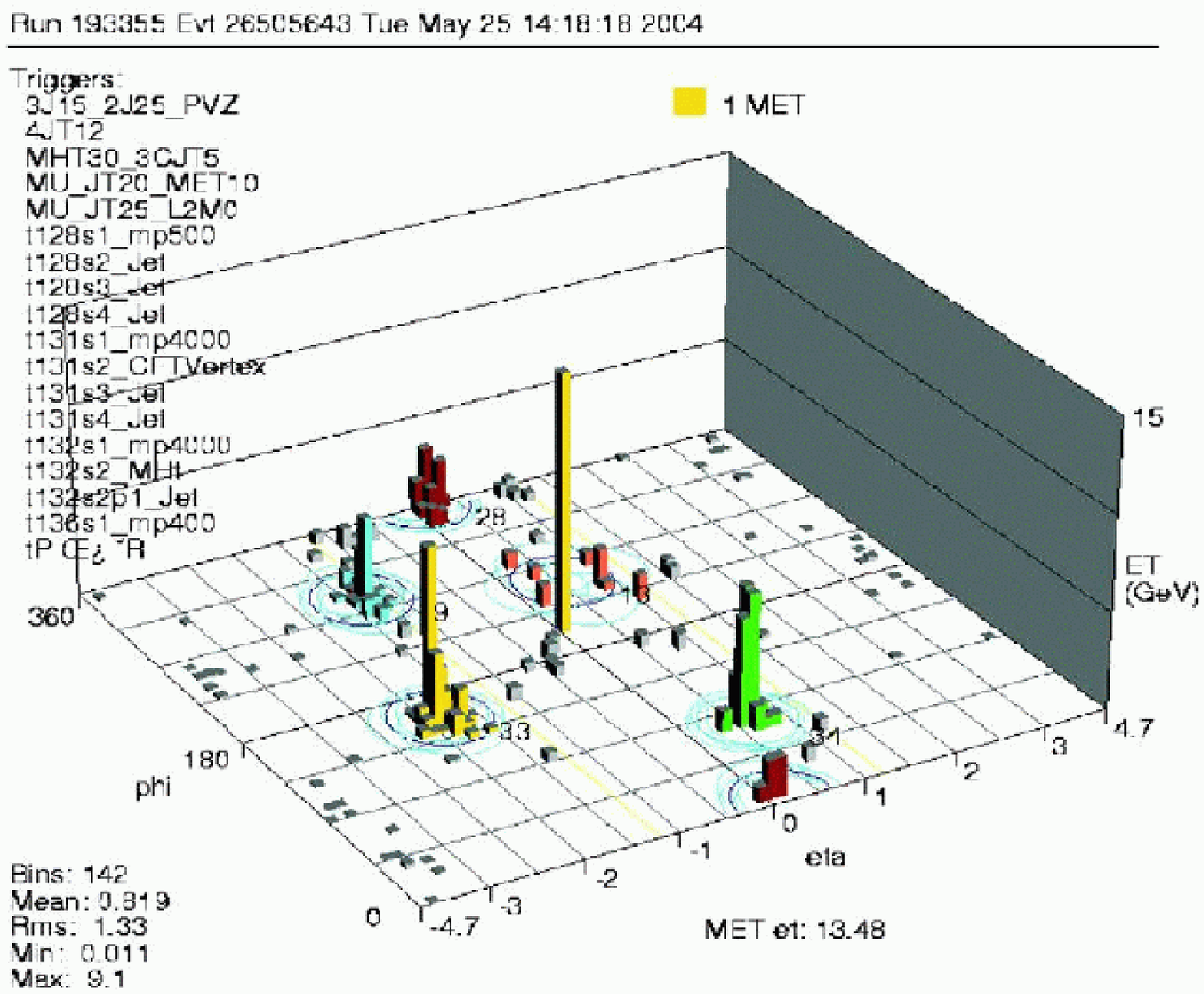} 
\end{minipage}
%\vspace*{-0.2cm}
\caption{D\O\ example of b-tagged event. Left: reconstructed tracks near the
interaction point. 
Upper right: jets clearly visible in the calorimeter.
Lower right: jets well separated in the $\eta$ versus $\phi$ plane.}
\label{fig:d0-b-event}
\end{figure}

The b-tagging capabilities are most important for the low-mass
Higgs boson searches and 
a critical parameter is the impact parameter resolution of the vertex detector. 
The improvement of the impact parameter resolution with a sensitive layer very 
close to the interaction point is illustrated in Fig.~\ref{fig:vertex} 
(from~\cite{cdf-l00} and~\cite{d0-l0}).
In CDF this layer is called L00 and in D\O\ it is called L0.
The figure shows also the b-quark tagging performance~\cite{d0-btag}.
An example of a quadruply b-tagged event is shown in 
Fig.~\ref{fig:d0-b-event} (from~\cite{d0-b-event}).

\begin{figure}[bp]
%\vspace*{-0.8cm}
\includegraphics[width=0.48\textwidth,height=6cm]{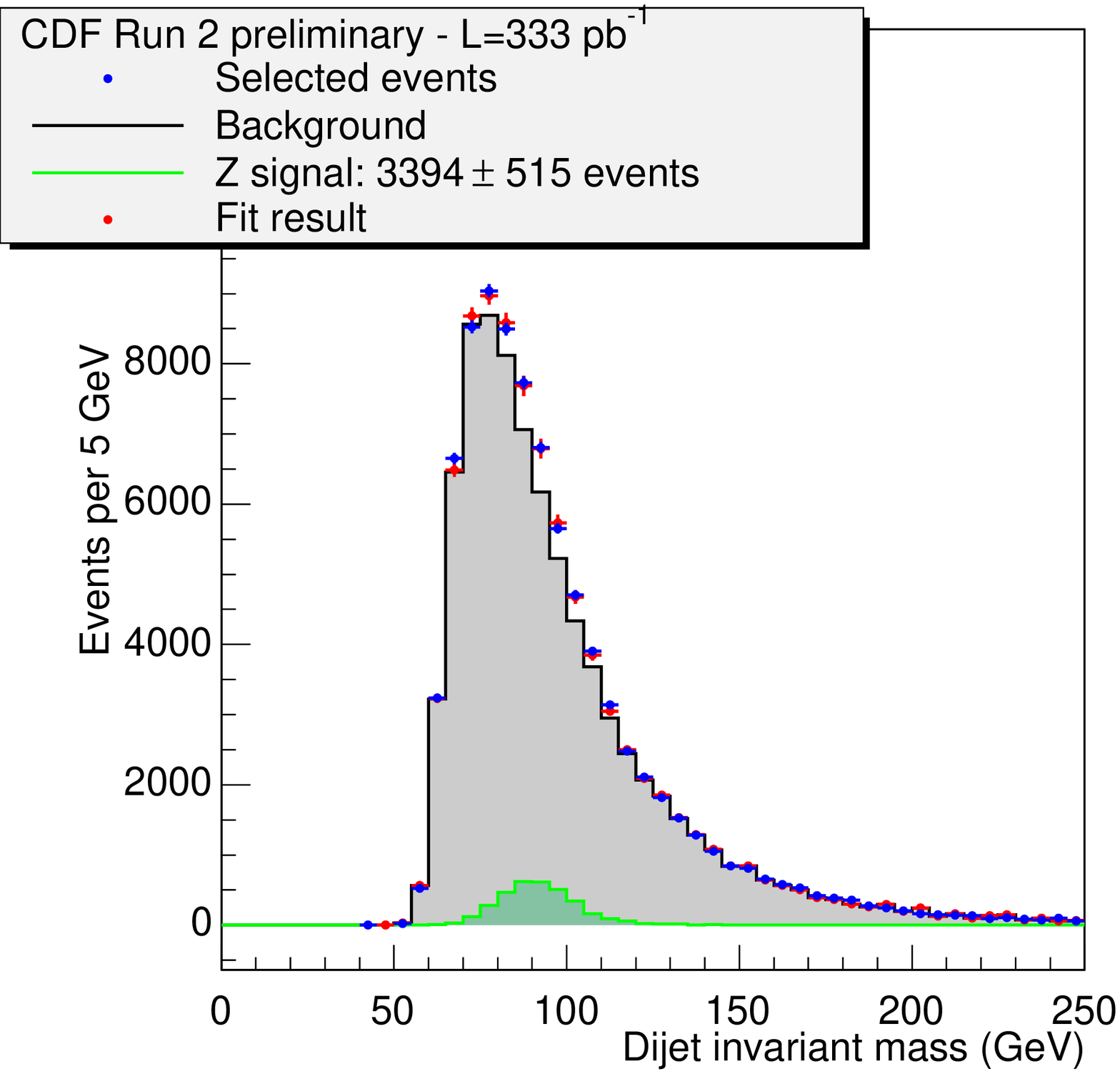} \hfill
\includegraphics[width=0.48\textwidth,height=6cm]{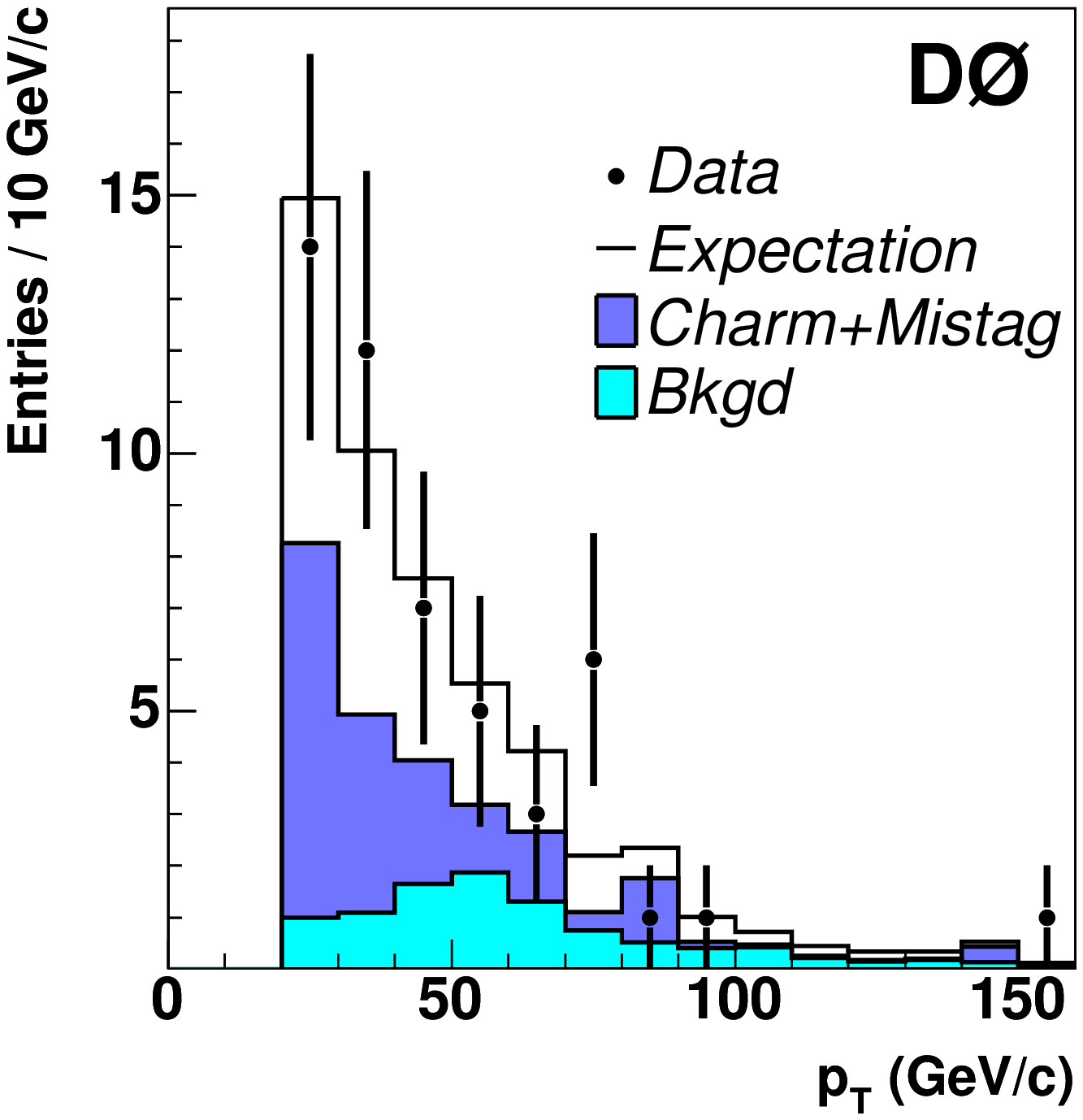}
%\vspace*{-0.5cm}
\caption{
Left: CDF $\rm Z\to \bb$ reconstructed invariant mass for $\rm H\to \bb$ searches.
Right: D\O\ $P_t$ distribution for b-tagged jets of $\rm Z+jets$ events.
}
\label{fig:zbb}
\end{figure}

Efficient B hadron tagging has already been demonstrated in data with $\rm Z \to\bb$
events. These measurements also contribute to the energy resolution 
and energy scale determinations. 
Figure~\ref{fig:zbb} (from~\cite{cdf-gghbb} and~\cite{d0-gghbb}) shows the 
reconstruction of the $\rm Z\to \bb$ mass and 
the good agreement between data and simulation for b-tagged $\rm Z+jets$ events.

\clearpage
\subsection{Gluon Fusion $\rm gg\rightarrow H$}

\subsubsection{$\rm gg\rightarrow H$ ($ \rm H\to WW$)}

\begin{figure}[hcbp]
%\vspace*{-0.43cm}
\begin{minipage}{0.32\textwidth}
\includegraphics[width=\textwidth]{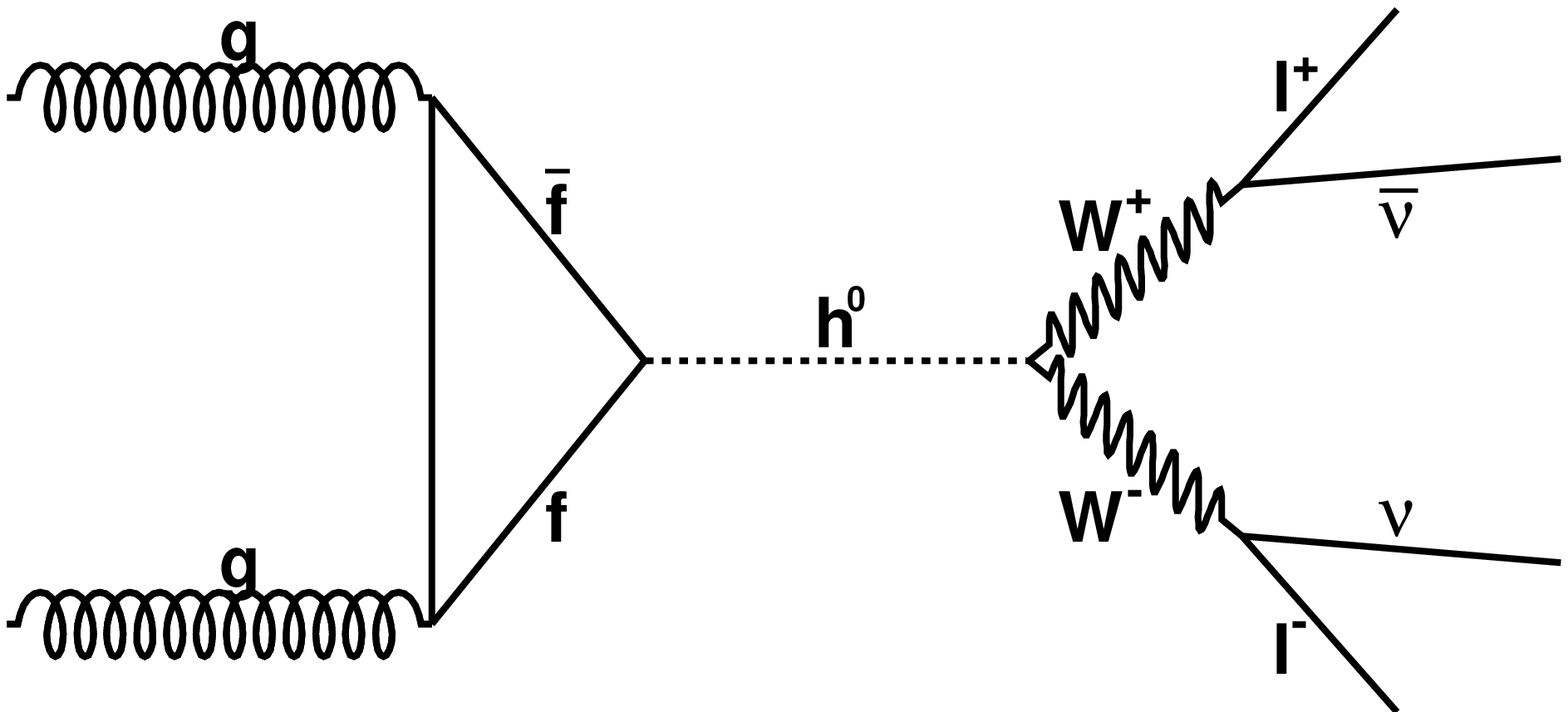} \hfill
\includegraphics[width=\textwidth]{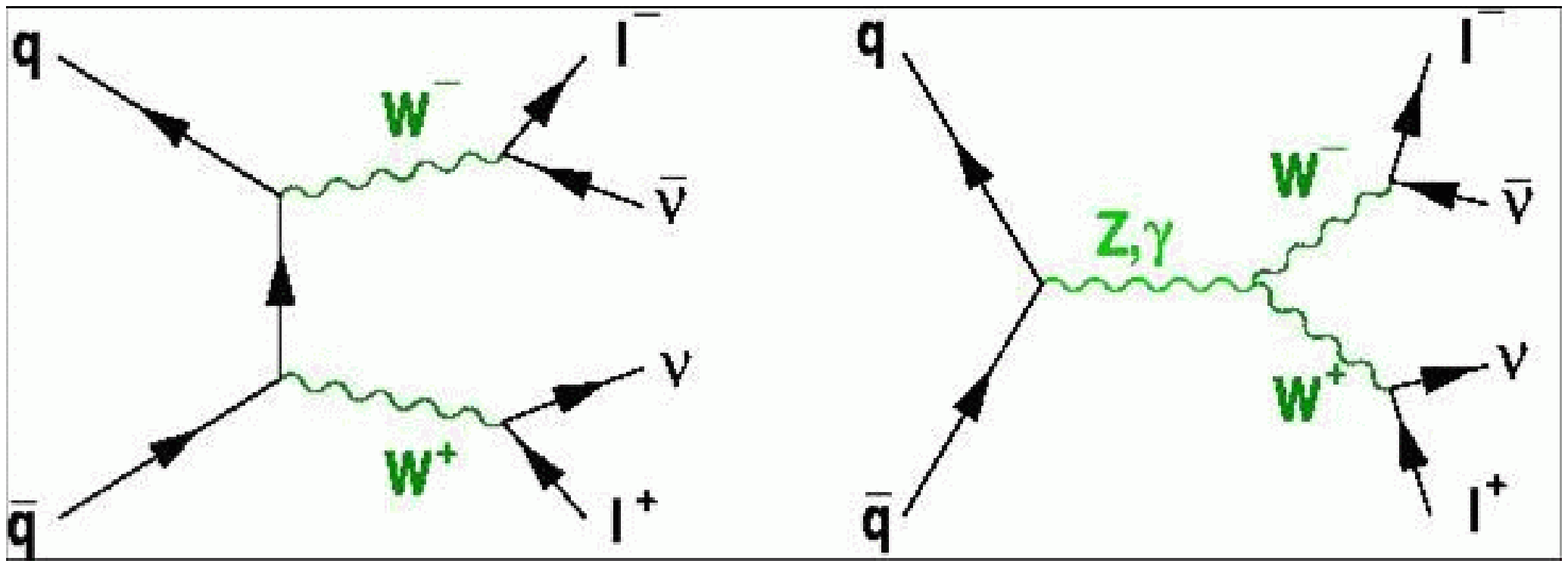}
\end{minipage}\hfill
\begin{minipage}{0.32\textwidth}
\includegraphics[width=0.6\textwidth]{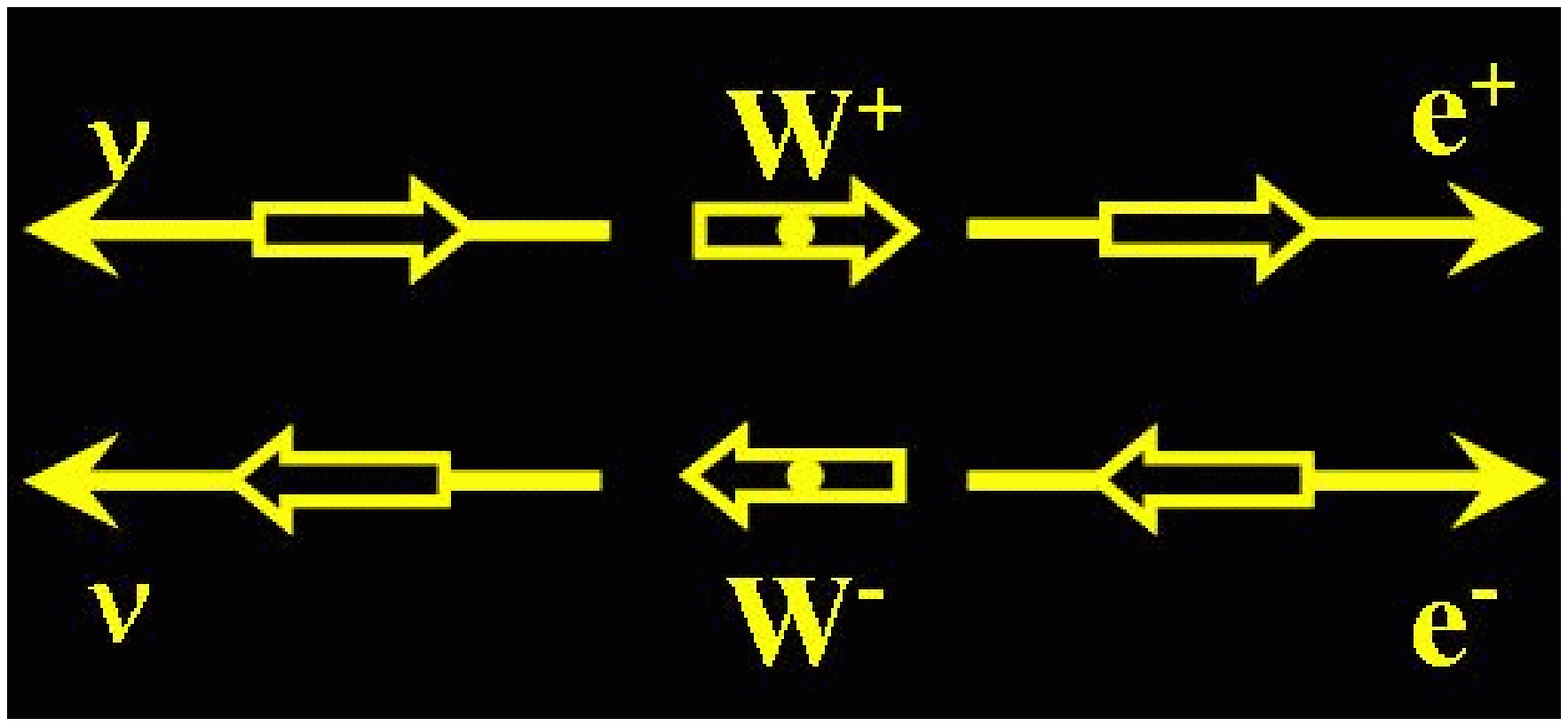}
\end{minipage}
\begin{minipage}{0.32\textwidth}
\includegraphics[width=\textwidth,height=4.8cm]{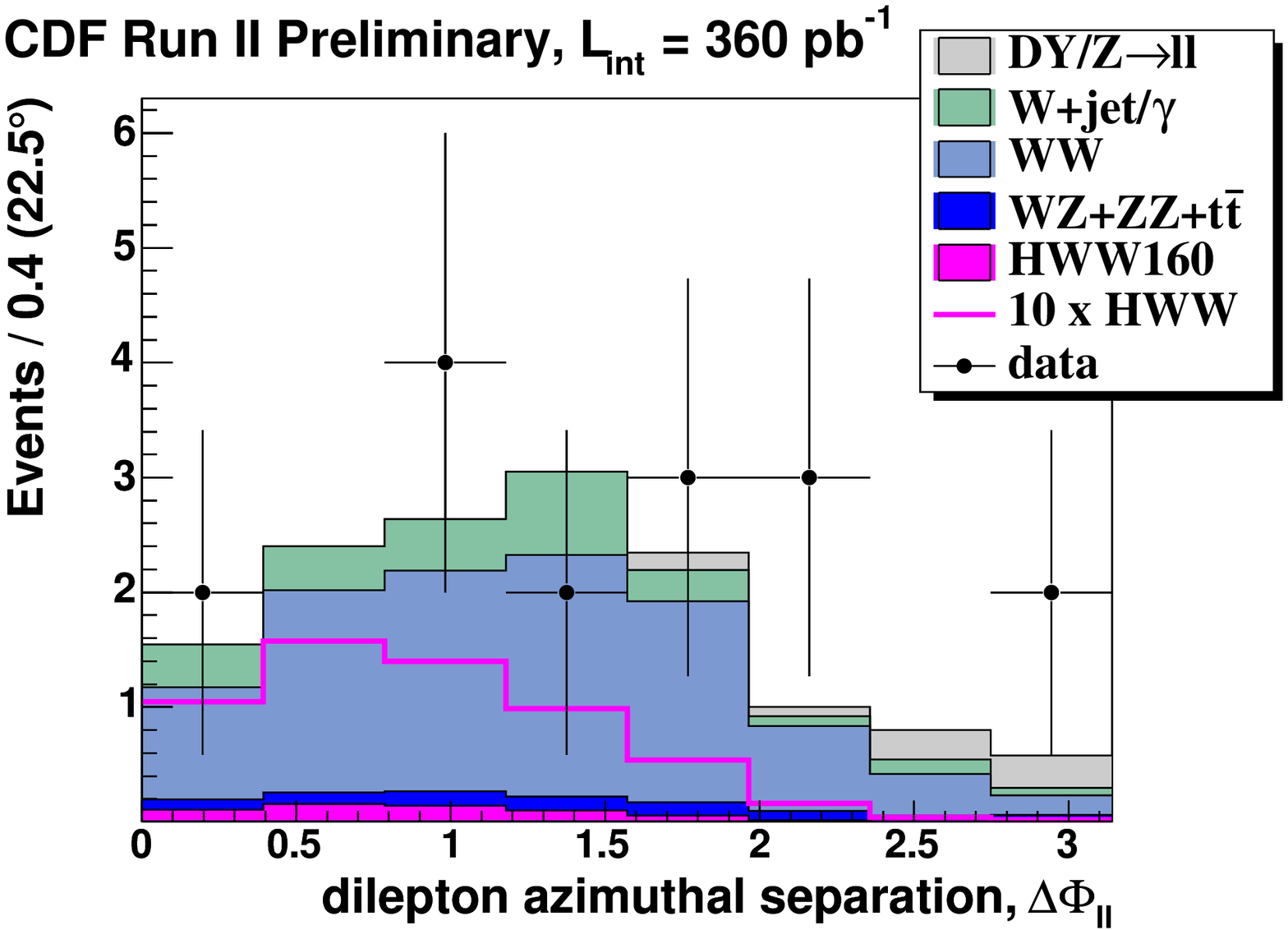}
\end{minipage}
%\vspace*{-0.35cm}
\caption{
Left: $\rm gg\rightarrow H$ ($\rm H\to WW$) signal and background processes.
Center: indication of spin correlations between final state leptons and W pairs, 
        which lead to different dilepton azimuthal angular ($\Delta \Phi_{\rm ll}$)
        distributions for signal and background.
Right: $\Delta \Phi_{\rm ll}$ distribution for data, and simulated signal and background. 
$\Delta \Phi_{\rm ll}$ is predicted to be smaller for the signal.
}
\label{fig:cdf-ggHww}
\end{figure}

\begin{figure}[hcp]
%\vspace*{-0.6cm}
\includegraphics[width=0.32\textwidth,height=5cm]{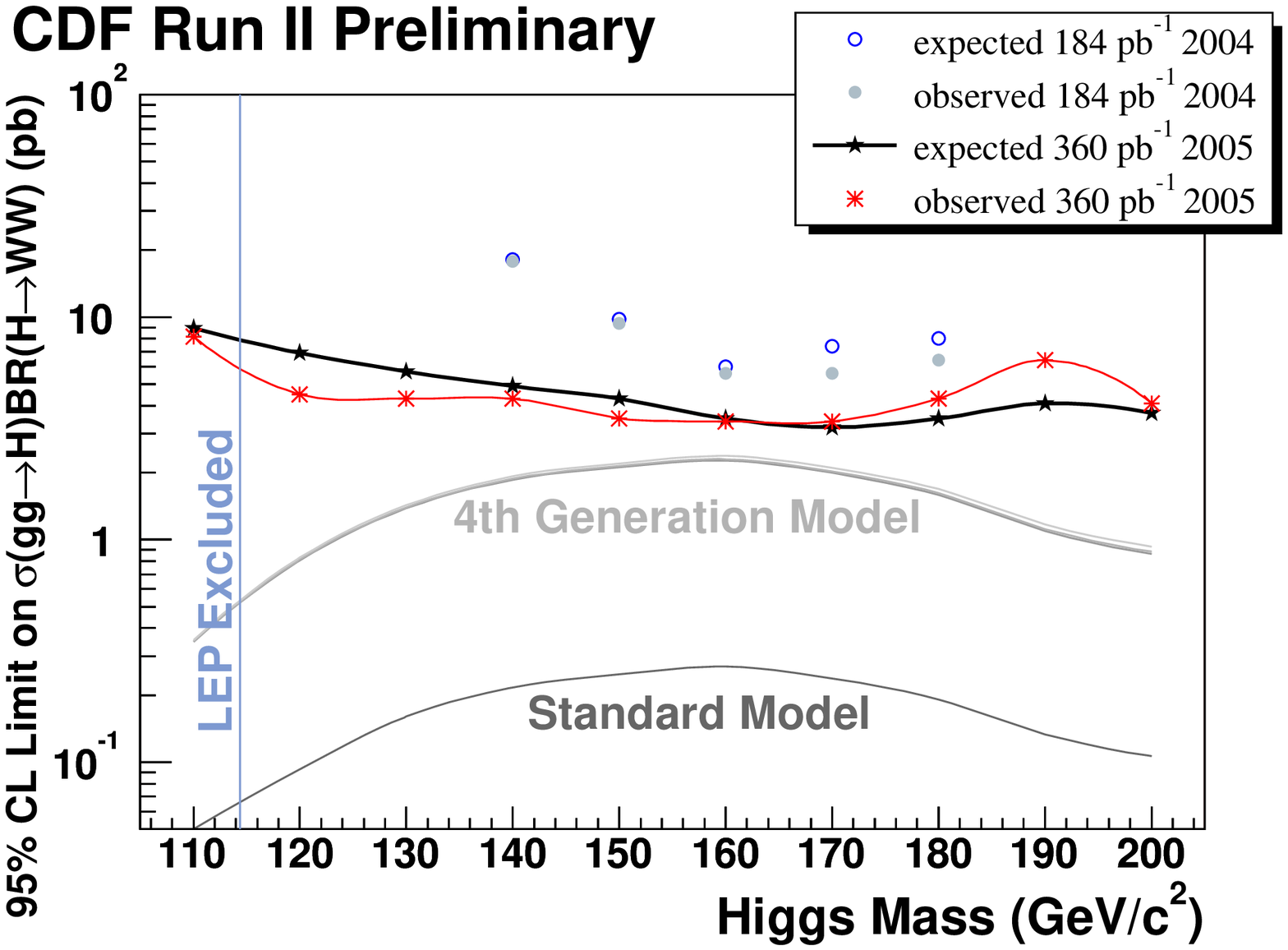}\hfill
\includegraphics[width=0.32\textwidth,height=5cm]{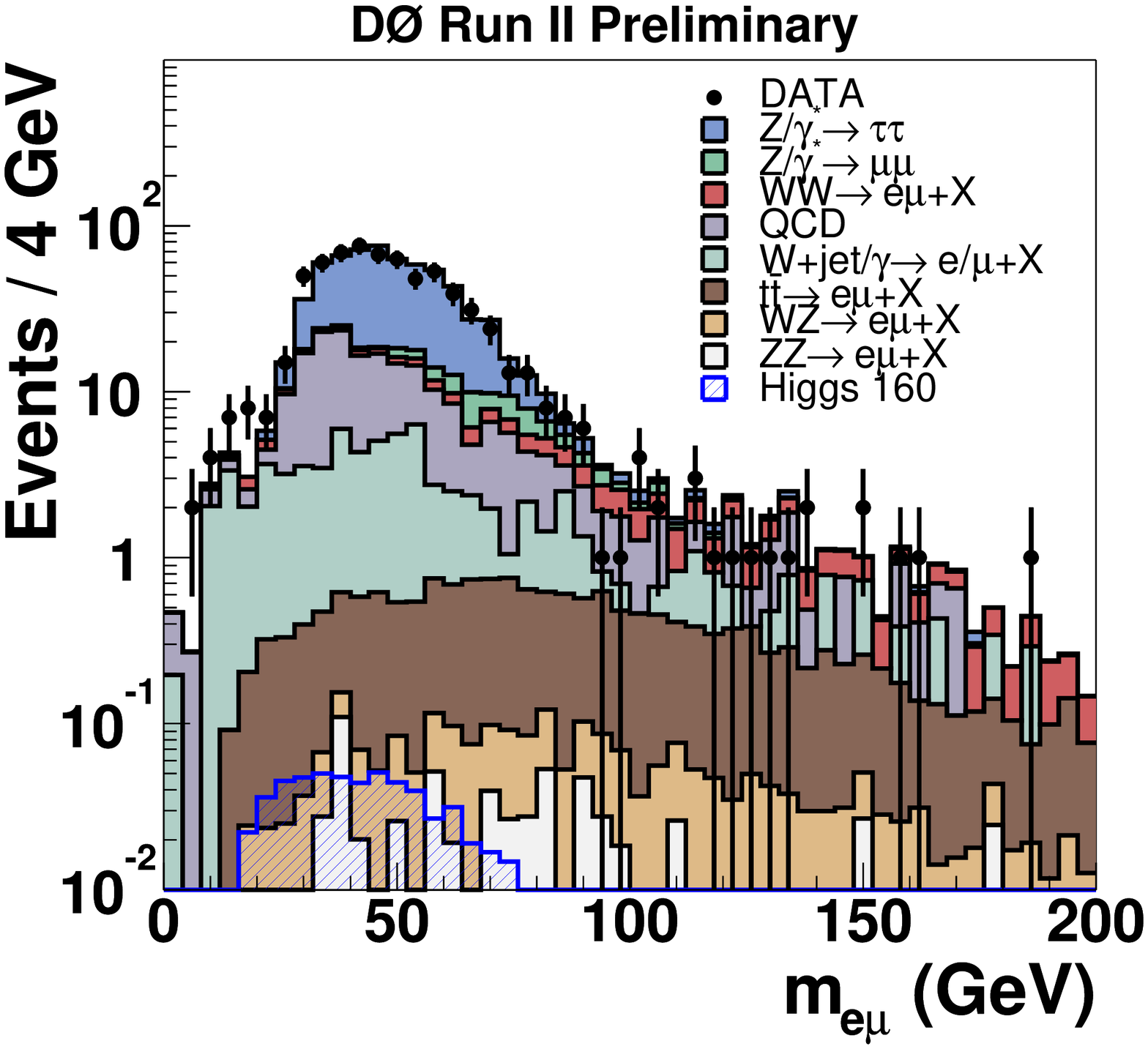} \hfill
\includegraphics[width=0.32\textwidth,height=5cm]{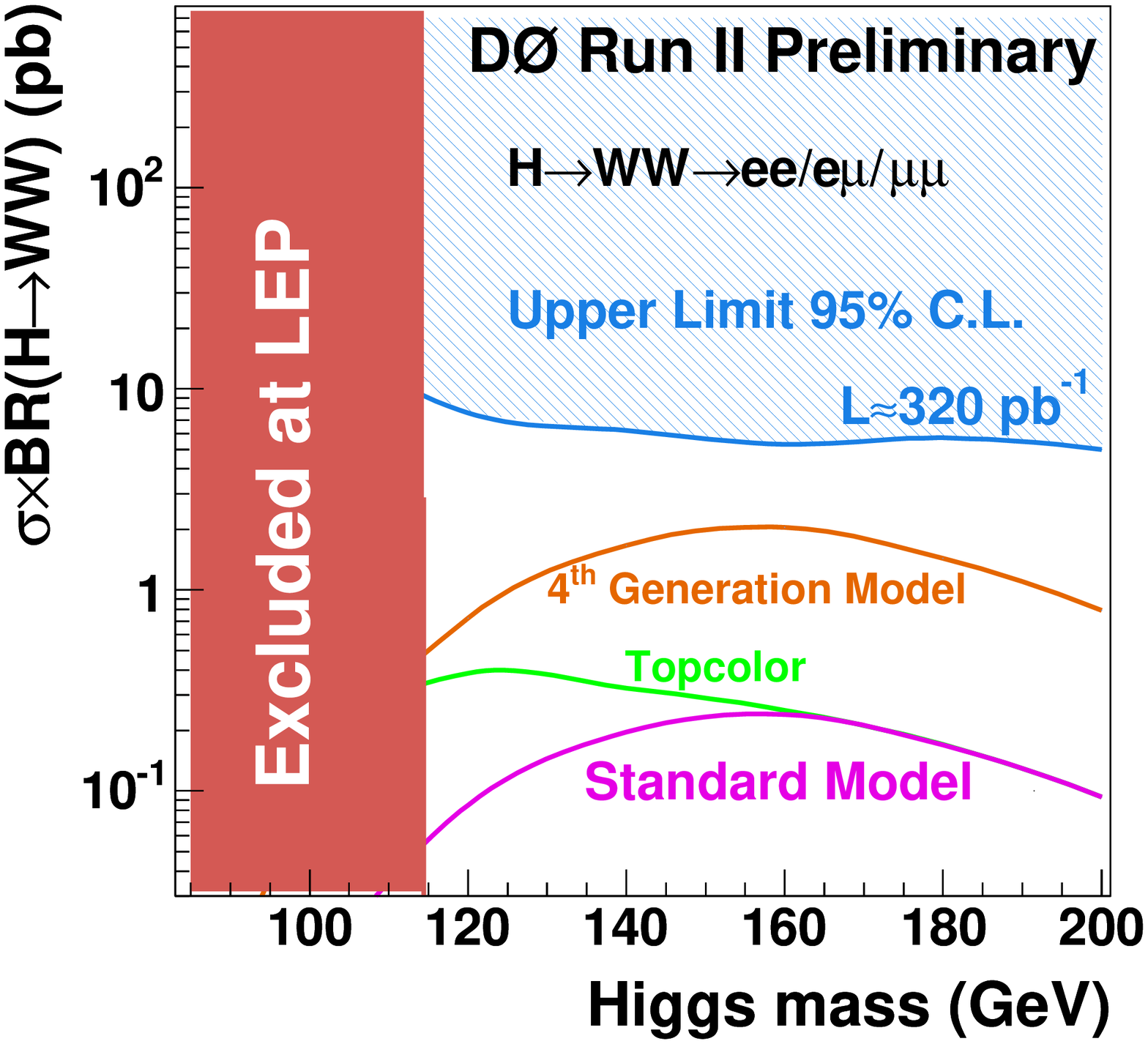}
%\vspace*{-0.5cm}
\caption{Left: CDF limits on the cross section of the $\rm gg\rightarrow H$ ($\rm H\to WW$) process.
         Center: D\O\ reconstructed invariant mass of $\rm e\mu$ pairs.
         Right: D\O\ limits on the cross section of the $\rm gg\rightarrow H$ ($\rm H\to WW$) process.
}
\label{fig:d0-ggHww}
\vspace*{-0.7cm}
\end{figure}

For Higgs boson masses above about 135 GeV, the process ggH ($\rm H\to WW$) becomes important.
The production and decay process is illustrated in Fig.~\ref{fig:cdf-ggHww}, also shown
is a background process leading to the same final state particles. The spin information allows 
separation of signal and background. The angle between the opposite charged leptons $\Delta\Phi_{\rm ll}$
tends to be smaller for the signal than for the background as shown in 
Fig.~\ref{fig:cdf-ggHww} (from~\cite{cdf-ggHww}) and corresponding limits in Fig.~\ref{fig:d0-ggHww}.
The good understanding of the expected background and the limits on the Higgs boson
production cross section are shown in Fig.~\ref{fig:d0-ggHww} (from~\cite{d0-ggHww} and~\cite{cdf-ggHww}).
Owing to the overwhelming $\rm b \bar b$ background, the $\rm gg\rightarrow H$ ($ \rm H\to \bb$)
channel is not feasible at the Tevatron.

\clearpage
\subsection{Associated Production}

\subsubsection{WH ($\rm H\to \bb$)}

\begin{figure}[hbcp]
%\vspace*{-0.7cm}
\includegraphics[width=0.32\textwidth,height=6cm]{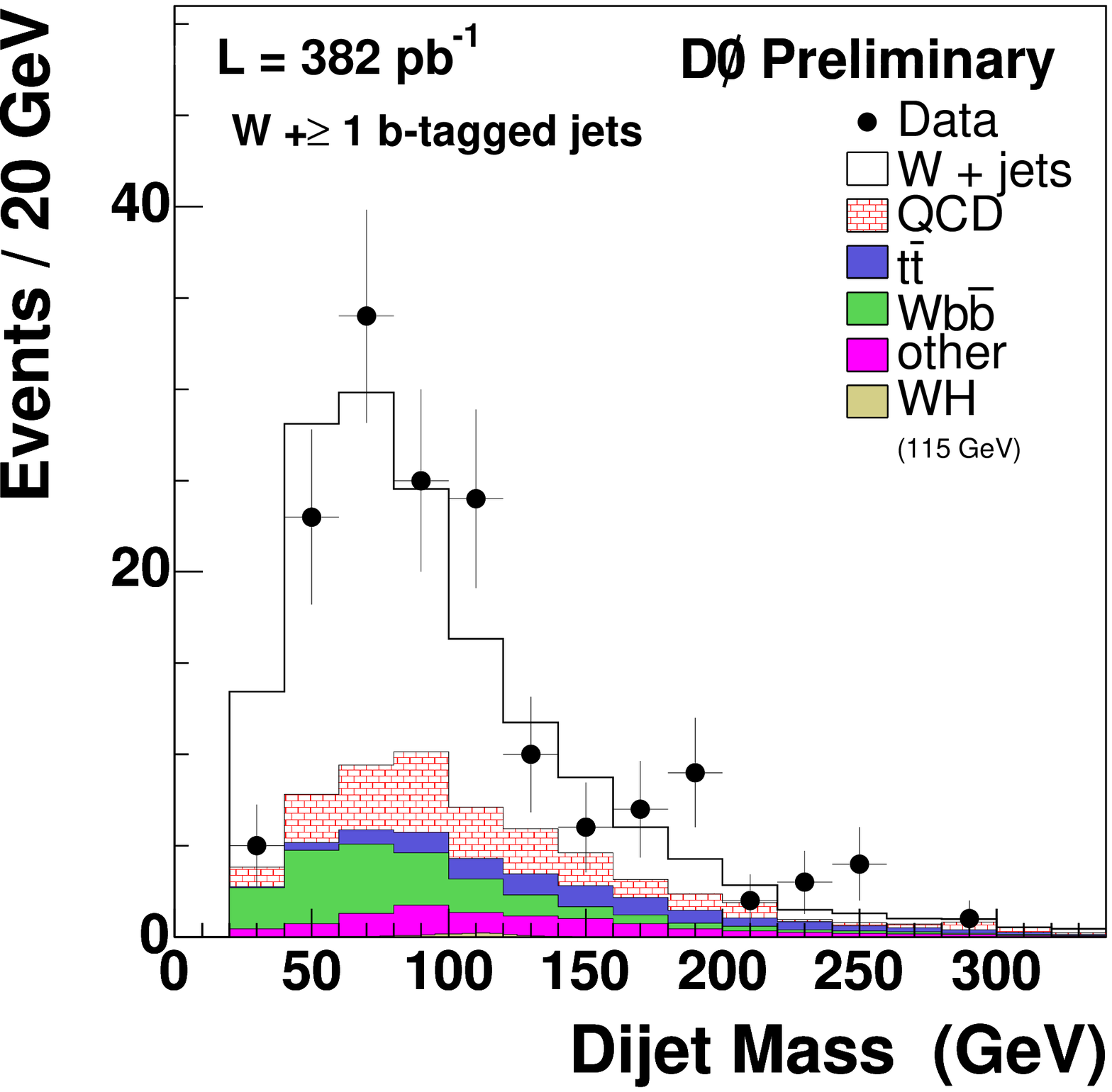} \hfill
\includegraphics[width=0.32\textwidth,height=6cm]{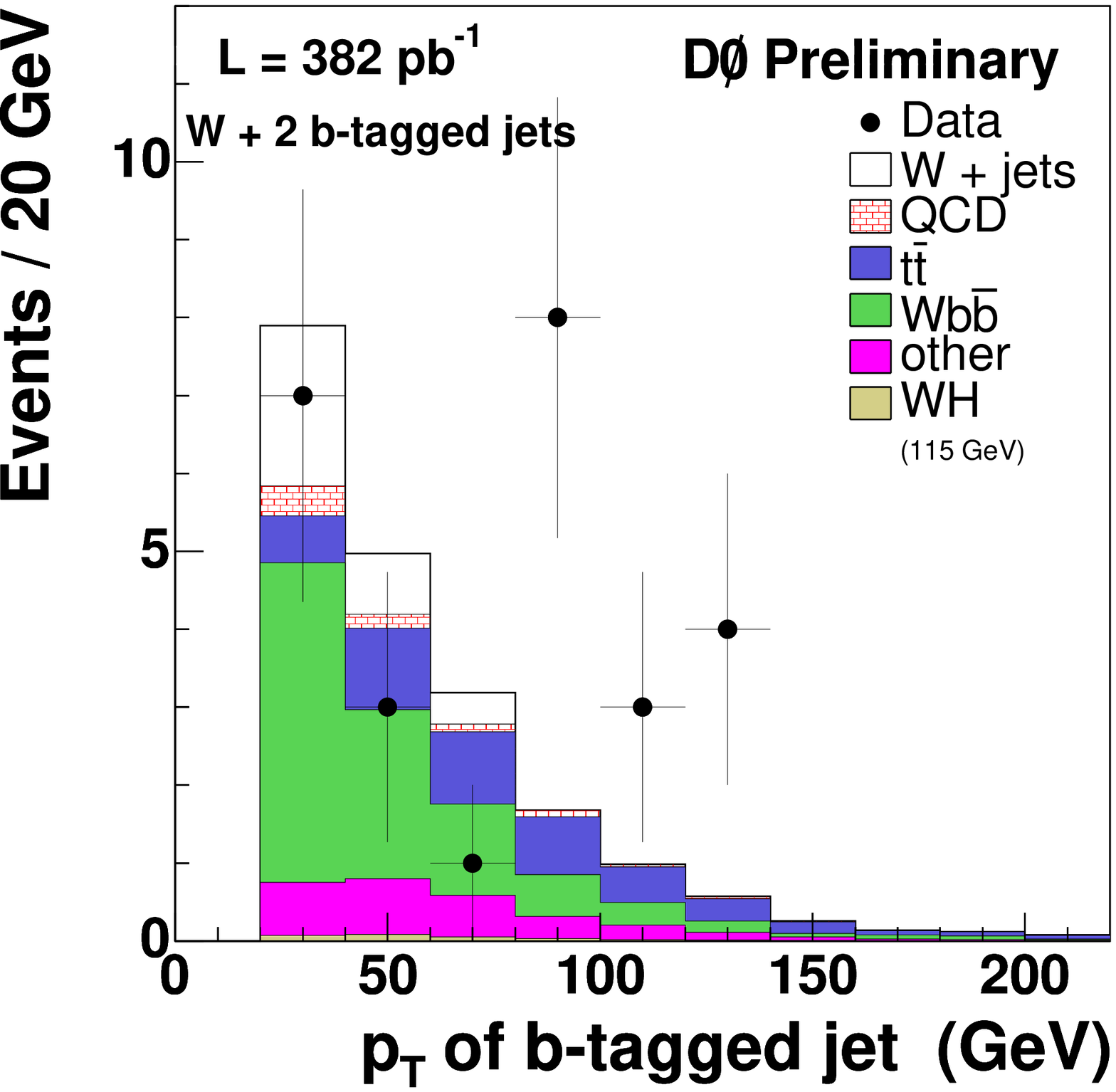} \hfill
\includegraphics[width=0.32\textwidth,height=6.3cm]{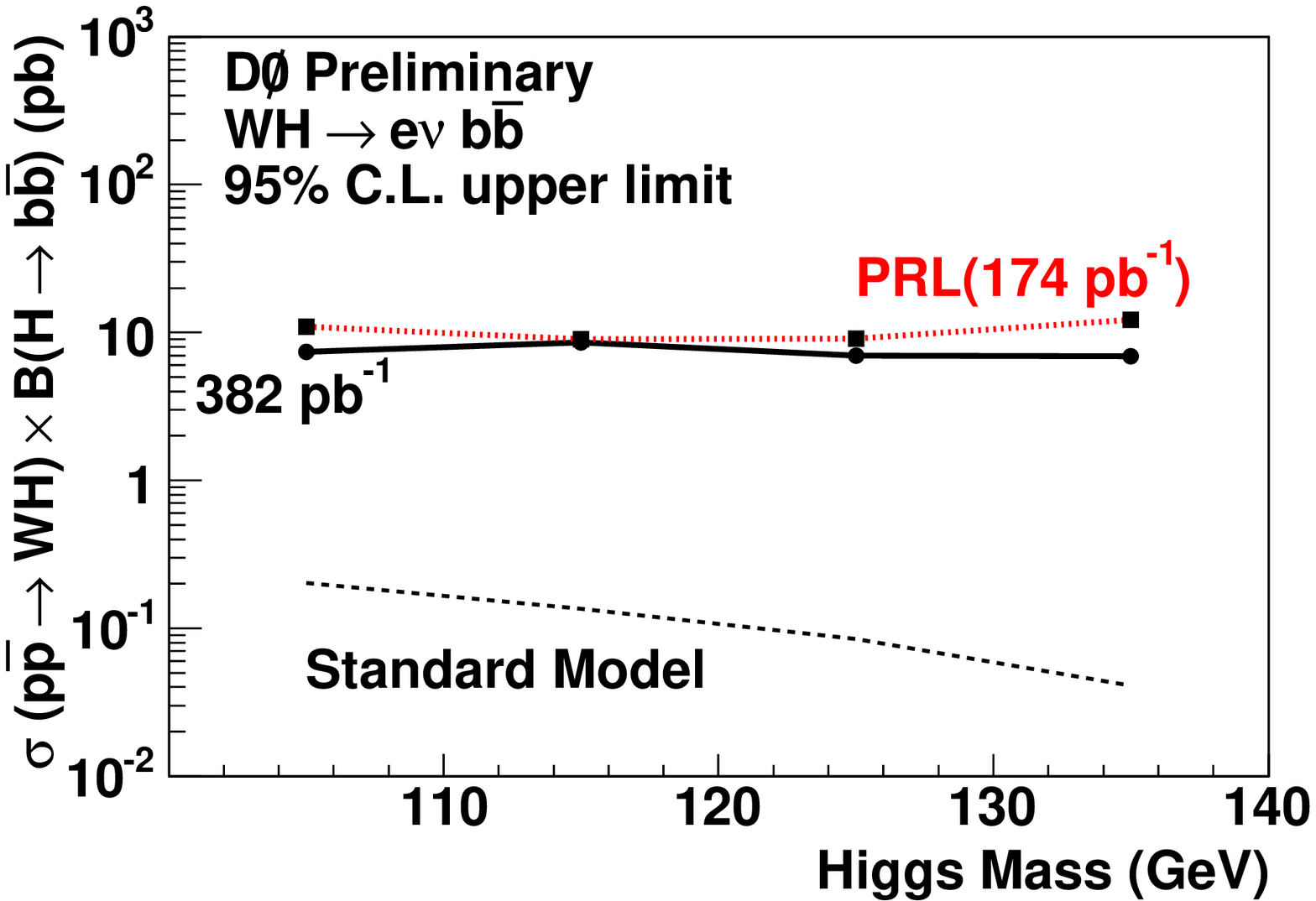}
%\vspace*{-0.3cm}
\caption{D\O\ WH ($\rm H\to \bb$).
Left: single b-tagging.
Center: double b-tagging.
Right: production cross section limit.}
\label{fig:d0-wHbb}
%\vspace*{-1cm}
\end{figure}

\begin{figure}[bp]
\vspace*{1cm}
\includegraphics[width=0.32\textwidth,height=6cm]{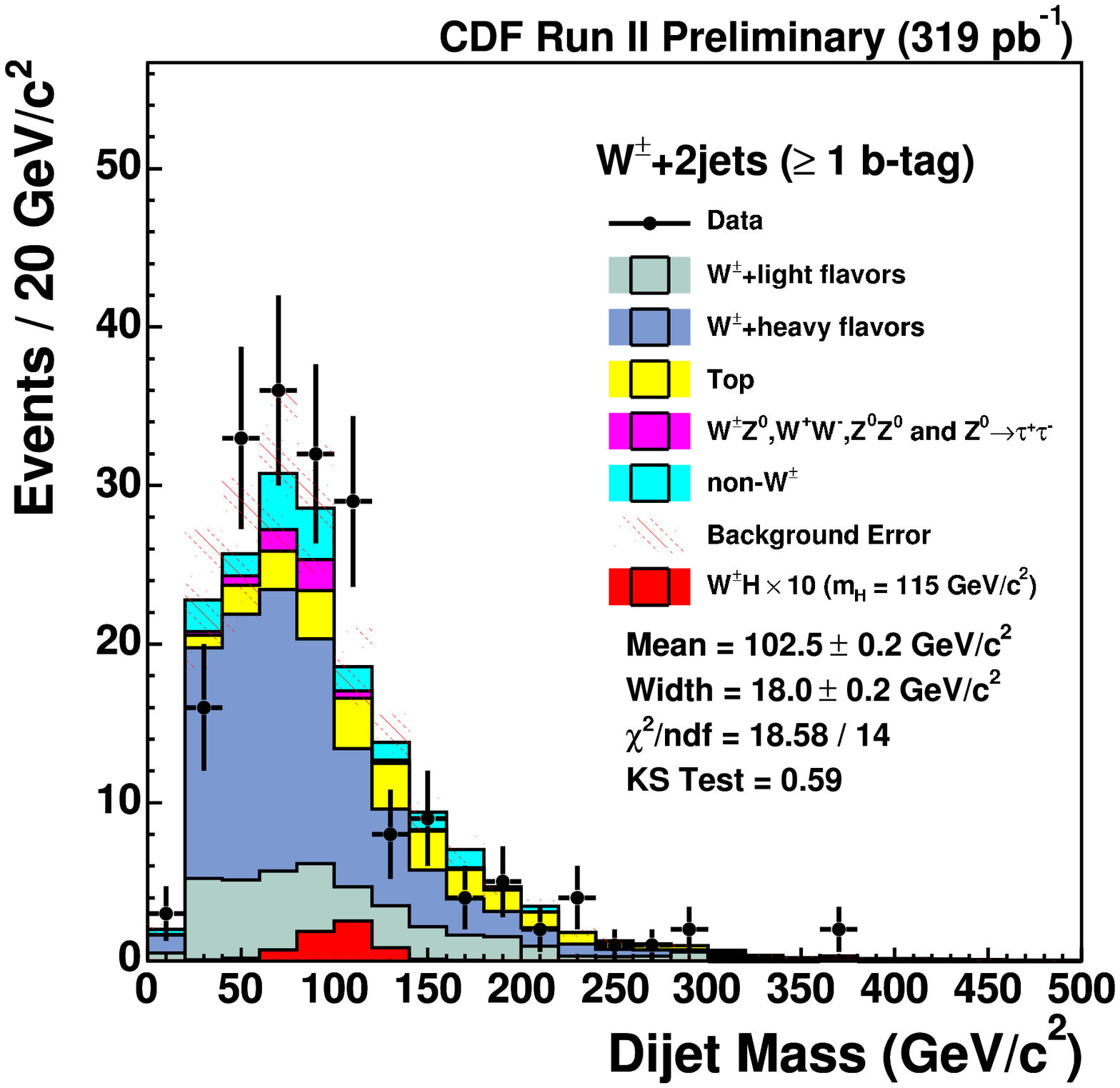} \hfill
\includegraphics[width=0.32\textwidth,height=6cm]{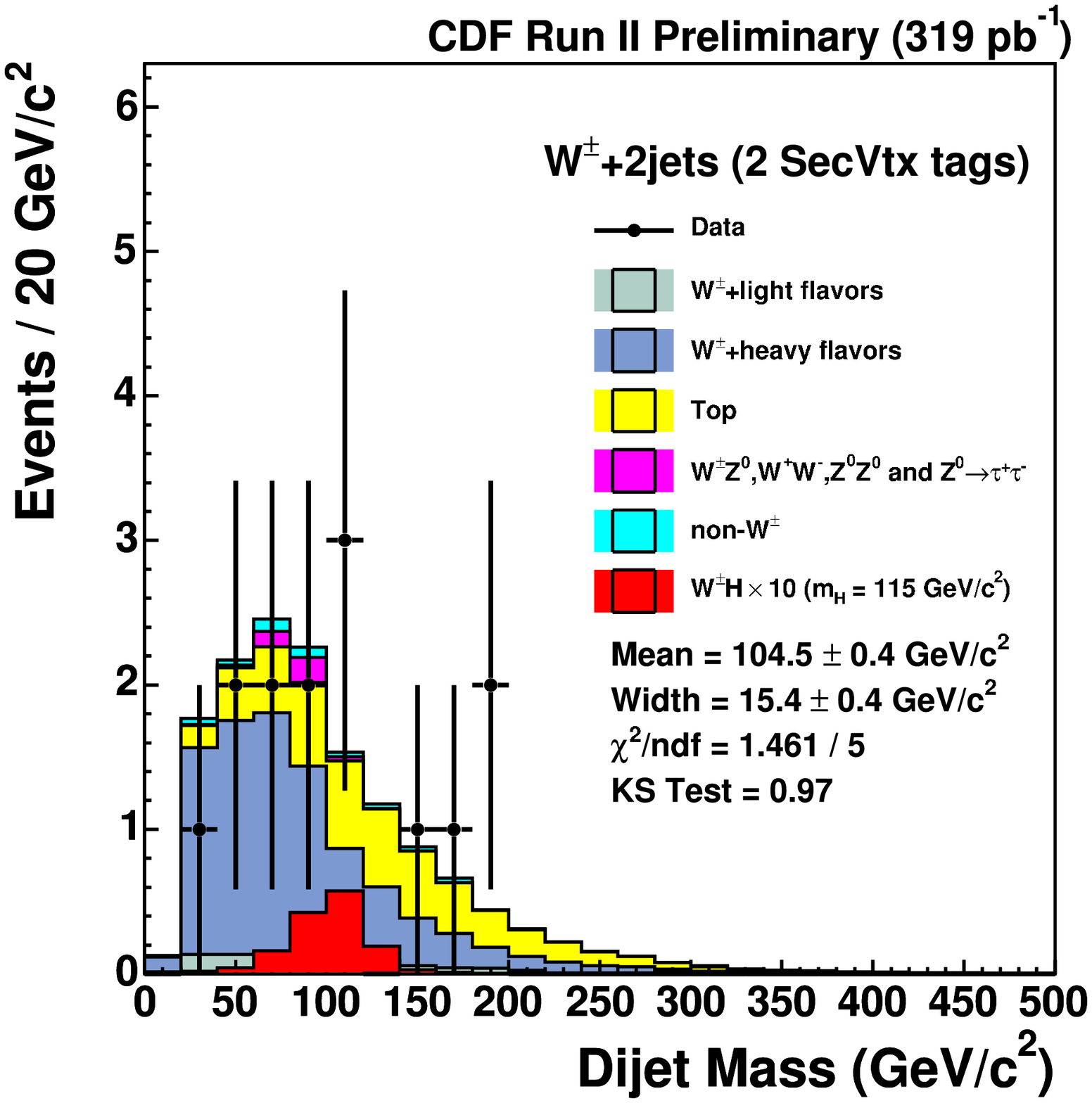} \hfill
\includegraphics[width=0.32\textwidth,height=6cm]{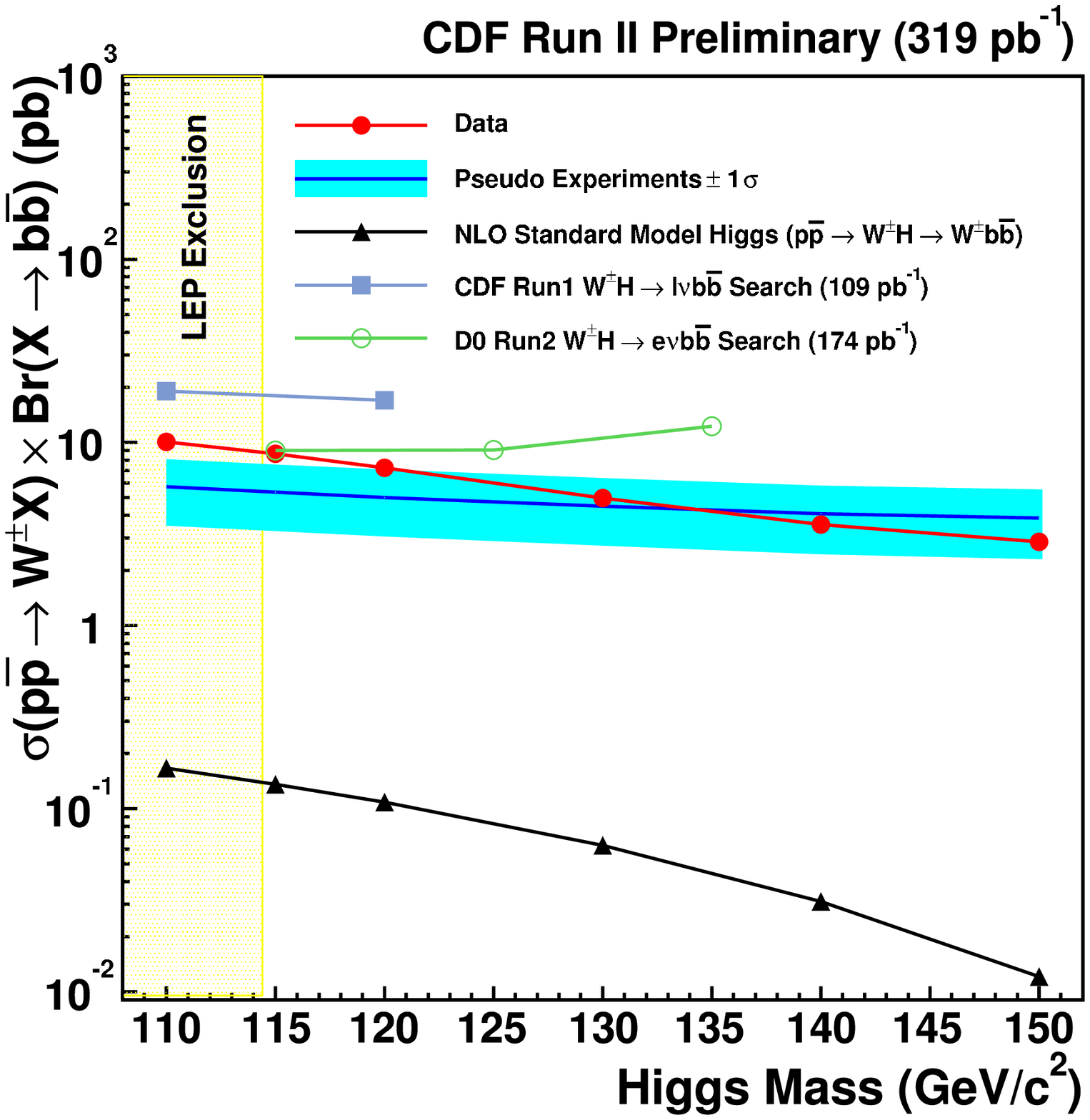}
\vspace*{-0.1cm}
\caption{
CDF WH ($\rm H\to \bb$).
Left: single b-tagging.
Center: double b-tagging.
Right: production cross section limit.
Note that the limit is based on the single b-tagged sample 
which has a slightly stronger expected limited.
}
\label{fig:cdf-wHbb}
\end{figure}

An important discovery channel is the reaction WH ($\rm H\to \bb$), where the
W decays either to $\rm e\nu$ or $\rm \mu\nu$. The tagging of two b-quarks
improves the signal to background ratio as shown in Fig.~\ref{fig:d0-wHbb}
(from~\cite{d0-WHbb}). Similar results are obtained from CDF, as shown in
Fig.~\ref{fig:cdf-wHbb} (from~\cite{cdf-WHbb}).

\subsubsection{WH ($\rm H\to WW$)}

\begin{figure}[hp]
\vspace*{0.2cm}
\includegraphics[width=0.66\textwidth,height=6cm]{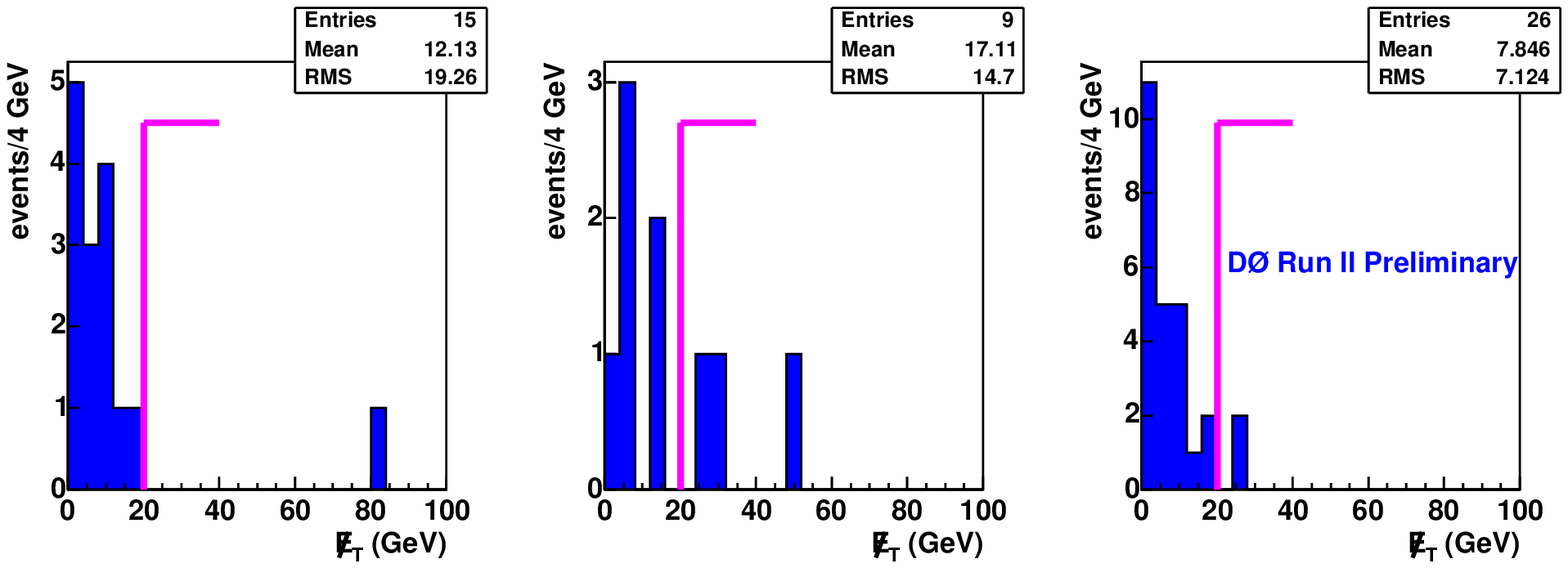} \hfill
\includegraphics[width=0.33\textwidth,height=6cm]{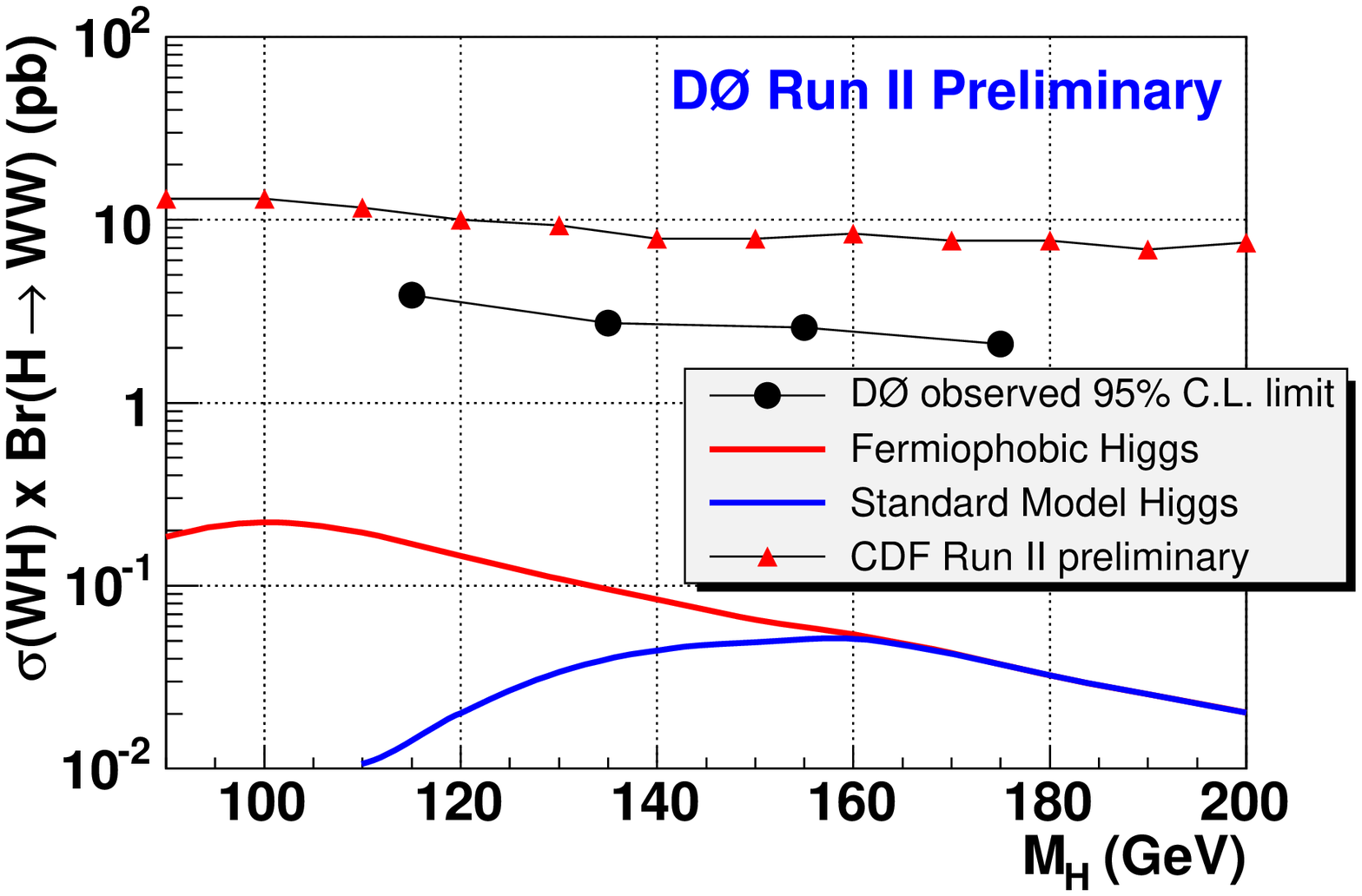} 
\vspace*{-1cm}
\caption{D\O\ WH ($\rm H\to WW$).
Left: missing transverse energy for like-sign ee (left), e$\mu$ (center) and $\mu\mu$ (right). 
Right: cross section limit.
}
\label{fig:d0-wHww}
\vspace*{-0.2cm}
\end{figure}

\begin{figure}[hbtp]
\includegraphics[width=0.49\textwidth,height=8cm]{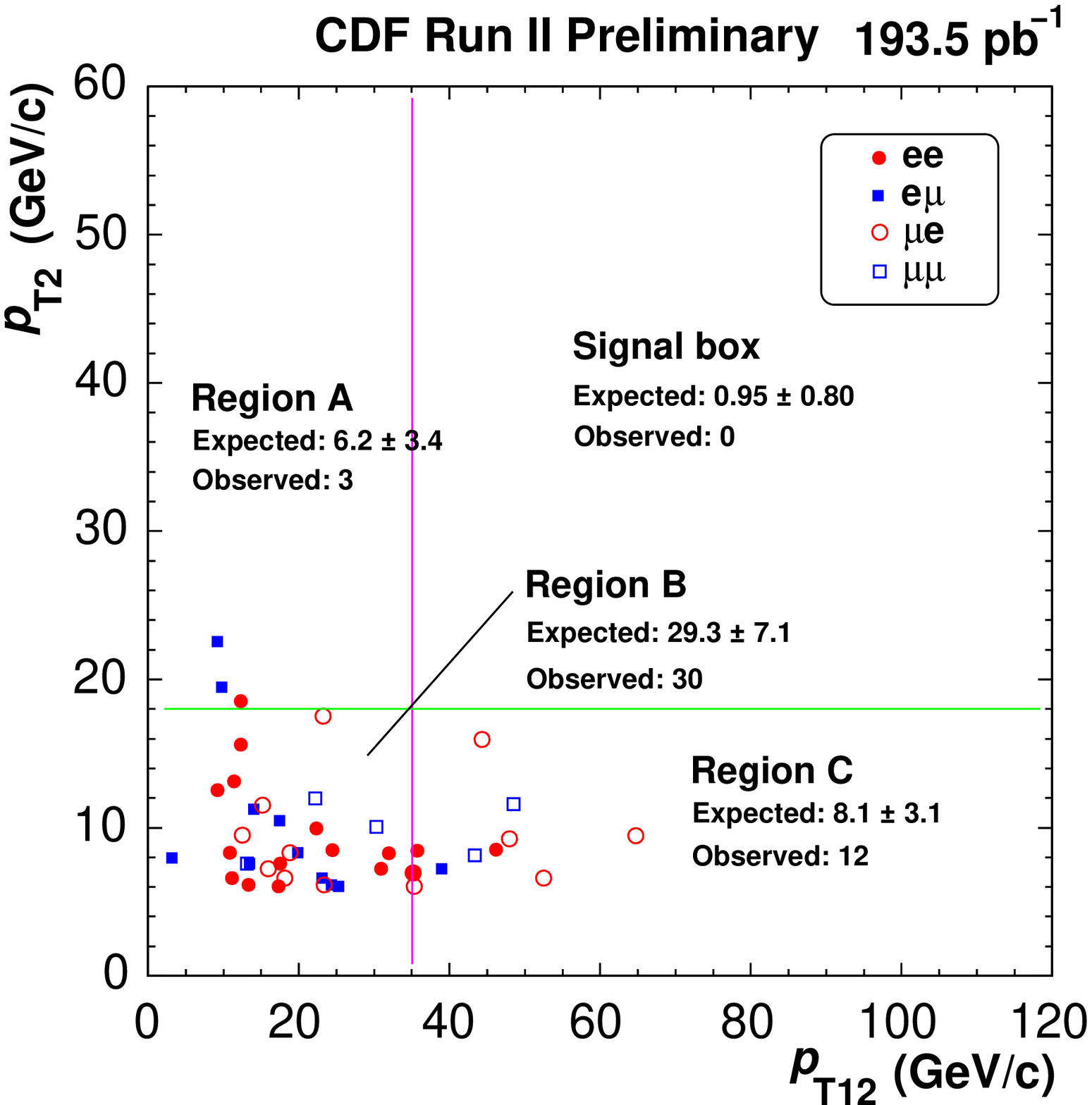} \hfill
\includegraphics[width=0.49\textwidth,height=8cm]{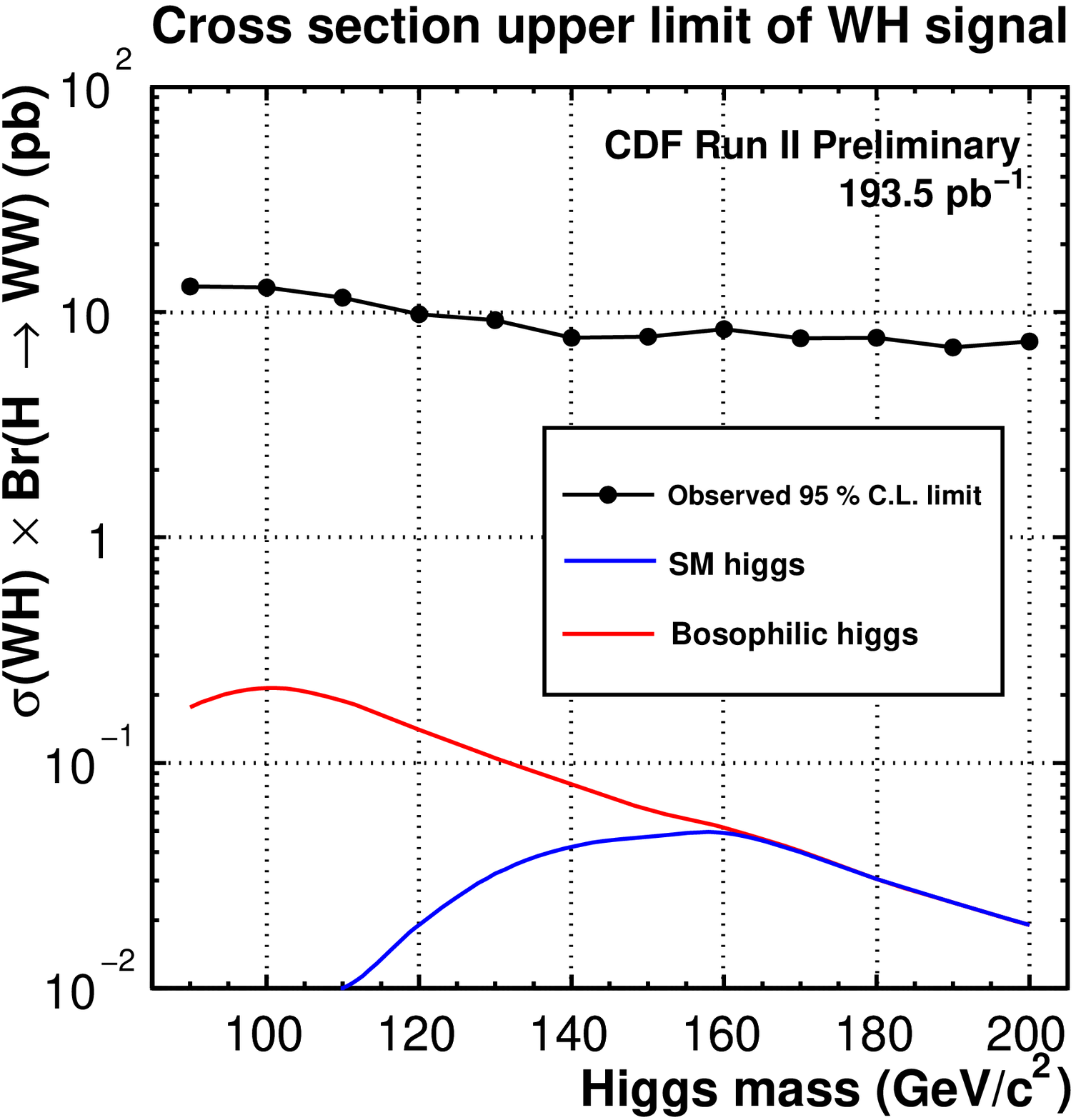}
%\vspace*{-0.4cm}
\caption{CDF WH ($\rm H\to WW$).
Left: comparison of simulated background and observed number of events in the plane of
the second most energetic lepton $p_{\rm T2}$ versus di-lepton vector sum $p_{\rm T12}$.
Right: cross section limit.
}
\label{fig:cdf-wHww}
\vspace*{-0.5cm}
\end{figure}

Recent results for the search WH ($\rm H\to WW$) in the like-signed charged lepton final state
are shown in Figs.~\ref{fig:d0-wHww} (from~\cite{d0-WHww}) and~\ref{fig:cdf-wHww} (from~\cite{cdf-WHww}), 
which lead to similar sensitivity as in the $\rm H\to \bb$ decay mode.

\clearpage

\subsubsection{$\rm ZH\to \nn\bb$}

\begin{figure}[hp]
\vspace*{-0.4cm}
\includegraphics[width=0.32\textwidth,height=5cm]{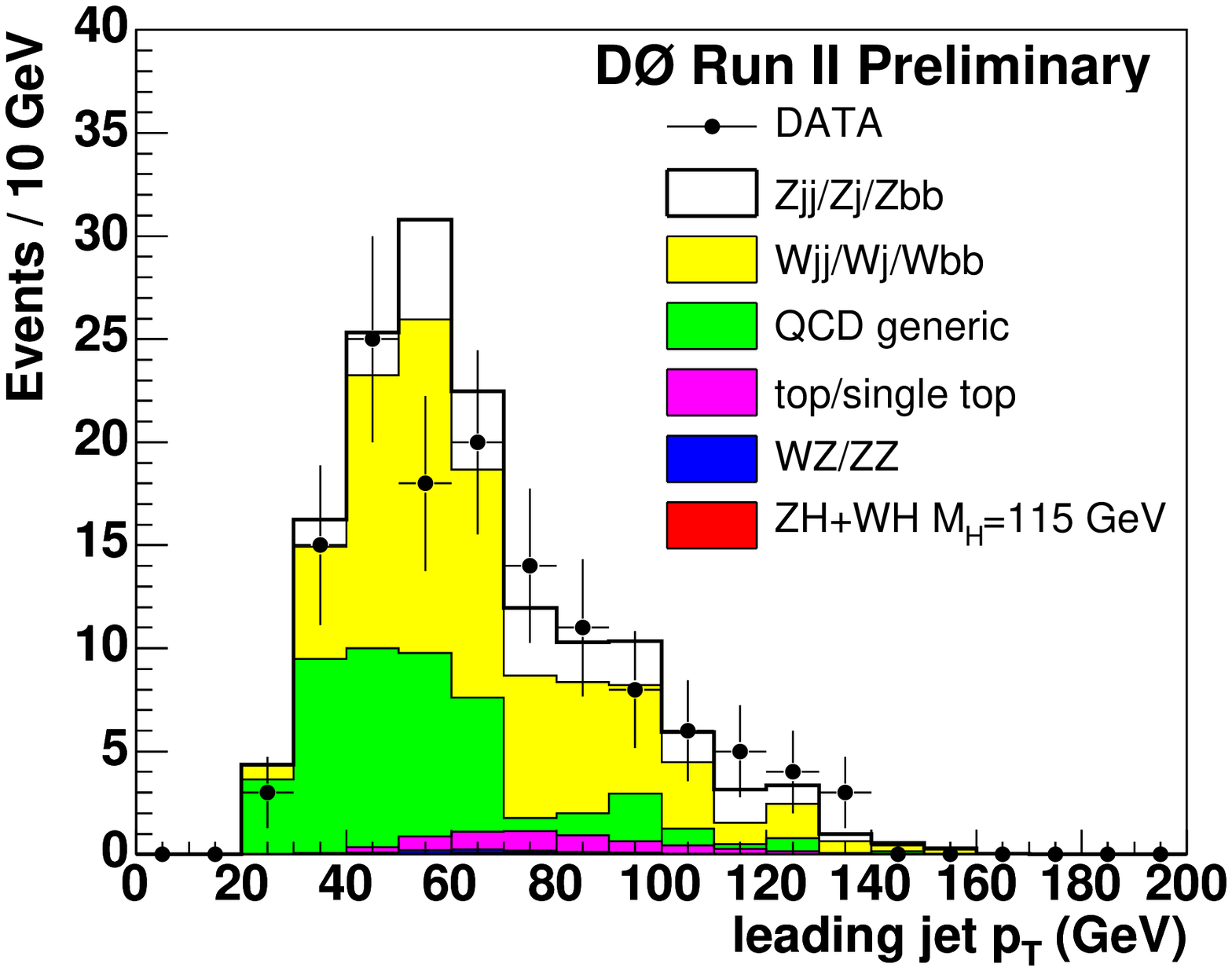} \hfill
\includegraphics[width=0.32\textwidth,height=5cm]{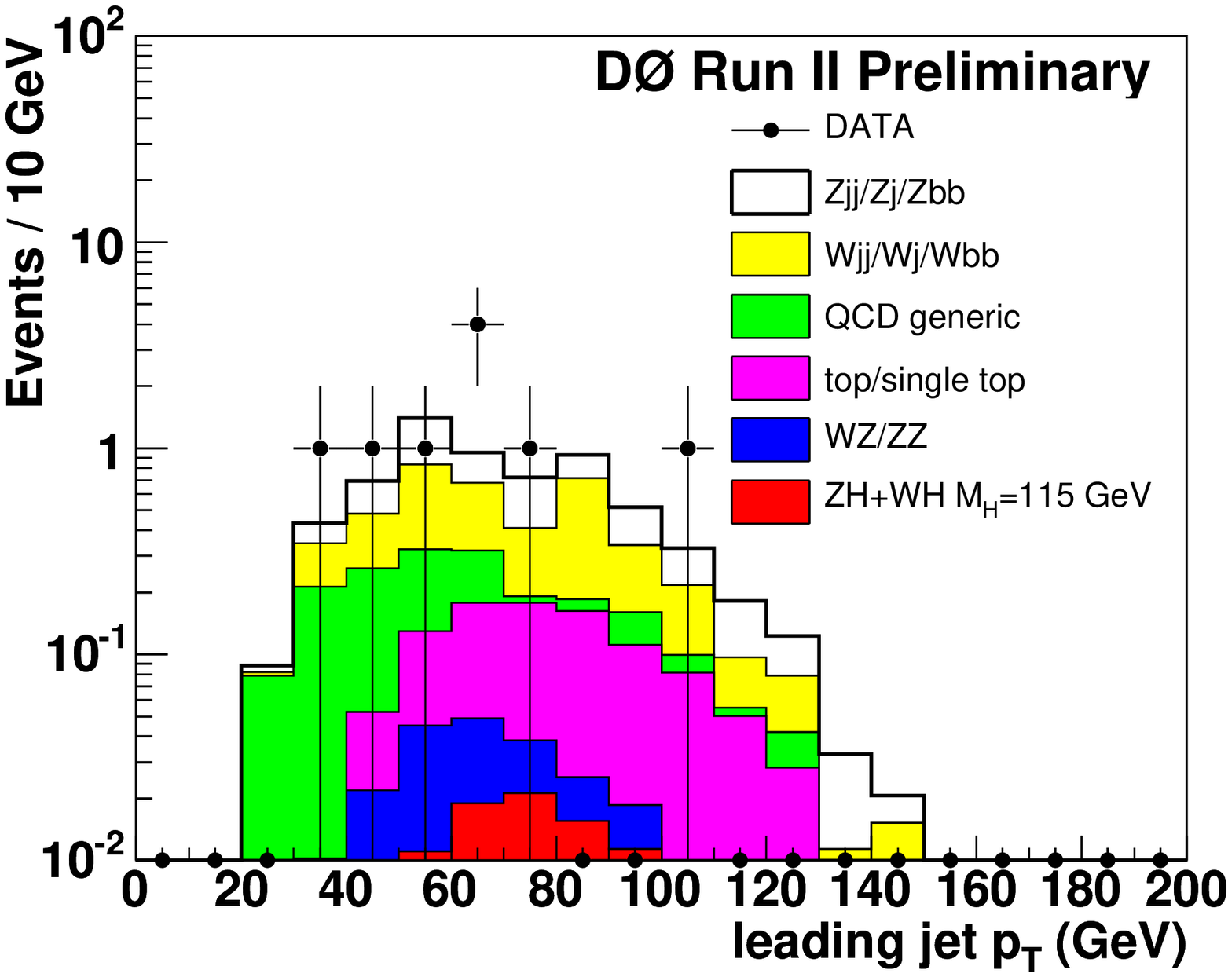} \hfill
\includegraphics[width=0.32\textwidth,height=5cm]{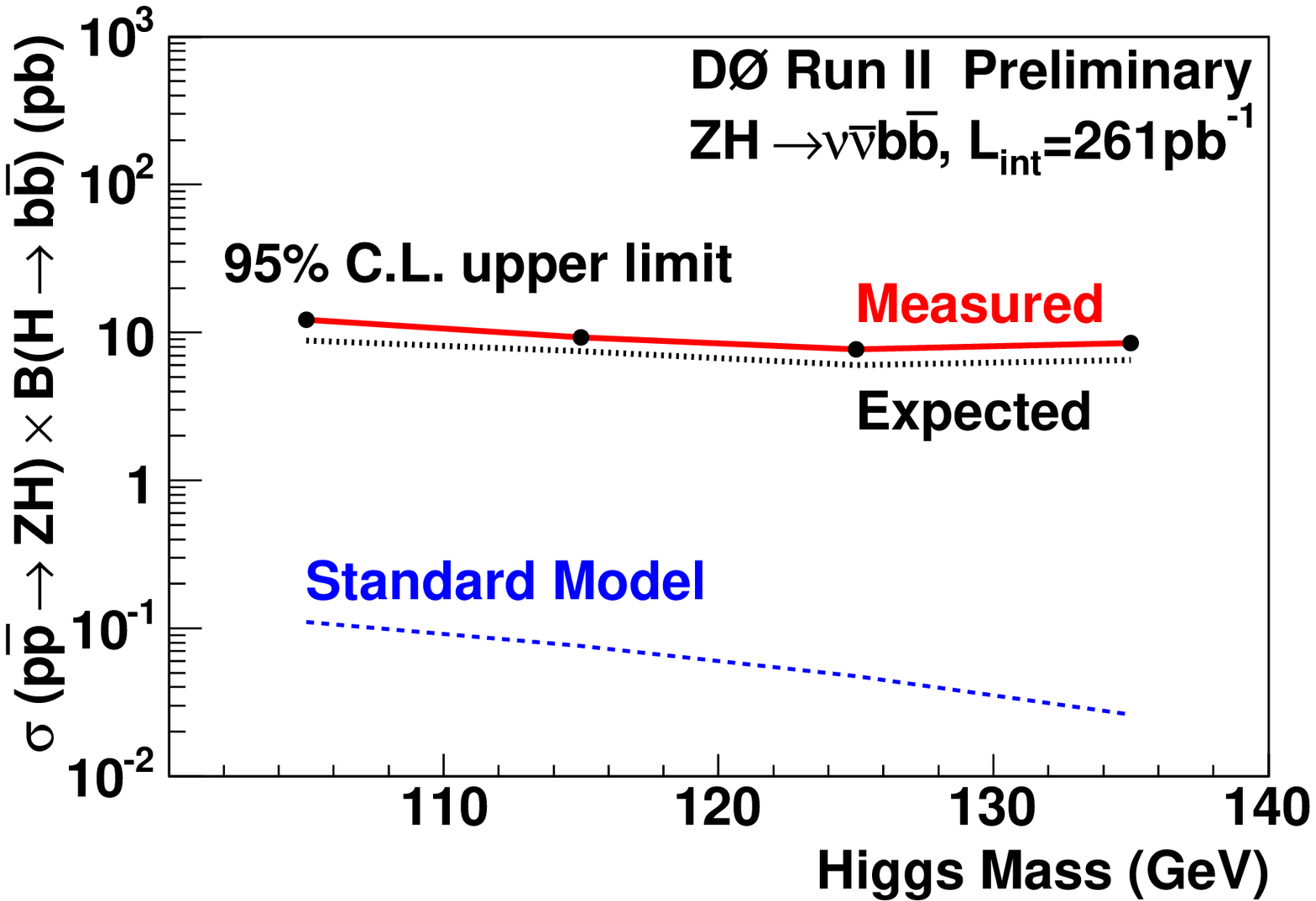}
%\vspace*{-0.5cm}
\caption{D\O\ ZH ($\rm H\to\bb$).
Left: single b-tagging. Center: double b-tagging. Right: cross section limit.
}
\label{fig:d0-zhbb}
\end{figure}

\begin{figure}[bhp]
%\vspace*{-0.5cm}
\includegraphics[width=0.49\textwidth]{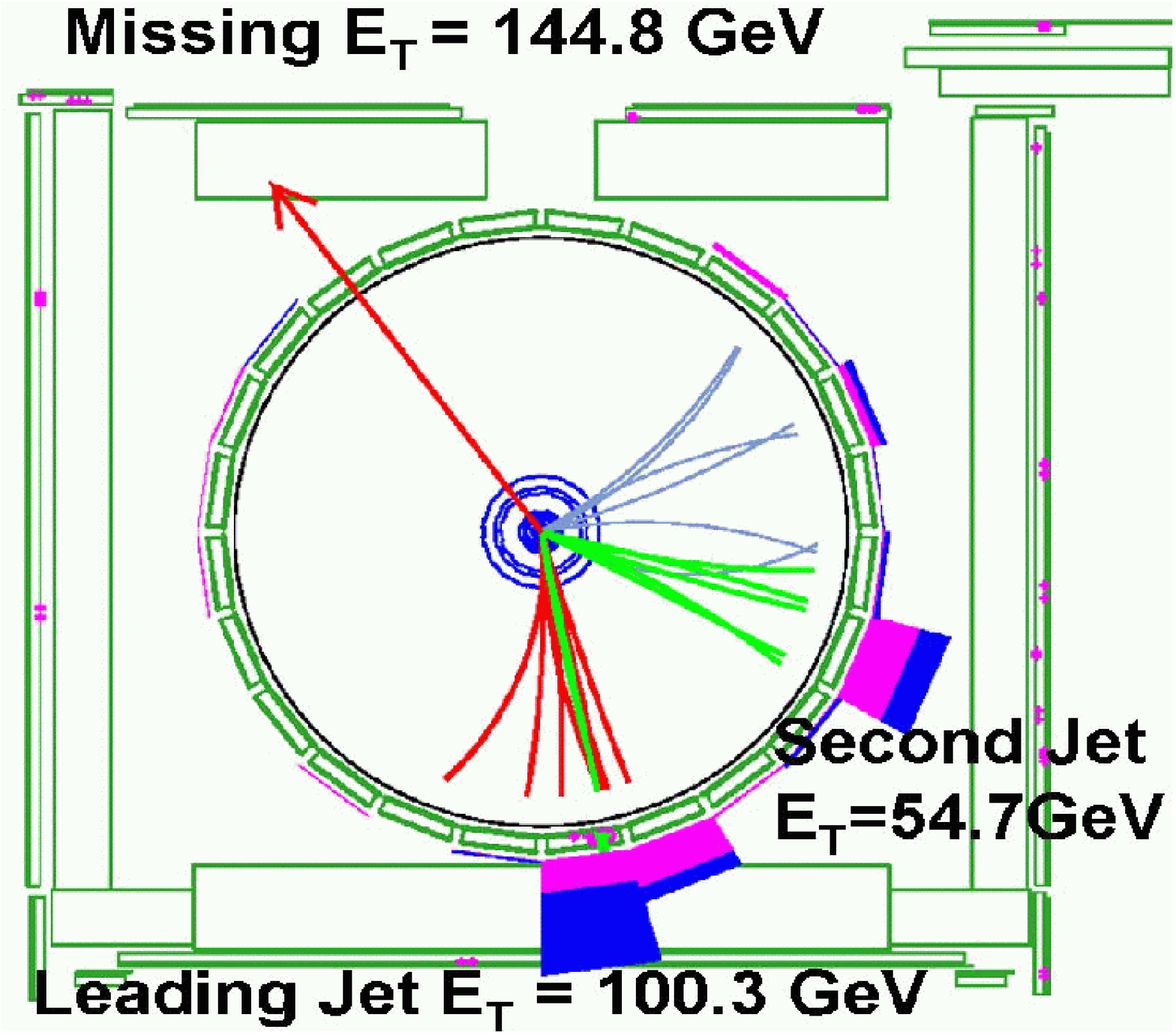}\hfill
\includegraphics[width=0.49\textwidth,bbllx=-36pt,bblly=126pt,bburx=578pt,bbury=669pt,clip=]{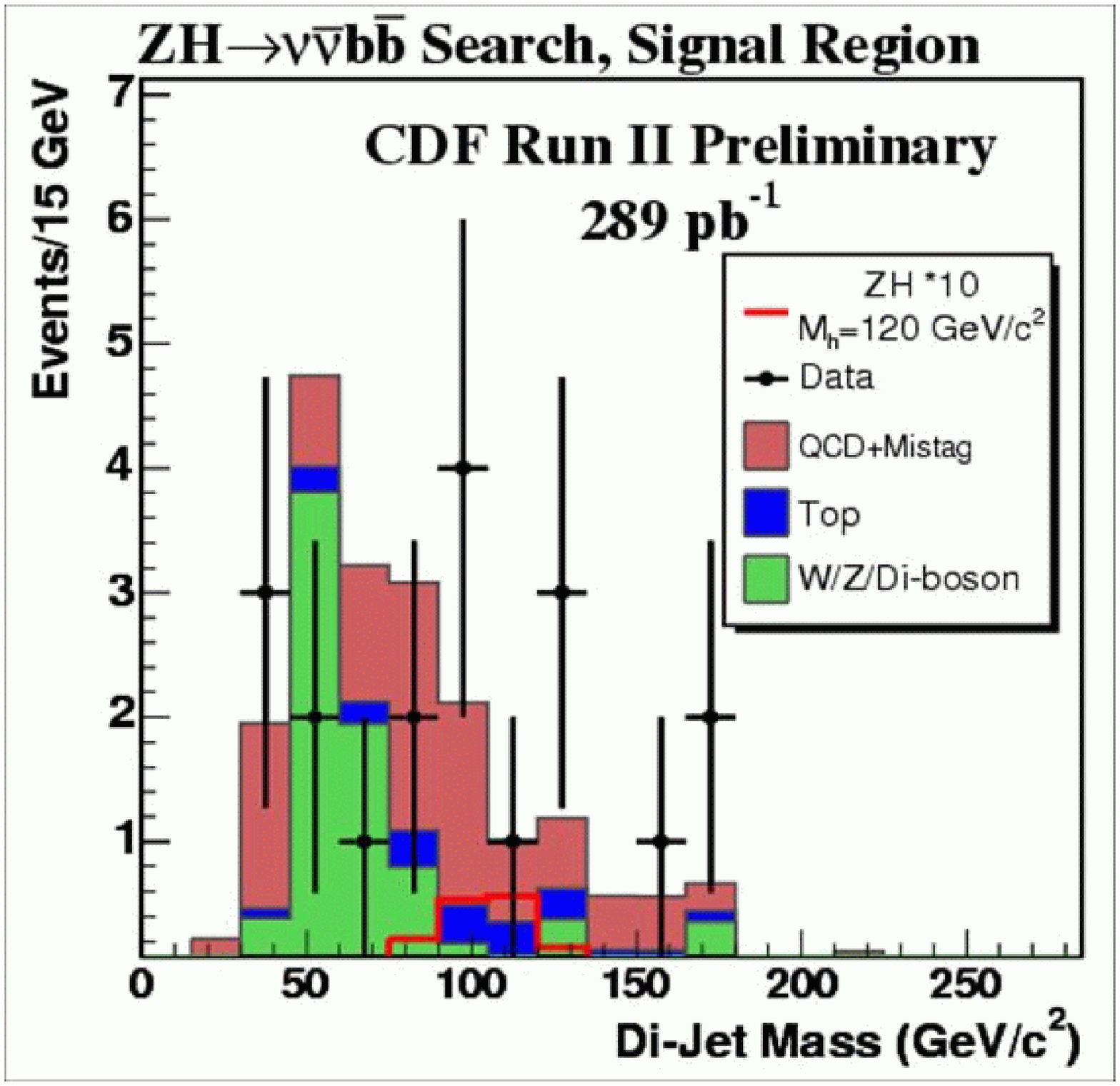}
%\vspace*{-0.4cm}
\caption{CDF ZH ($\rm H\to\bb$). Left: candidate event. 
         Right: di-jet mass for data and simulated signal and background.
}
\label{fig:cdf-zHbb}
%\vspace*{-0.65cm}
\end{figure}

Both Tevatron experiments have searched for a $\rm ZH\to \nn\bb$ signal, which has a
smaller cross section compared to WH production.
The results from the expected missing energy and b-jet signal 
are shown in Figs.~\ref{fig:d0-zhbb} (from~\cite{d0-zHbb}) and~\ref{fig:cdf-zHbb} (from~\cite{cdf-zHbb}). 
The CDF cross section limits are shown in Fig.~\ref{fig:cdf-d0-sm} (from~\cite{cdf-zHbb}).

\clearpage
\subsection{Summary of SM Higgs Cross Section Limits}

\begin{figure}[tp]
\vspace*{1cm}
\begin{center}
%\begin{minipage}{0.49\textwidth}
%\includegraphics[width=\textwidth,height=6cm]{cdf290905overview-limits.ps}
\includegraphics[width=0.8\textwidth]{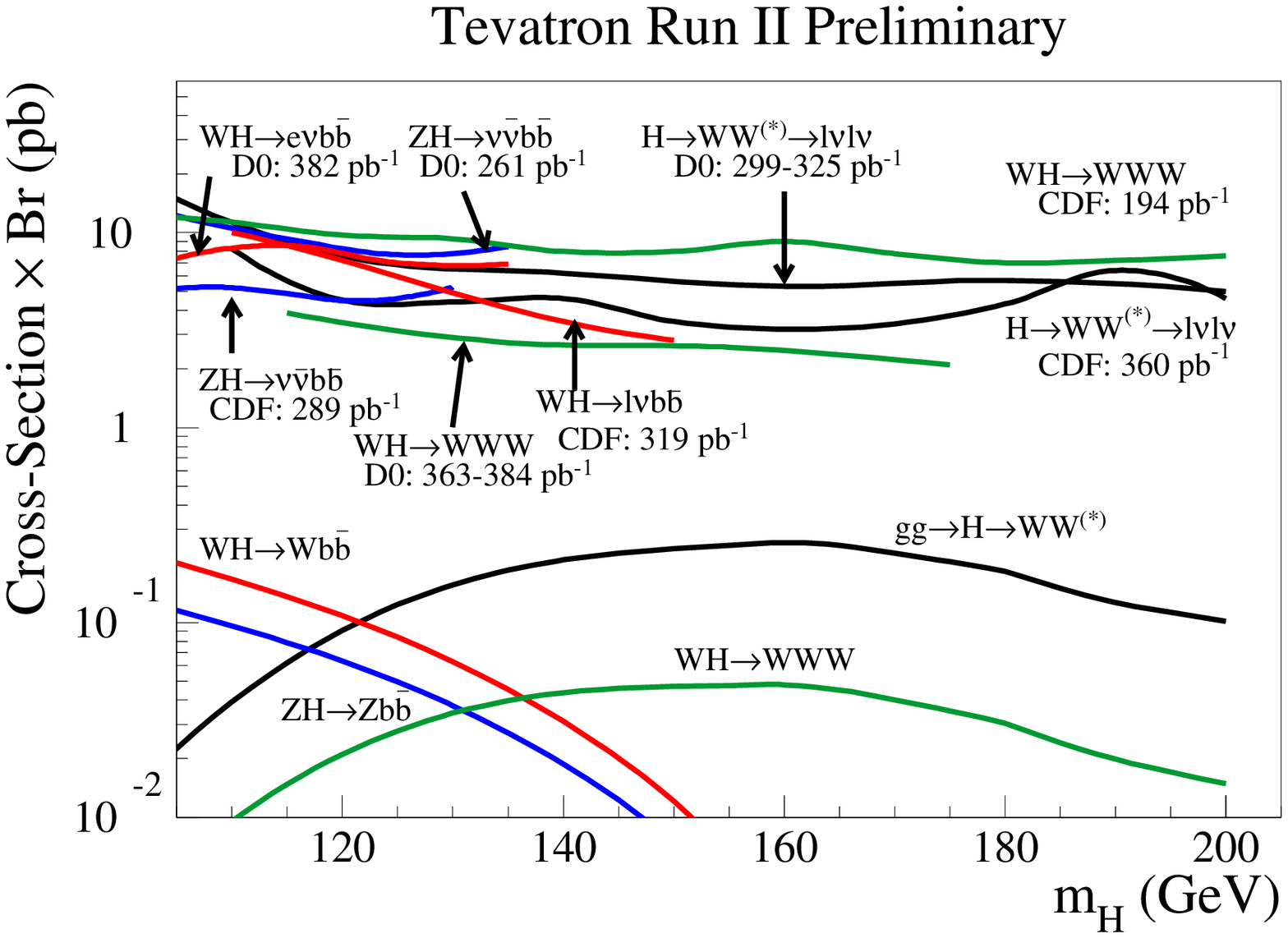}
%\vspace*{-1.1cm}
%\end{minipage}\hfill 
%\begin{minipage}{0.49\textwidth}
%\includegraphics[width=\textwidth,height=6cm]{cdf290905-norm-limits.ps}
\includegraphics[width=0.8\textwidth]{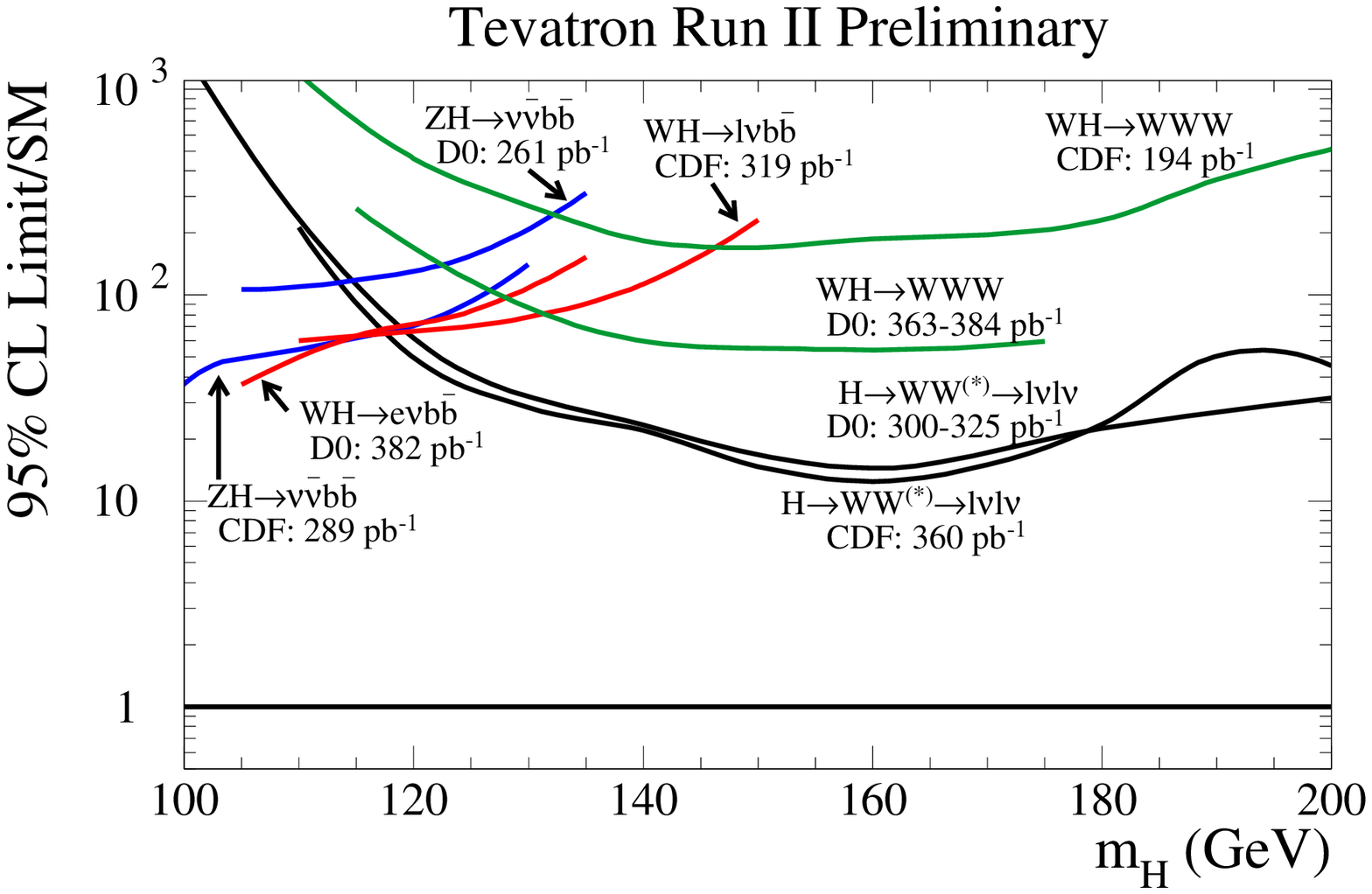}
%\end{minipage}
%\vspace*{-0.2cm}
\end{center}
\caption{Left: summary of SM Higgs boson limits from CDF and D\O\ Higgs boson searches.
         Right: ratio of observed cross section limit and exected SM cross section.
}
\label{fig:cdf-d0-sm}
\vspace*{0.4cm}
\end{figure}

Cross section limits for a 120 GeV SM Higgs boson production are summarized in Table~\ref{tab:limits}. 
At this mass the SM expectation 
is about $\rm \sigma\times BR=0.1~pb$ for the processes $\rm WH\to W\bb$ and $\rm gg\to H\to \bb$
processes. 
The current CDF and D\O\ limits as a function of the SM Higgs boson mass 
are given in Fig.~\ref{fig:cdf-d0-sm} (from~\cite{summary-limits}),
and also shown are the ratios of observed cross section limits and the
expected SM cross section.

\begin{table}[htb]
%\vspace*{-0.2cm}
\renewcommand{\arraystretch}{1.2} % enlarge line spacing
\caption{Summary of observed and expected limits at 95\% CL from CDF and D\O\ for a 120 GeV SM Higgs boson . 
         Recently, through optimizations, the expected D\O\ $\rm \nn\bb$ limit has been reduced by 
         about a factor 2~\cite{gregorio}.
\label{tab:limits} }
{
%\begin{minipage}{0.48\textwidth}
\begin{center}
\begin{tabular}{l|c|c|c|c}
Channel              & experiment & luminosity  & \multicolumn{2}{c}{limit $\sigma\times$BR (pb)} \\ 
                     &            & (pb$^{-1}$) & ~~~obs.~~~        &  exp. \\ \hline
$\rm H$$\to$WW$\to$$\rm \ell\nu\ell\nu$ & CDF  & 360       & 4.5 & 6.9   \\
                                   & D\O  & 299-325        & 5.6 & 9.5   \\
$\rm WH\to l\nu\bb$        & CDF  & 319                    & 7.2 & 5.0   \\
                           & D\O   & 382                   & 7.8 & 5.7  \\
$\rm WH\to WWW$            & CDF  & 194                    & 9.7  & 15  \\
                           & D\O   & 363-384               & 3.5 & 4.2\\ 
$\rm ZH\to \nn\bb$         & CDF  & 289                    & 4.5  & 3.6 \\
                           & D\O   & 261                   & 8.5 & 6.5 
\end{tabular}
\end{center}
}
\end{table}

These limits are about a factor two weaker compared to the estimates~\cite{prospects99,prospects03}
for the luminosity of about 300~pb$^{-1}$. Improvements will come from optimized b-quark tagging, 
and also from larger e/$\mu$ acceptance, better jet mass resolution, and from using 
advanced analysis techniques.
Higher Higgs boson sensitivities will also result from the luminosity increase over the next years. 
Currently already about 1~fb$^{-1}$ is recorded per experiment, and the total luminosity will increase 
up to about 8~fb$^{-1}$. 
Reaching the sensitivity of LEP (114 GeV), and extending beyond, possibly to reach 180 GeV, 
will be a large experimental challenge over the next few years. 
The combination of the $\rm H\to \bb$ and $\rm H\to WW^*$ searches give lower sensitivity 
reduction around 140 GeV compared to the estimates shown in Fig.~\ref{fig:prospects}.
Nevertheless, particular attention needs to be devoted to the mass region around 140~GeV for 
example by investigating new search reactions (e.g. $\rm H\to \tau\tau$~\cite{tao}) which were 
not included in the previous estimates~\cite{prospects99,prospects03}.

\clearpage

\clearpage
\subsection{Beyond the SM}

\subsubsection{$\rm \bb h$, $\rm \bb H $ and $\rm \bb A$ in the General 2-Doublet Higgs Model or MSSM}

\begin{figure}[htbp]
\vspace*{0.6cm}
\begin{minipage}{0.4\textwidth}
\includegraphics[width=\textwidth]{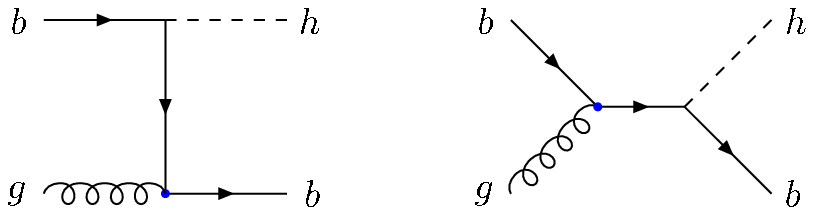}
\includegraphics[width=\textwidth]{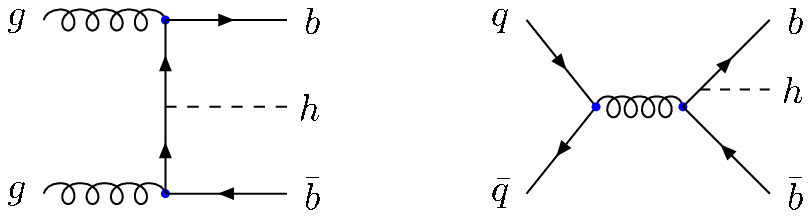}
\end{minipage} \hfill
\begin{minipage}{0.4\textwidth}
\includegraphics[width=\textwidth]{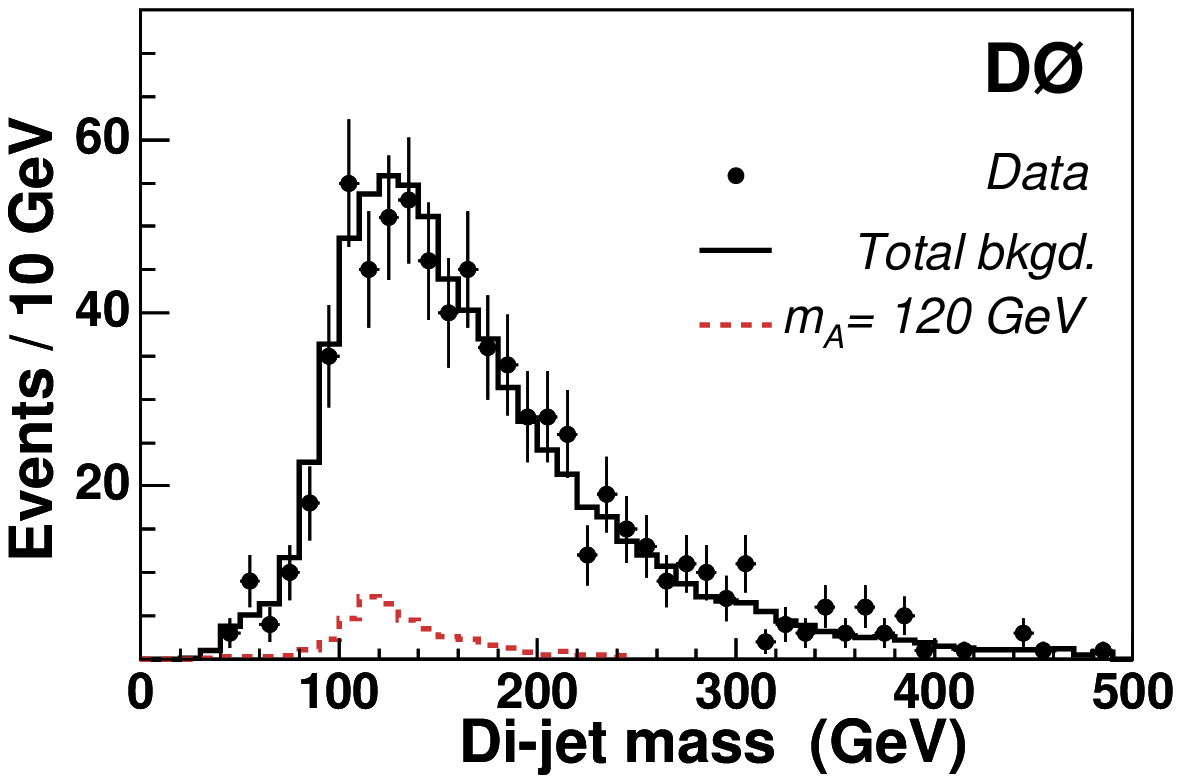}
\end{minipage}
\vspace*{-0.2cm}
\caption{D\O. Left: leading-order Feynman diagrams for 
               neutral Higgs boson production in the five-flavor scheme (top) and 
               four-flavor scheme (bottom). 
        Right: invariant mass spectrum of two leading jets in events with
               at least three b-tagged jets, estimated background, and the
               signal simulation for a 120~GeV Higgs boson that can be excluded at the 95\% CL 
               (for $\tan\beta\approx 30$).
        } \label{fig:bba}
\vspace*{1cm}
\end{figure}

Higgs boson production process in association with b-quarks in $\rm p \bar p$
collisions has been calculated in two ways: in the five-flavor scheme~\cite{5fns}, 
where only one b-quark has to be present in the final state, while in the four-flavor 
scheme~\cite{4fns}, two b-quarks are explicitly required in the calculation. 
Both calculations are now
available at next-to-leading order (NLO QCD), and agree taking into account the theoretical uncertainties.
Figure~\ref{fig:bba} (from~\cite{d0-bba}) illustrates these processes for h production at leading order (LO), 
and analogous diagrams can be drawn for the H and A bosons.
The cross section depends mostly on $\tan^2\beta$ and to a smaller extent on other 
Supersymmetric parameters as given by: 
$$
\sigma\times BR_{\rm SUSY} \approx 2 \sigma_{\rm SM} \frac{\tan^2\beta}{(1+\delta_b)^2}\frac{9}{9+(1+\delta_b)^2},
$$
where $\delta_b = k \tan\beta$ with $k$ depending on the SUSY parameters, in particular also on $A_t$, the 
mixing in the scalar top sector, the gluino mass, the $\mu$ parameter, stop and sbottom masses.

\begin{figure}[htbp]
\centering
\includegraphics[width=0.49\textwidth]{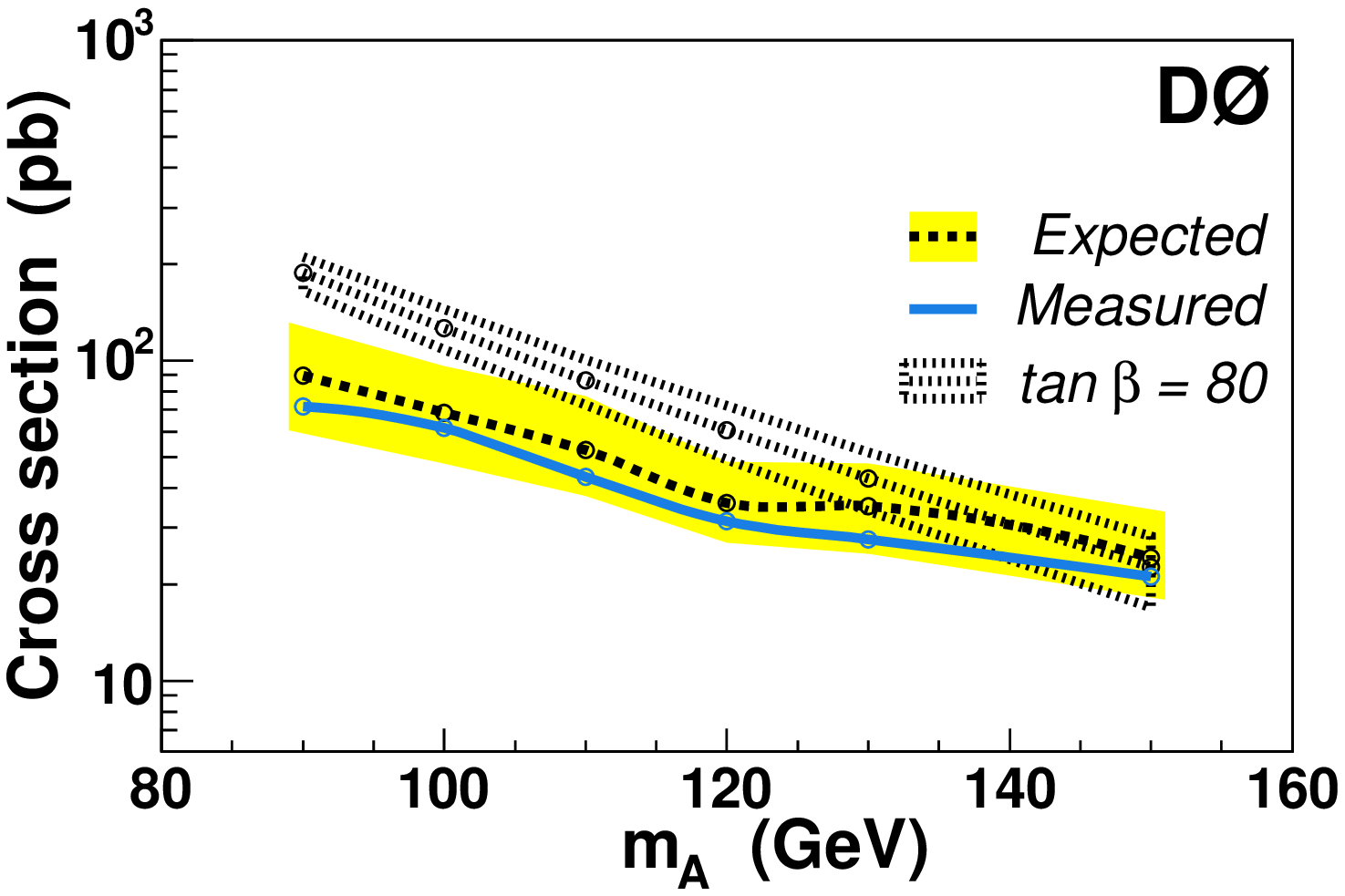} \hfill
\includegraphics[width=0.49\textwidth]{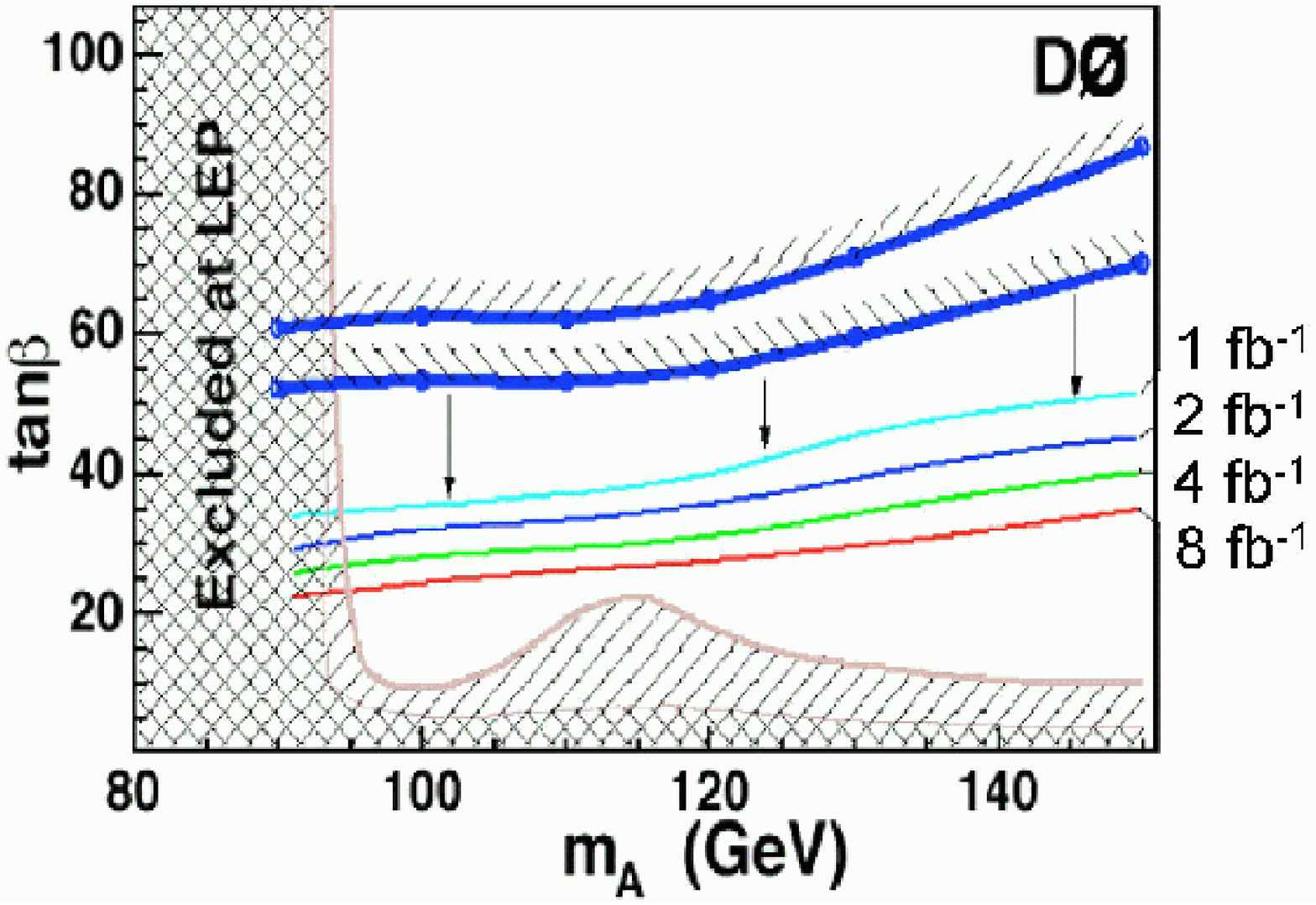}
\vspace*{-0.3cm}
\caption{
D\O. Left: expected (in absence of a signal) and measured 95\% CL upper limits on the signal
cross section as a function of $m_A$. The band
indicates the $\pm 1\sigma$ range on the expected limit. 
Also shown is the cross section for the signal at $\tan \beta = 80$ in the ``no
mixing'' scenario of the MSSM, with the theoretical uncertainty
indicated by the overlaid band.
Right: 95\% CL upper limit on $\tan \beta$ as a function of $m_A$
for two scenarios of the MSSM, ``no mixing'' and ``maximal mixing.''
In addition, the expected sensitivities at 95\% CL for the maximal mixing scenario are shown.
The limits obtained by the LEP experiments for the same two scenarios of the MSSM
are given also.} 
\label{fig:d0-bba-xsec}
\end{figure}

There is no indication of a $\rm \bb A$ production in the data (Fig.~\ref{fig:bba}).
The resulting limits on the production cross section, and in the MSSM are given in
Fig.~\ref{fig:d0-bba-xsec} (from~\cite{d0-bba}). 
Previous results by the CDF Collaboration~\cite{cdf-bbA} were not confirmed with the D\O\ data. In the CDF
analysis gluon parton distribution function PDFs were used that have been superseded, resulting in more stringent limits than 
would have been obtained with the more recent PDFs used in the D\O\ analysis.
Figure~\ref{fig:d0-bba-xsec}  (from~\cite{projections}) includes also the sensitivity 
prospects for larger luminosities.

\subsubsection{$\rm h,H,A\to \tautau$}
\begin{figure}[htbp]
%\vspace*{-0.2cm}
\includegraphics[width=0.49\columnwidth,height=6cm]{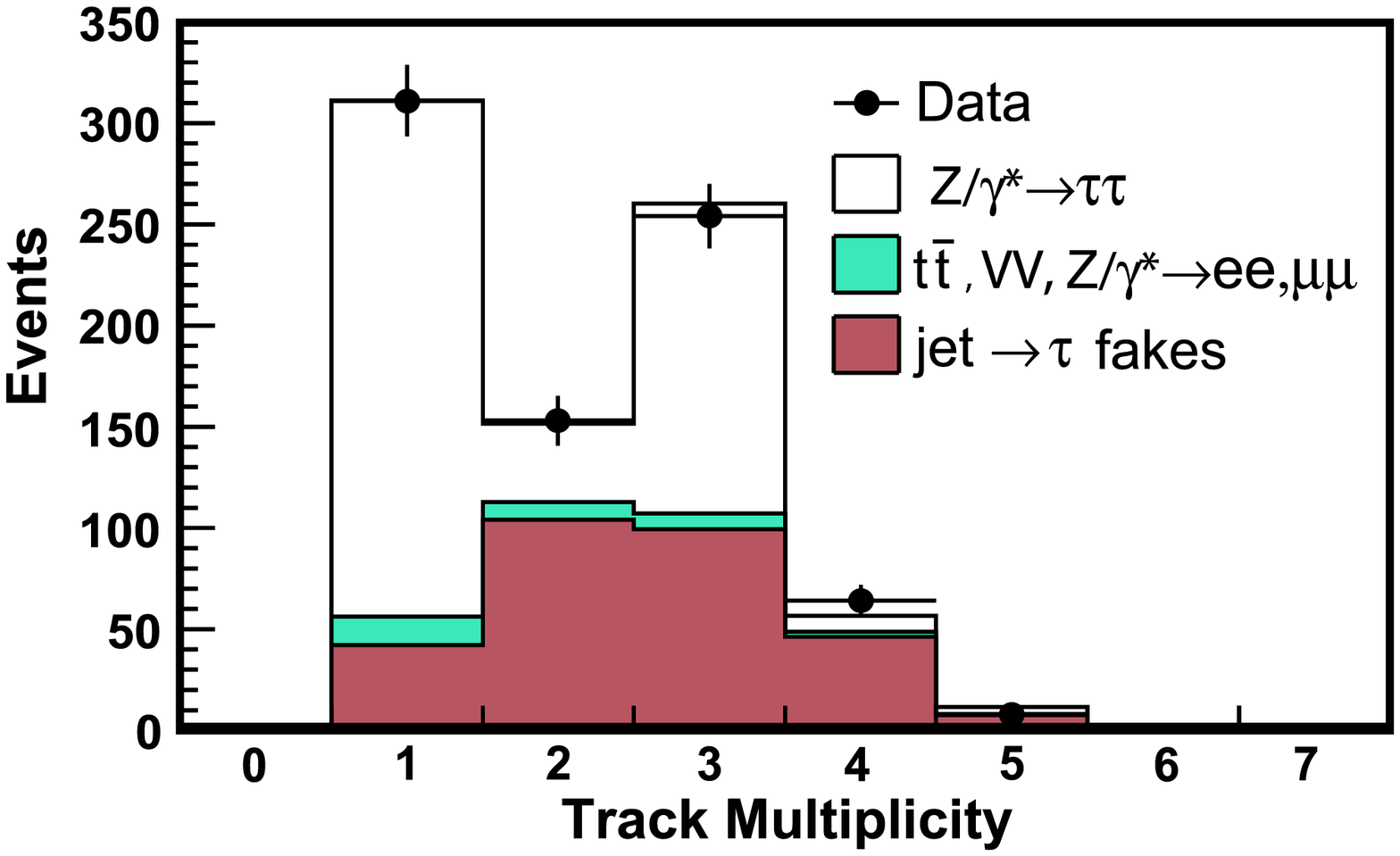} \hfill
\includegraphics[width=0.49\columnwidth,height=6cm]{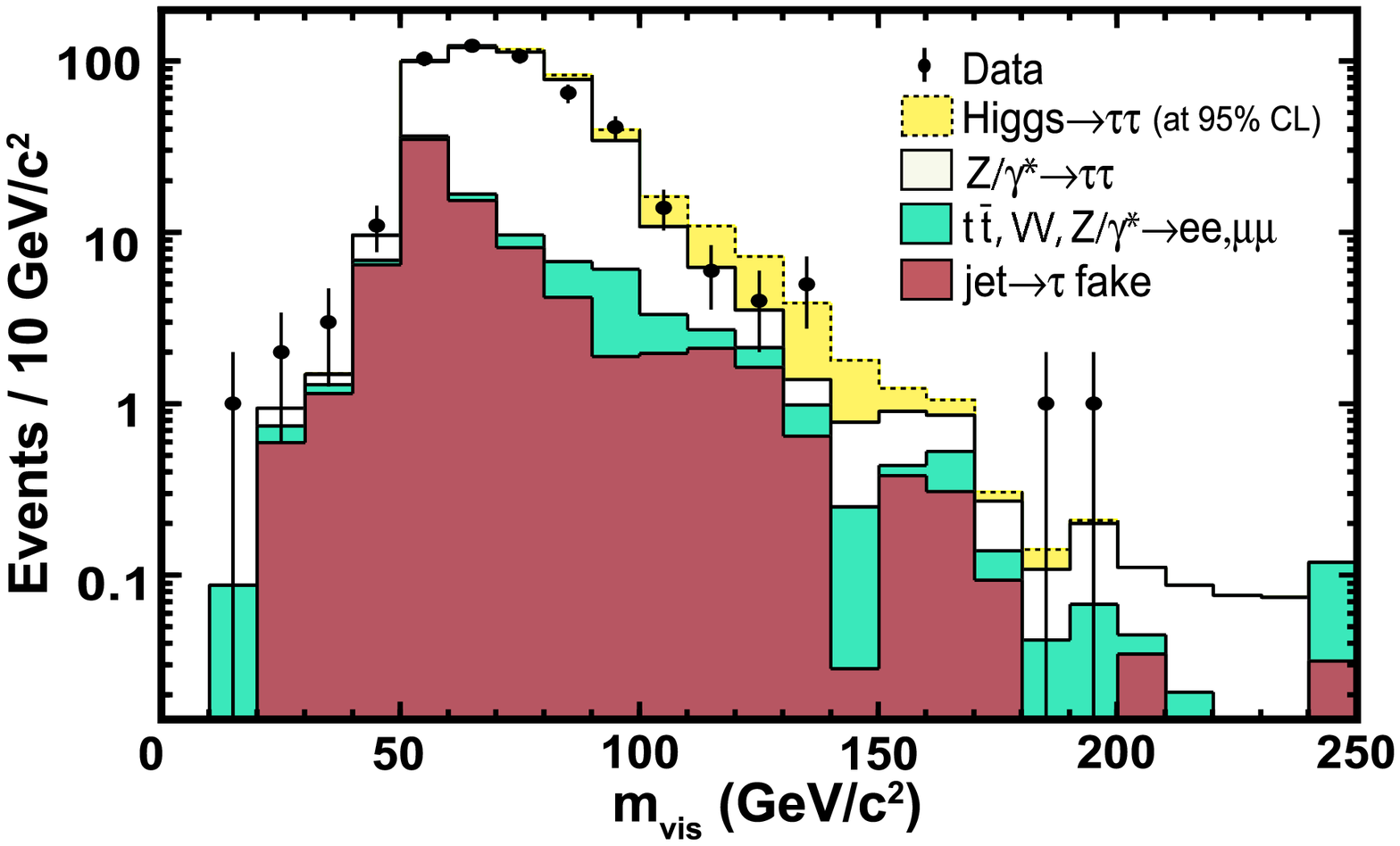}
\vspace*{-0.2cm}
\caption{
CDF. Left: track multiplicity for hadronically decaying tau candidates.
Right: visible mass distribution for a simulated signal of
$m_{\rm A}$ = 140~GeV, background and data.
} \label{fig:nTracks}
%\vspace*{-0.5cm}
\end{figure}

\begin{figure}[htbp]
\includegraphics[width=0.32\textwidth,height=5cm]{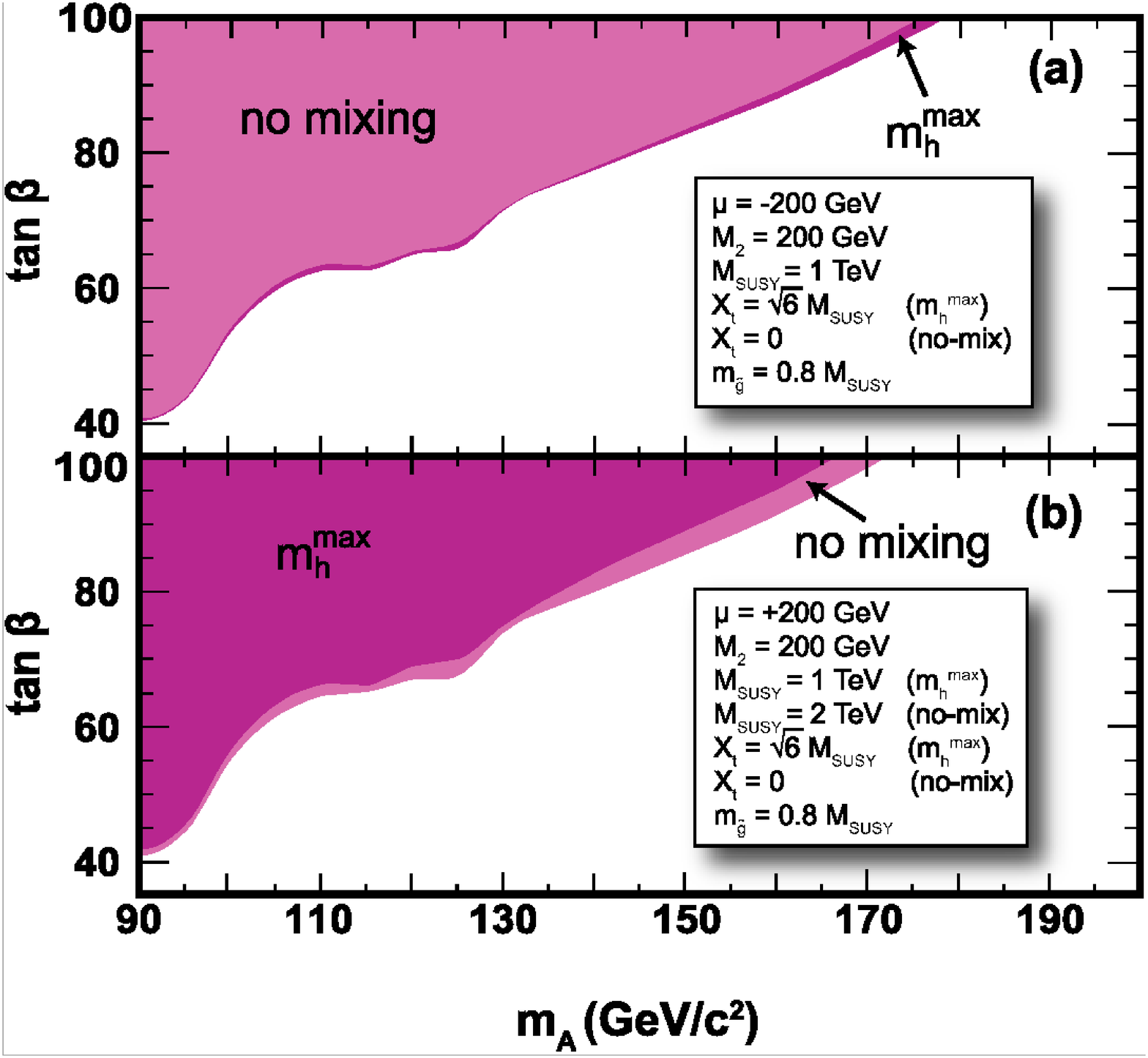}
\includegraphics[width=0.32\textwidth,height=5cm]{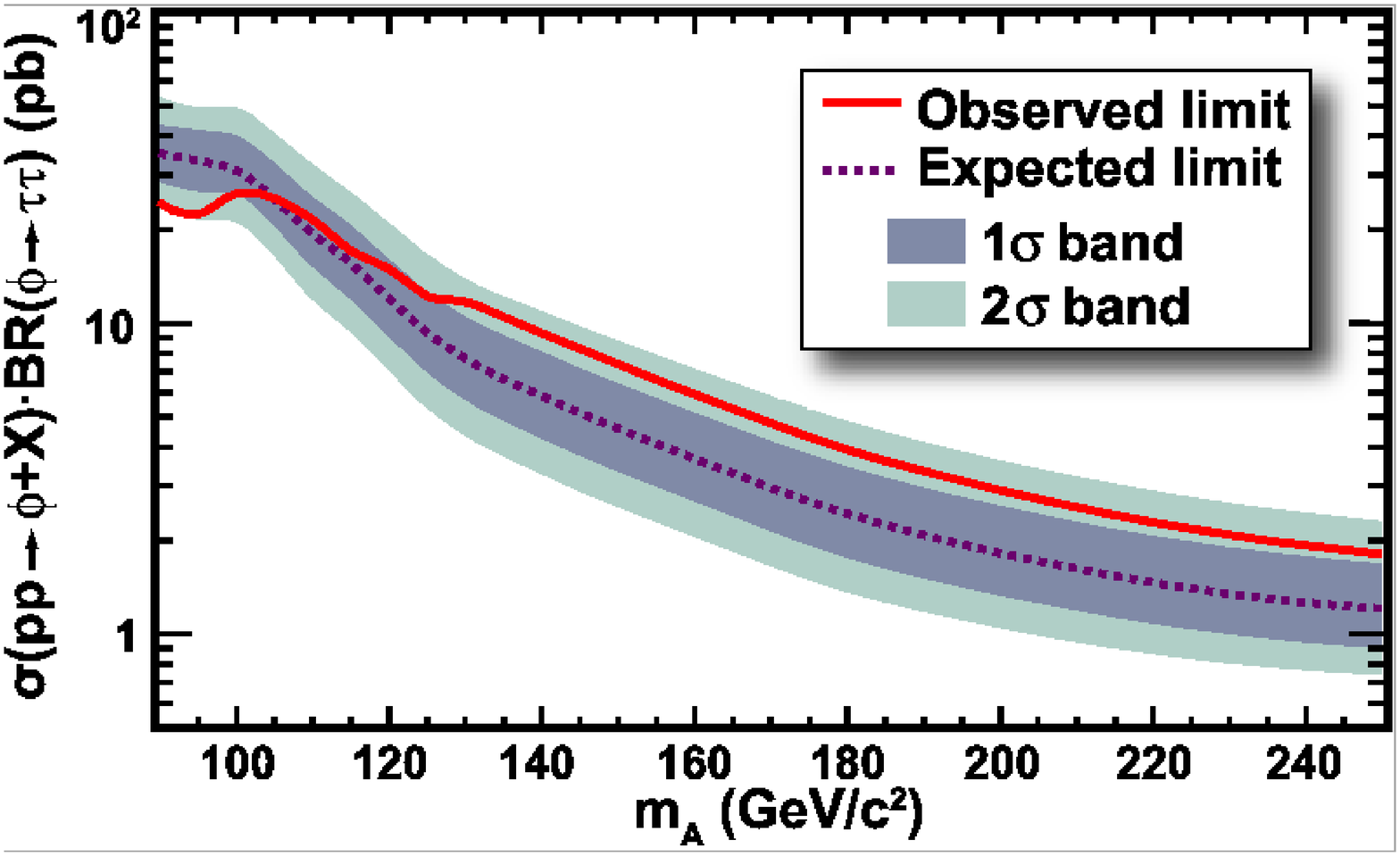}
\includegraphics[width=0.32\textwidth,height=5cm]{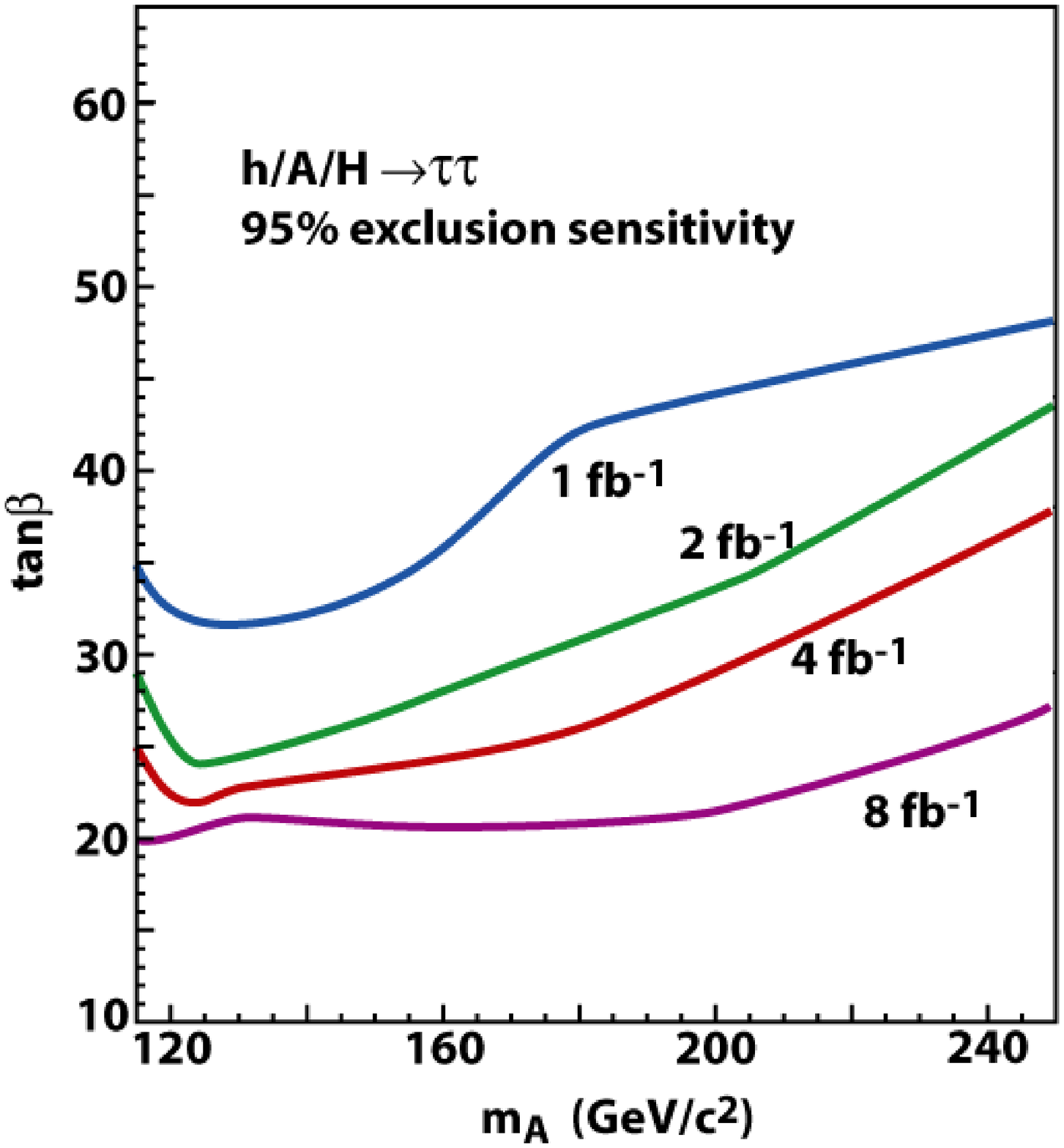}
\vspace*{-0.3cm}
\caption{
CDF. Left: excluded regions in the $\tan\beta$ versus $m_A$ plane for
the maximum-mixing and no-mixing scenarios with (a) $\mu<0$ and (b)
$\mu>0$.
Center: upper limits at 95\% CL on Higgs production cross section
times branching fraction to $\tau$ pairs.
Right: expected $\tan\beta$ sensitivities for larger luminosities.
} \label{fig:tanBeta}
%\vspace*{-1.5cm}
\end{figure}

The signature for $\rm h,H,A\to \tautau$ opens additional possibilities for Higgs 
boson discovery.
The characteristic enhancement in the one- and three-track bins from 
hadronic $\tau$ decays is shown in Fig.~\ref{fig:nTracks} (from~\cite{cdf-tau}). 
Tau leptons are identified as isolated jets with few tracks.
For $m_{\rm A}=140$~GeV, the visible mass for a simulated background and data is shown.
The background from misidentified jets, as determined by the simulation, is indicated.
Resulting limits in the MSSM and production cross section limits are given in 
Fig.~\ref{fig:tanBeta} (from~\cite{cdf-tau}), also shown are the sensitivity 
prospects with larger luminosity (from~\cite{projections}).

%\clearpage

\subsubsection{$\rm t\to H^+b$}

\begin{figure}[bp]
\begin{center}
\includegraphics[width=0.8\textwidth]{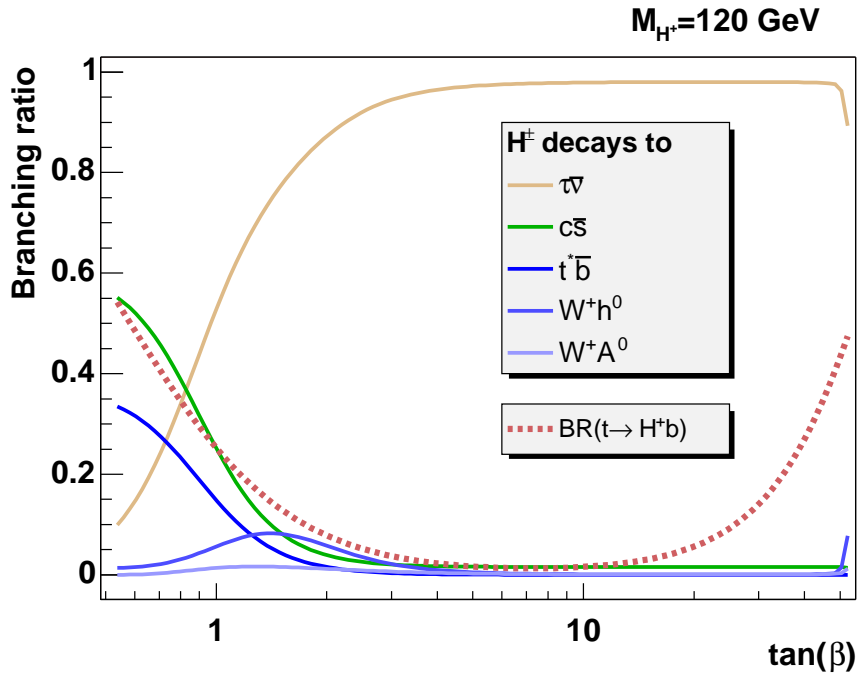}
\end{center}
\caption{
CDF. Branching ratios for a 120 GeV charged Higgs boson production in top decays and 
charged Higgs boson decays, which depends mostly on $\tan\beta$ in the general 2-doublet Higgs model.
} \label{fig:cdf-tbh}
%\vspace*{-1.cm}
\end{figure}

\begin{figure}[thp]
\begin{center}
\includegraphics[width=0.49\textwidth,height=10cm,bbllx=9pt,bblly=2pt,bburx=235pt,bbury=276pt,clip=]{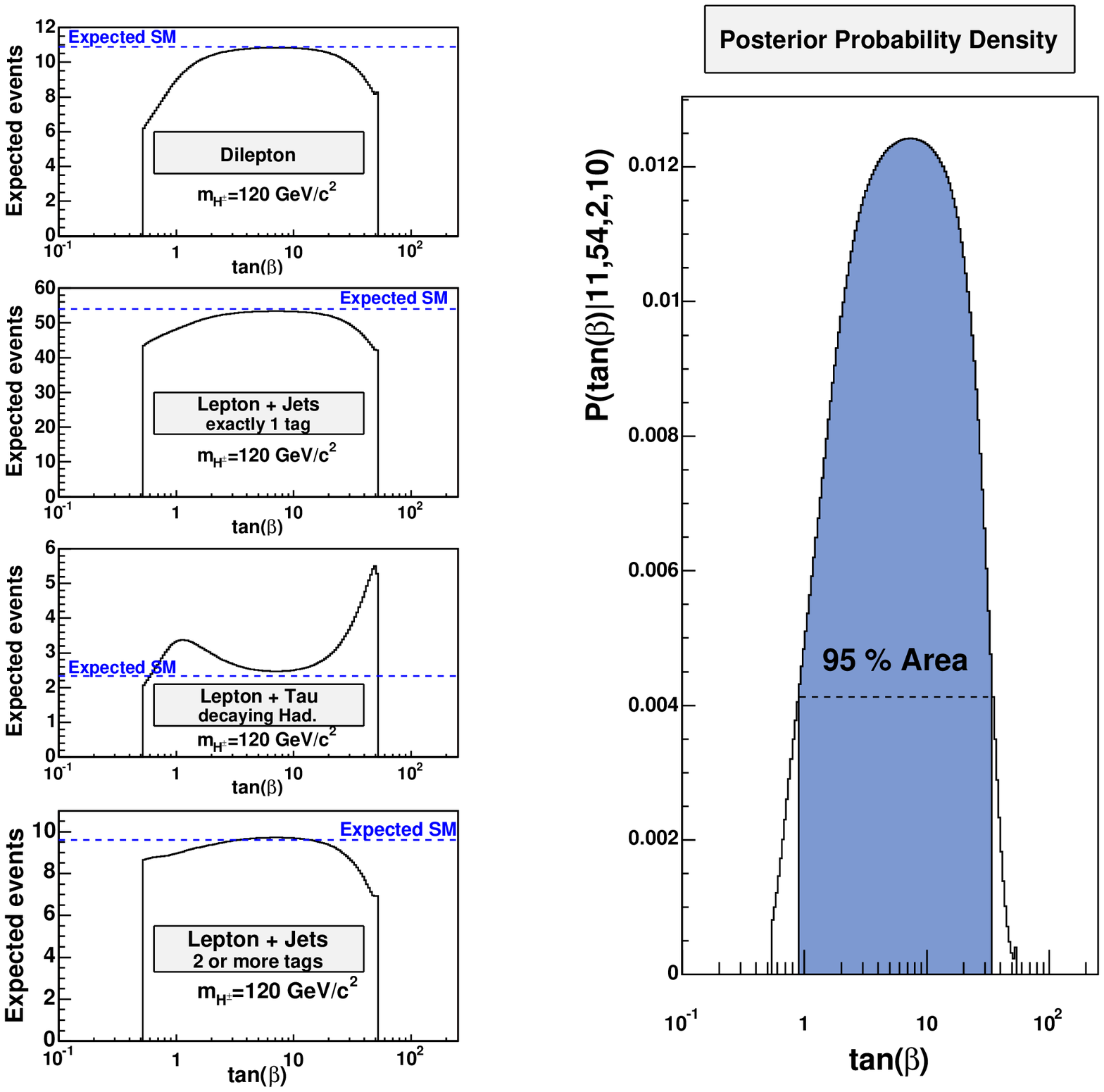}\hfill
\includegraphics[width=0.49\textwidth,height=10cm,bbllx=9pt,bblly=274pt,bburx=235pt,bbury=533pt,clip=]{cdf-ExpectedEventsDisplay.ps}
\end{center}
\caption{
CDF. Expected events from top decays in different analysis channels for the SM and as function of
$\tan\beta$ for charged Higgs boson production. The numbers of data events 2, 13, 8 and 49 agree well
with the SM expectations.
} \label{fig:cdf-tbh2}
%\vspace*{-1.cm}
\end{figure}

\begin{figure}[hcp]
\begin{center}
\includegraphics[width=0.8\textwidth,height=8cm]{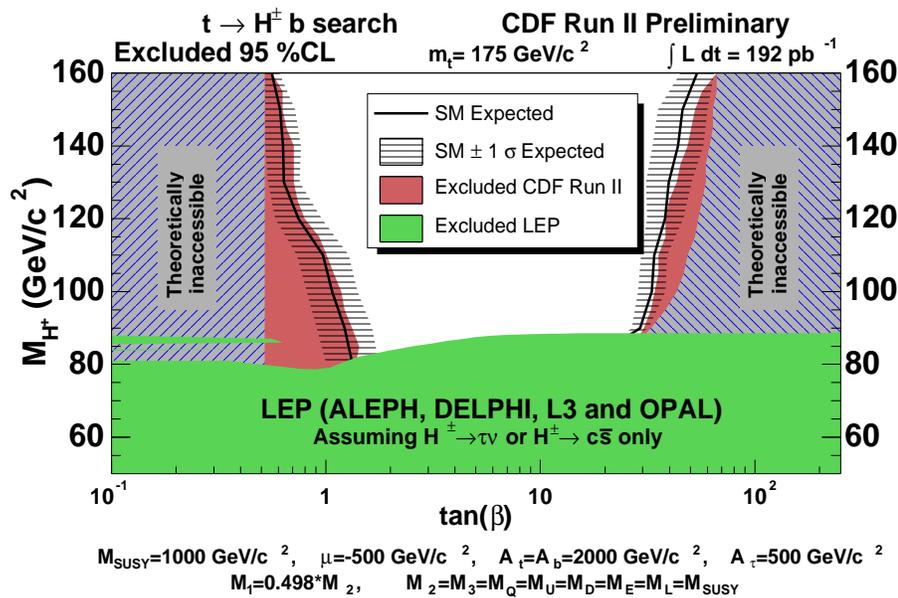}
\end{center}
\vspace*{-0.35cm}
\caption{
CDF. Limits on the charged Higgs boson mass as function of $\tan\beta$ for a specific set of SUSY parameters.
} \label{fig:cdf-tbh3}
\vspace*{-1.cm}
\end{figure}

The decay of top quarks $\rm t\to H^+b$ is possible in general Higgs boson models with two Higgs boson doublets.
The expected top and charged Higgs boson branching fractions are shown in Fig.~\ref{fig:cdf-tbh}
(from~\cite{cdf-tbh}) as a function of $\tan\beta$. The expected SM top decay rate would be modified. 
No deviation from the SM top decay rates are observed (Fig.~\ref{fig:cdf-tbh2}) and resulting limits
are given in Fig.~\ref{fig:cdf-tbh3}.

\clearpage
\subsubsection{$\rm H\to \gamma\gamma$}

In fermiophobic Higgs boson models, the dominant decay mode could be $\rm H\to \gamma\gamma$. 
No indication of a resonance has been observed in the $\gamma\gamma$ invariant mass spectrum
and limits on the  $\rm H\to \gamma\gamma$ production are set as shown in Fig.~\ref{fig:d0-h-gamma} 
(from~\cite{d0-gammagamma}). 

\subsubsection{$\rm H^{++}$}

\begin{figure}[htbp]
\includegraphics[width=0.49\textwidth,bbllx=16pt,bblly=373pt,bburx=437pt,bbury=710pt,clip=]{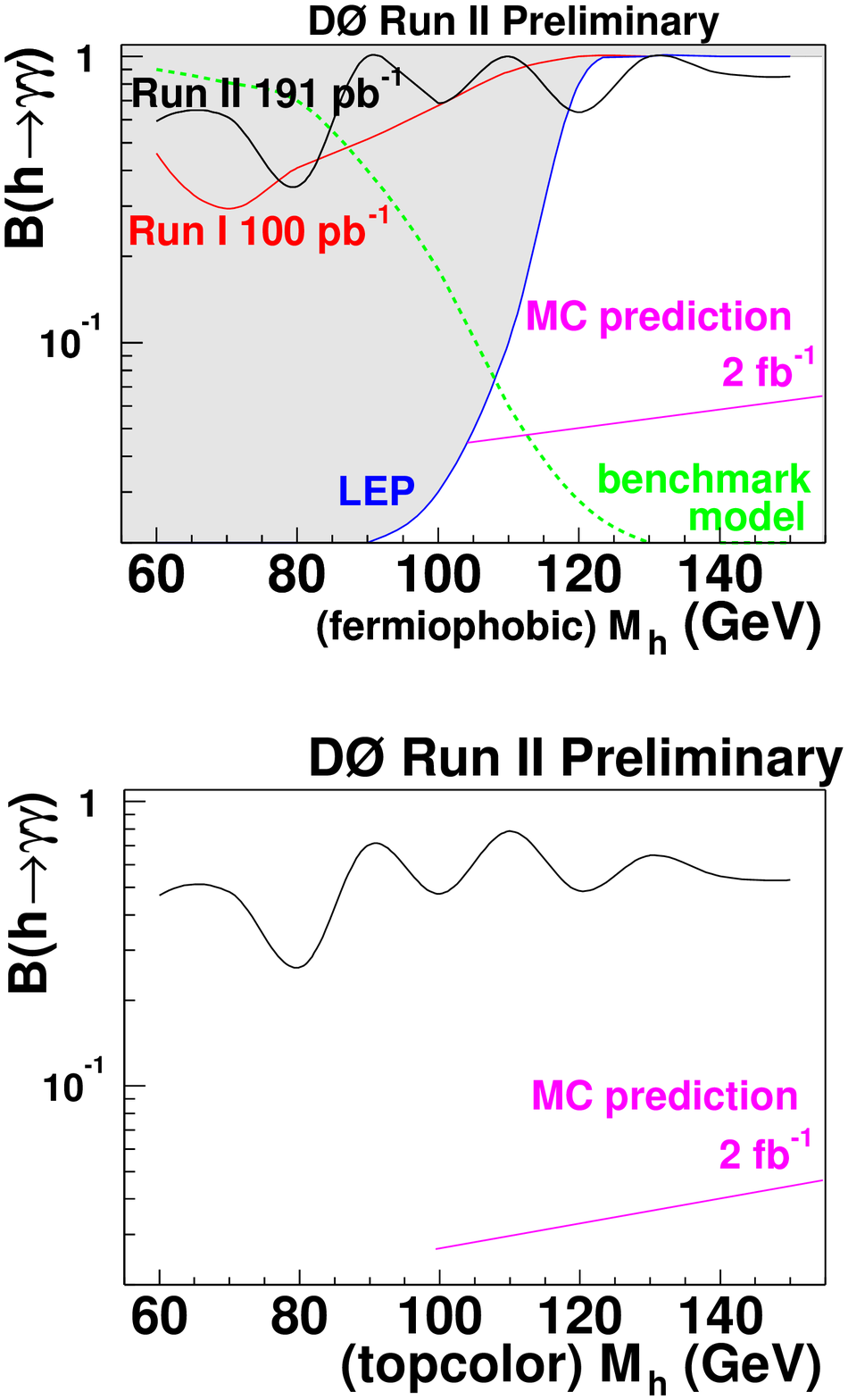} \hfill
\includegraphics[width=0.49\columnwidth]{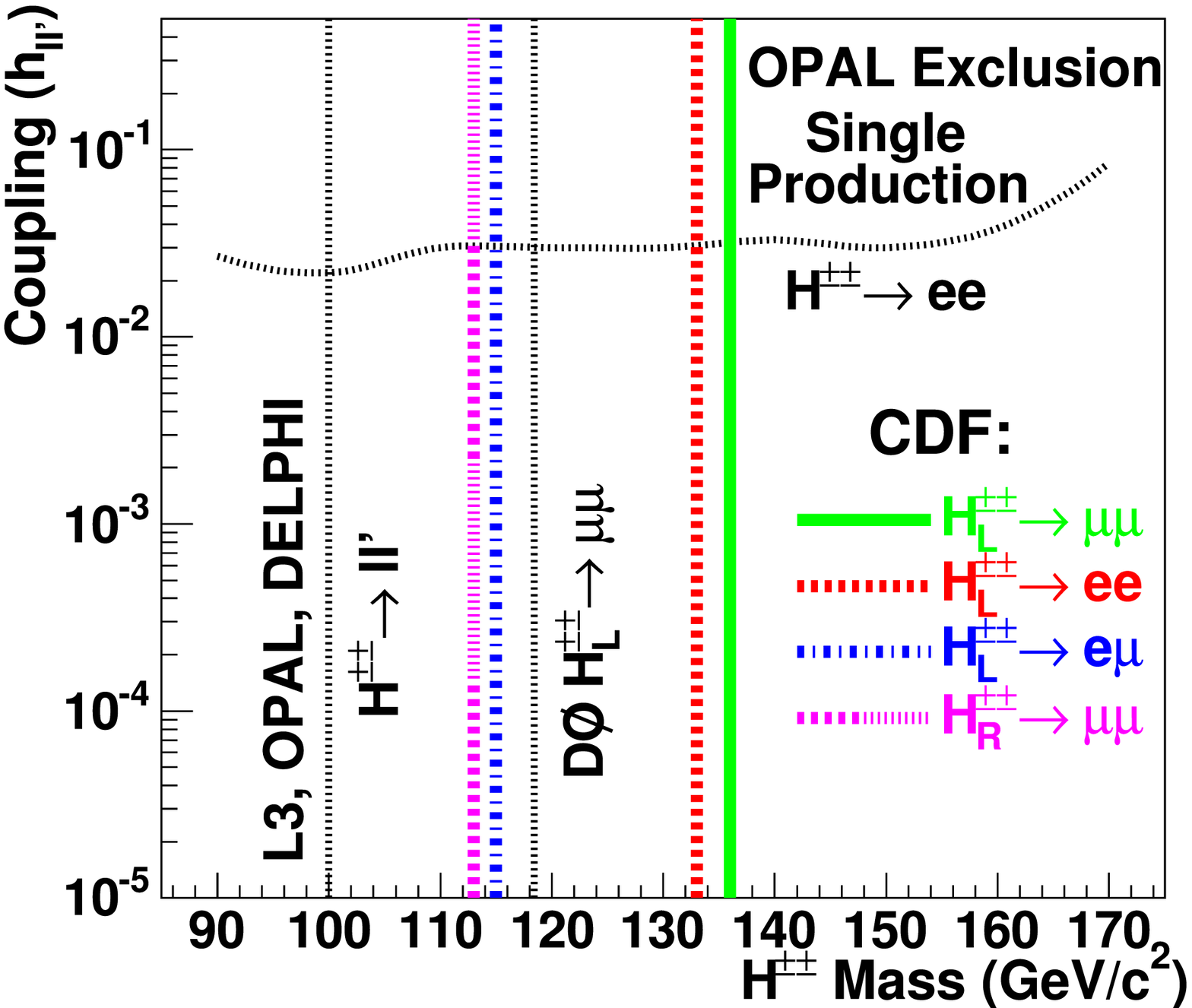} 
\vspace*{-0.1cm}
\caption{
Left: D\O\ limits on $\rm H\to \gamma\gamma$. 
Right: CDF and D\O\ doubly charged Higgs boson mass limits.
} \label{fig:d0-h-gamma}
\vspace*{2cm}
\end{figure}

The possibility of doubly-charged Higgs boson exists in models with Higgs boson triplets.
Pairs of like-sign charged leptons are expected from the decay of the doubly-charged
Higgs bosons. No indication has been observed in the data and limits on the double charged
Higgs boson mass from CDF~\cite{cdf-hpp} and D\O~\cite{d0-hpp} are given in Fig.~\ref{fig:d0-h-gamma}.

%\clearpage
\section{Supersymmetric Particle Searches at the Tevatron}

\begin{figure}[htbp]
%\vspace*{-0.5cm}
\includegraphics[width=0.49\textwidth]{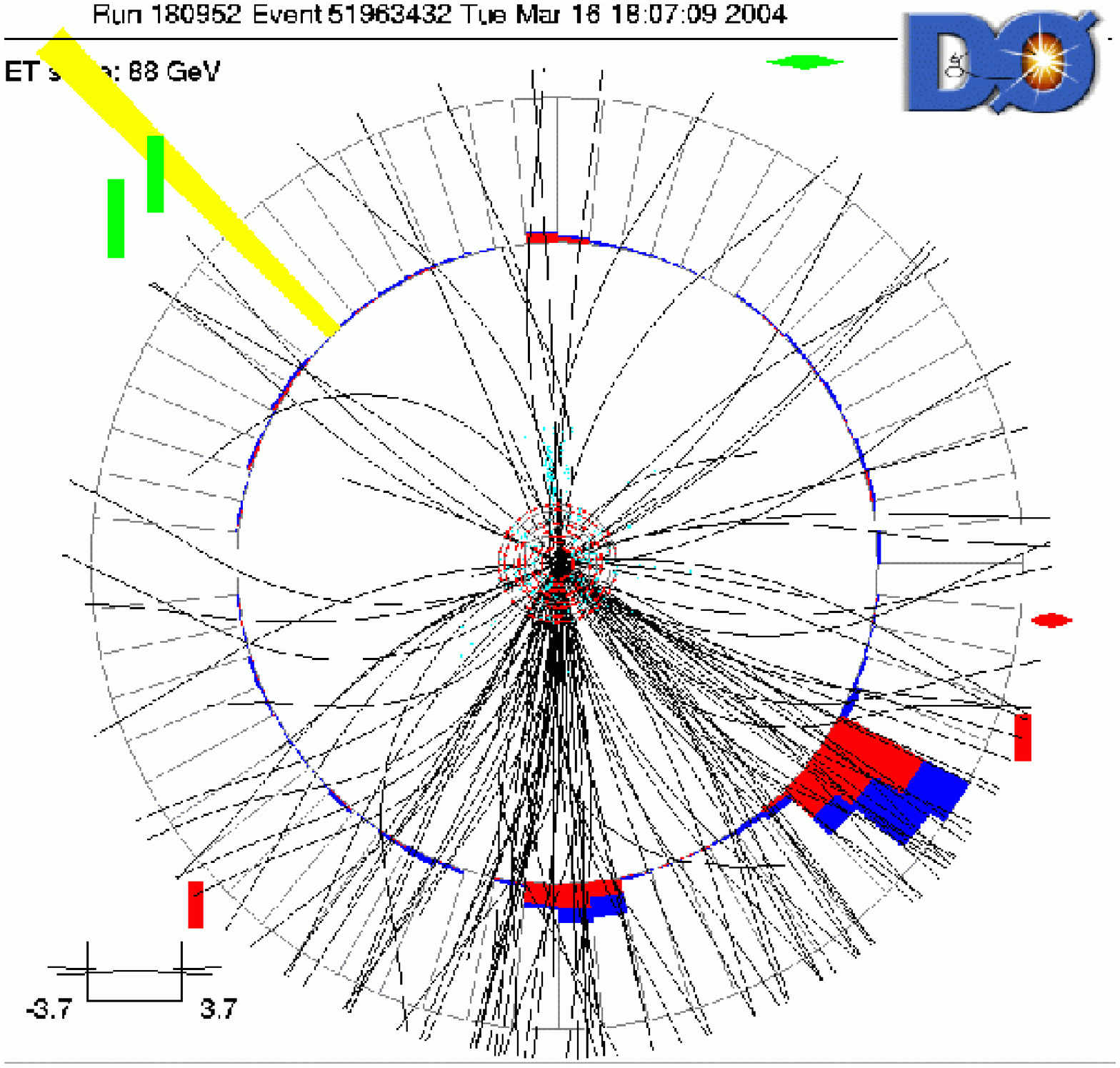} \hfill
\includegraphics[width=0.49\textwidth,bbllx=0pt,bblly=90pt,bburx=610pt,bbury=700pt,clip=]{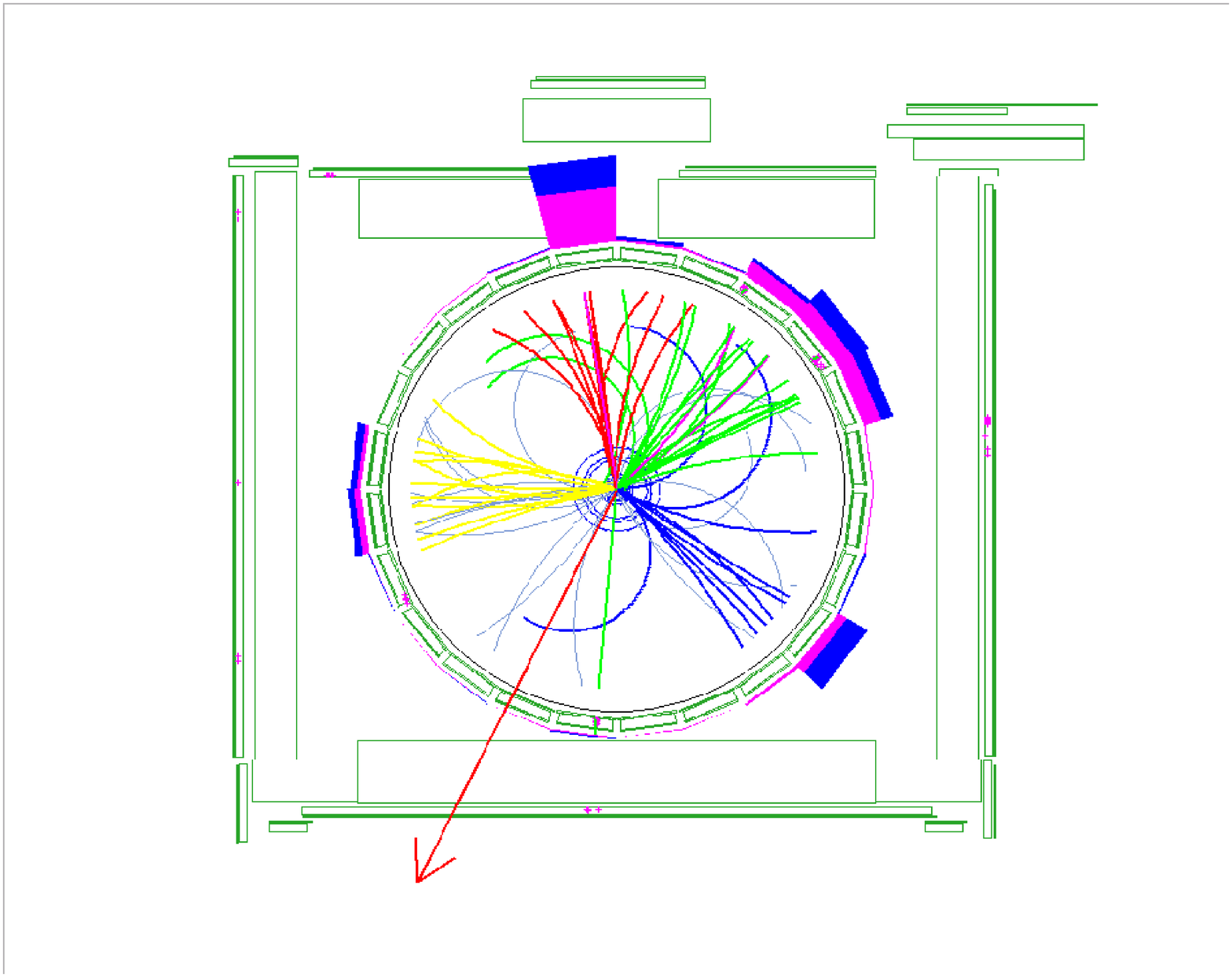} 
%\vspace*{-0.5cm}
\caption{Events with large missing momentum and energy. 
Left: D\O\ $H_T=410$~GeV. 
Right: CDF $H_T=404$~GeV. 
} \label{fig:ht}
\end{figure}

In the production of Supersymmetric particles (sparticles) the lightest supersymmetric
particle (LSP) is stable (if R-parity is conserved) 
and escapes detection, leading to missing momentum and missing 
energy in the recorded events. Figure~\ref{fig:ht} shows events with jets and large 
$H_T$ defined as the sum of the transverse jet momenta for D\O~\cite{d0-sq-gl} and 
defined as the sum of the transverse jet energies for CDF~\cite{cdf-ht}.  

\begin{figure}[h!]
\includegraphics[width=0.48\textwidth]{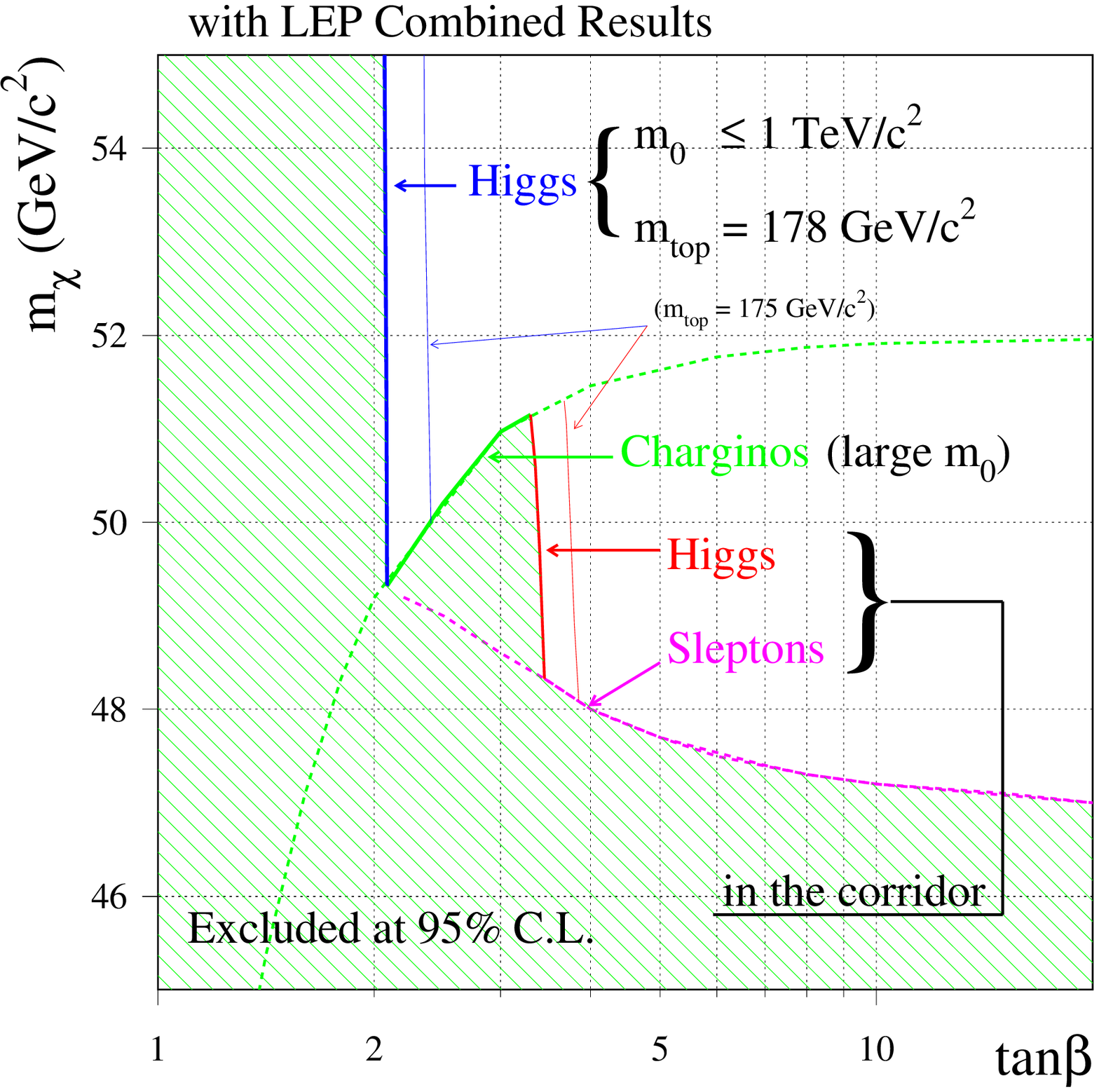} \hfill
\includegraphics[width=0.48\textwidth]{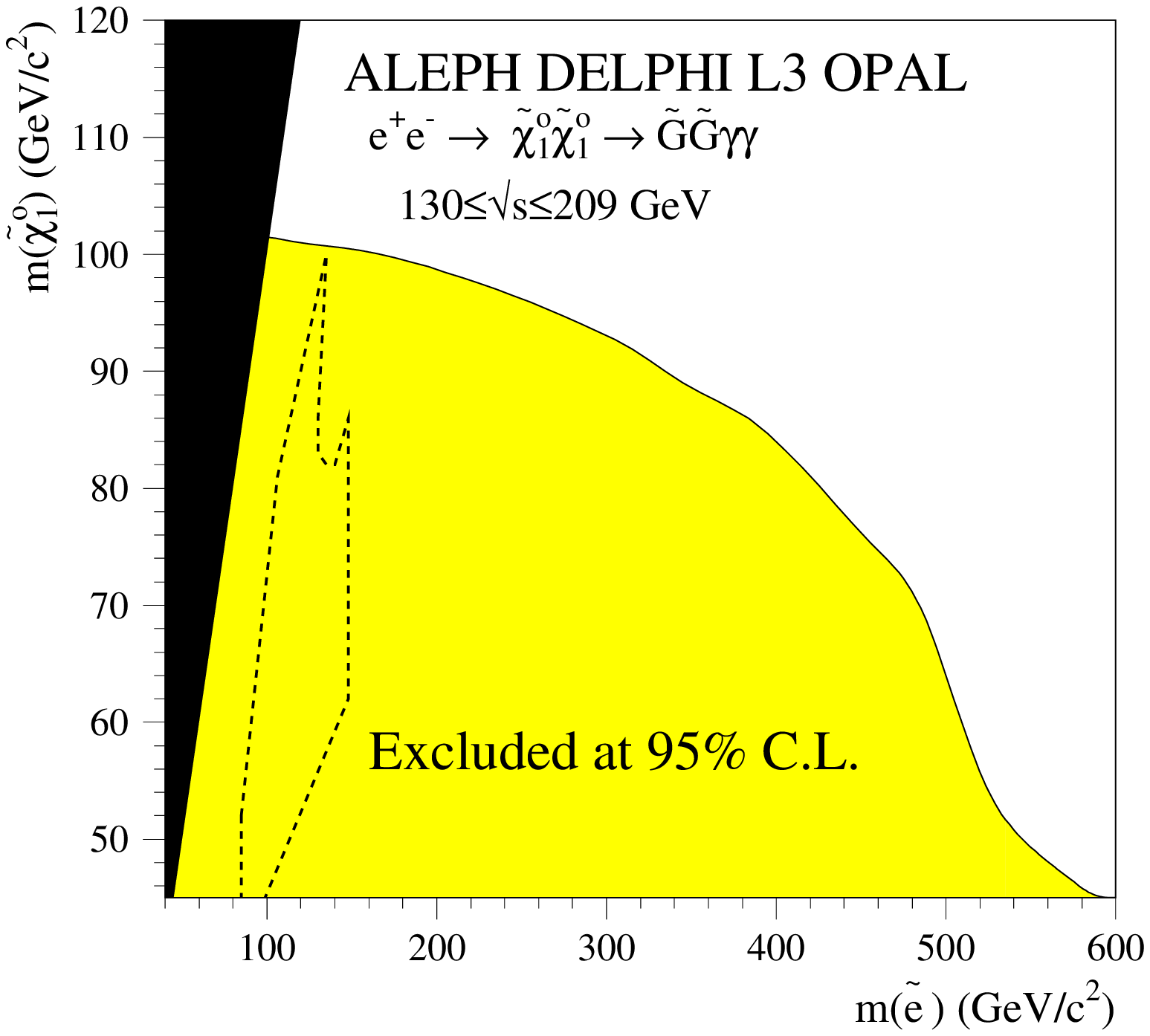} 
\vspace*{-0.5cm}
\caption{
Left: combined LEP limits in the $\tan\beta$ versus neutralino plane in a constrained MSSM.
Right: combined LEP exclusion in the selectron versus neutralino plane from acoplanar photon searches. 
The CDF Run-I $\rm ee\gamma\gamma$ candidate event is indicated in the excluded region.
} \label{fig:lep-limits}
%\vspace*{-1.5cm}
\end{figure}

Previous combined results from the LEP experiments have set limits on several sparticles close to the
kinematic reach. 
Figure~\ref{fig:lep-limits} (from~\cite{lep-susy})
shows the excluded region in the $\tan\beta$ versus LSP mass plane in the MSSM,
and in the Gauge Mediated Supersymmetry Breaking (GMSB) model. The position of an intriguing 
CDF $\rm ee\gamma\gamma$ candidate event~\cite{cdf-eegg} from the Tevatron Run-I is indicated 
(dashed line) and the GMSB interpretation for this event is excluded at 95\% CL.

%\clearpage
\subsection{Di-photon (GMSB Interpretation)}

\begin{figure}[htbp]
%\vspace*{-0.3cm}
\includegraphics[width=0.18\textwidth,height=4cm]{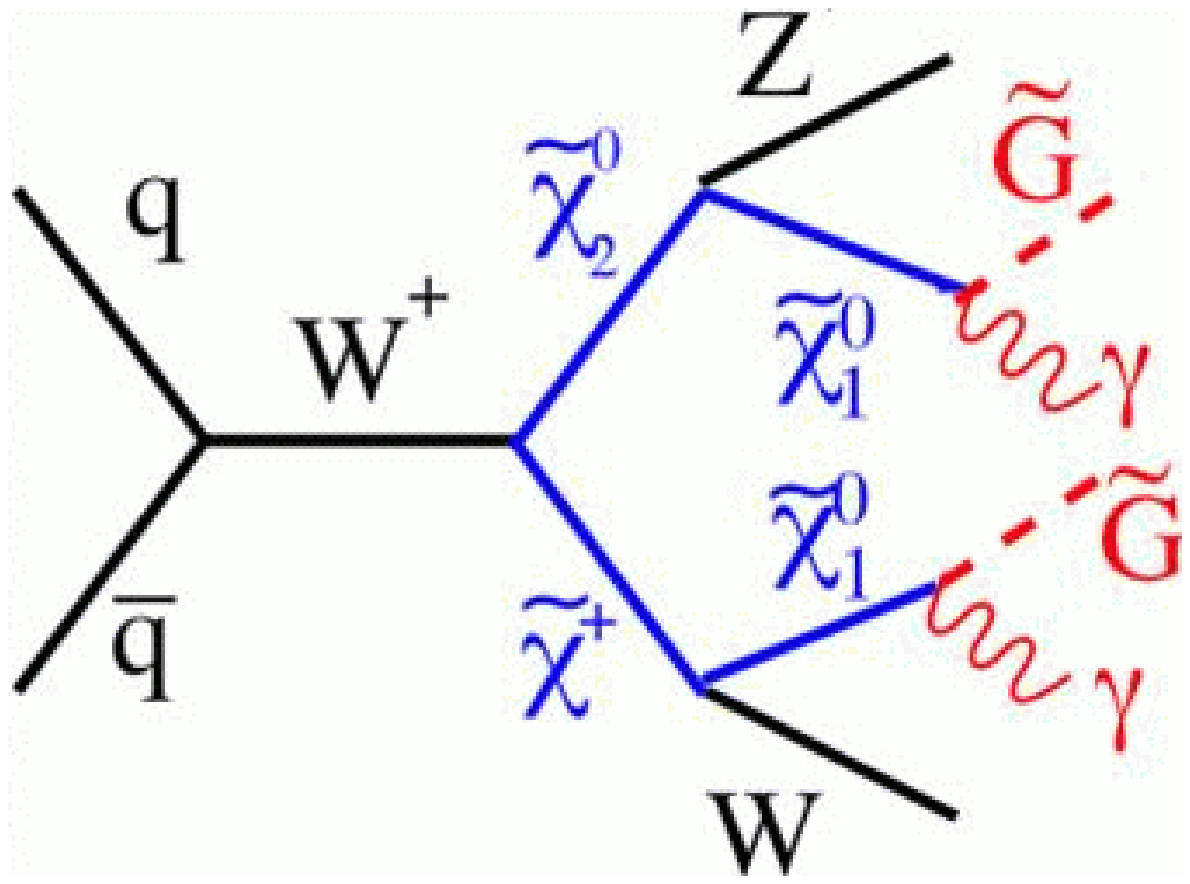}
\includegraphics[width=0.4\textwidth,height=6cm]{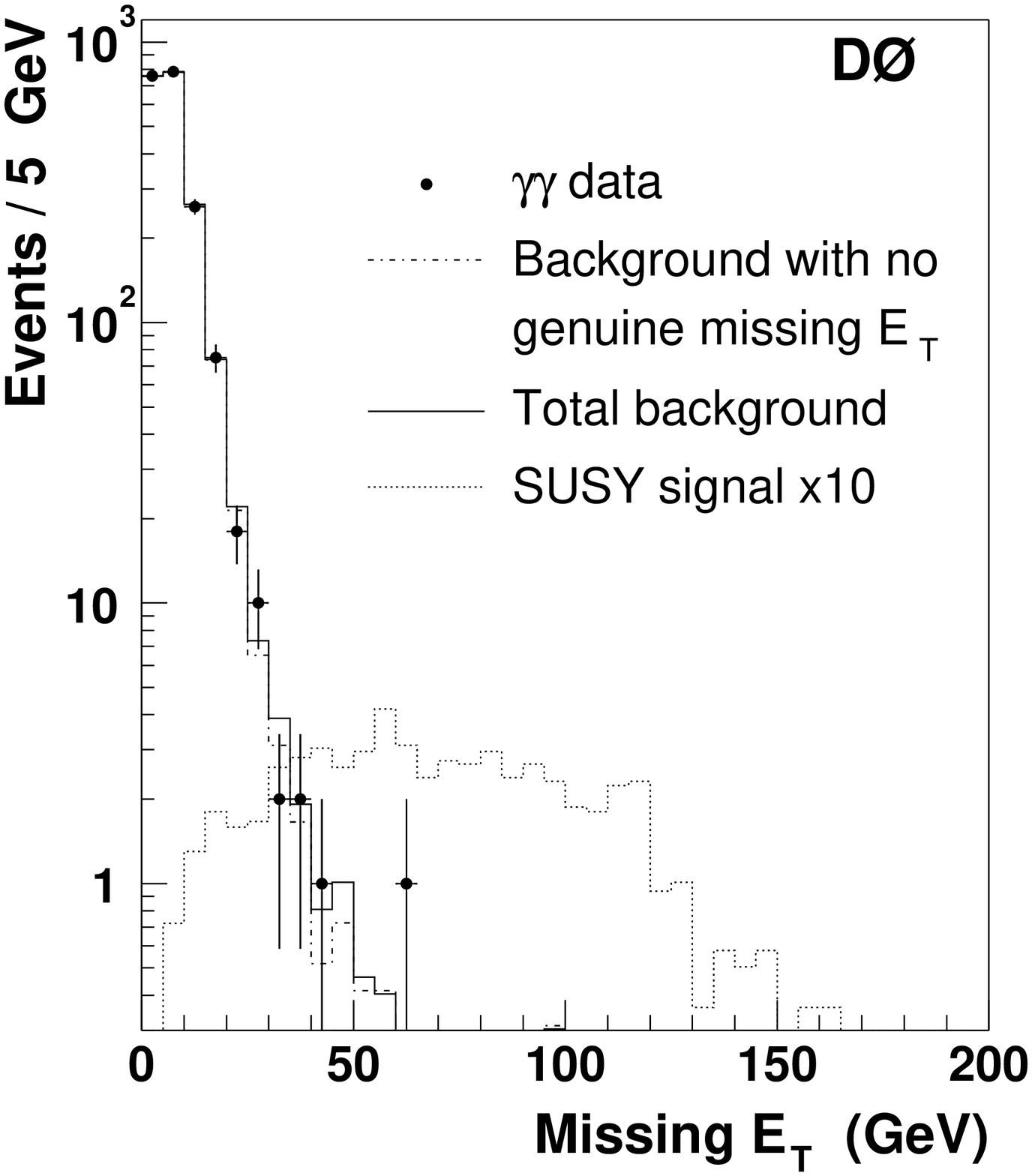} \hfill
\includegraphics[width=0.4\textwidth,height=6cm]{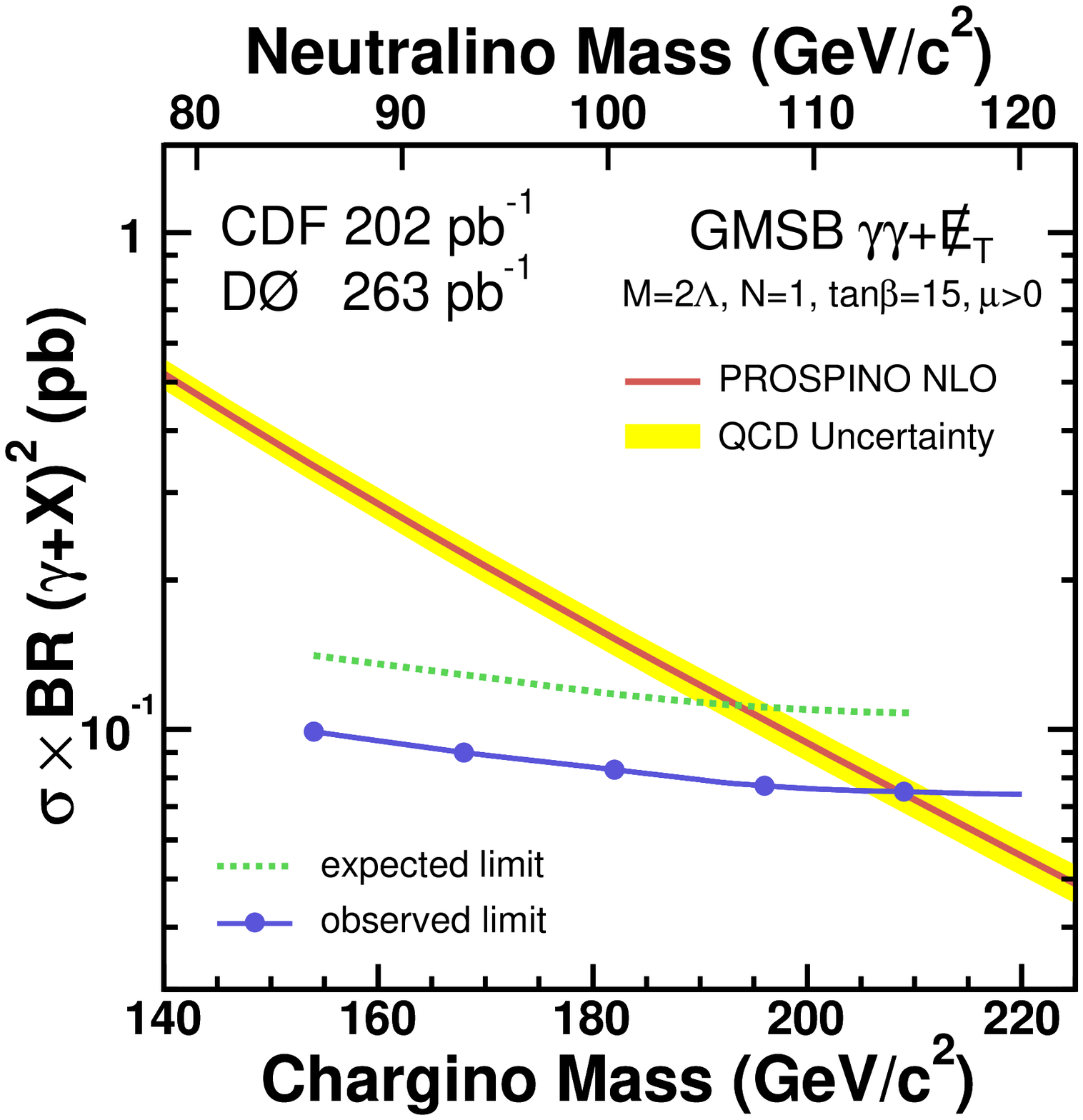} 
\vspace*{-1cm}
\caption{D\O\ GMSB. 
Left: production graph.
Center: missing transverse energy.
Right: combined CDF and D\O\ exclusion.
} \label{fig:d0-gmsb}
\vspace*{0.3cm}
\end{figure}

In the GMSB model, chargino-neutralino production has been searched for by CDF and D\O.
The production reaction is illustrated in Fig.~\ref{fig:d0-gmsb} (from~\cite{d0-gmsb}). The figure shows
also that the expected signal and background can be well separated.
There is no indication of a signal in the data and the combined limits on the chargino mass
from CDF and D\O\ are given~\cite{cdf-d0-gmsb}.

\subsection{Tri-Lepton Signatures}

\begin{figure}
%\vspace*{-0.5cm}
\includegraphics[width=0.32\textwidth,height=5cm]{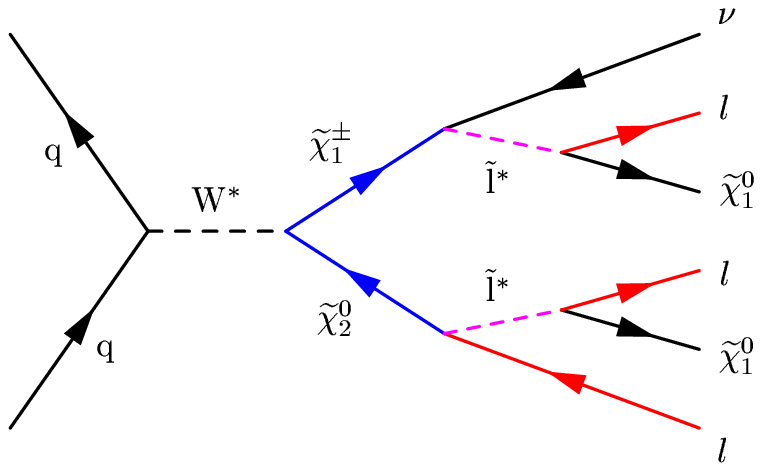}\hfill
\includegraphics[width=0.3\textwidth,height=5cm]{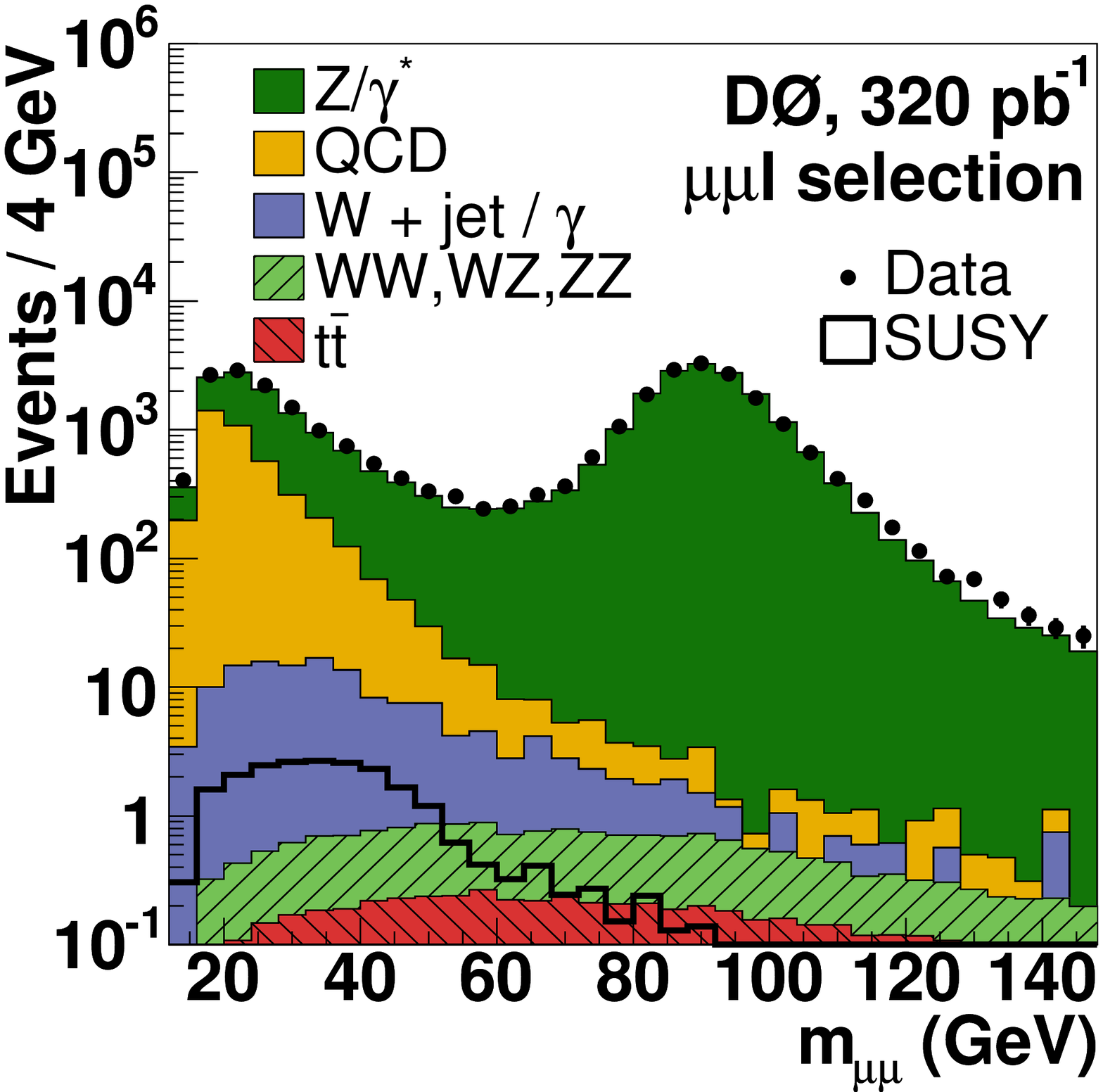}\hfill
\includegraphics[width=0.3\textwidth,height=5cm]{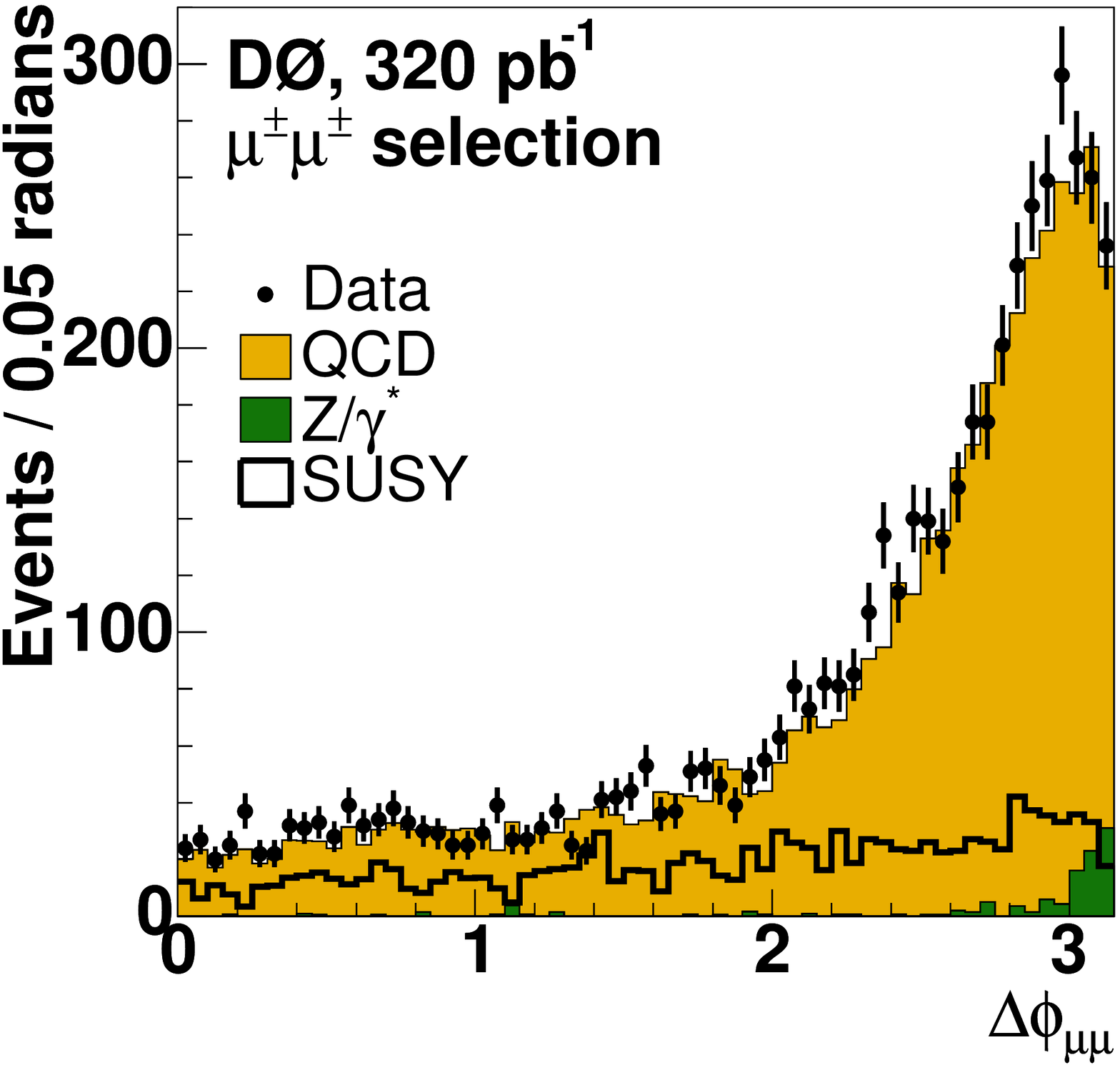}
\vspace*{-0.5cm}
\caption{\label{fig:d0-tri-l-1}
D\O. Left: tri-lepton production graph.
Center: invariant di-muon mass $m_{\mu\mu}$.
Right: di-muon opening angle~$\Delta\phi_{\mu\mu}$ for data, background
and simulated signal.
}
\end{figure}

\begin{figure}
\begin{minipage}{0.65\textwidth}
\includegraphics[width=0.48\textwidth]{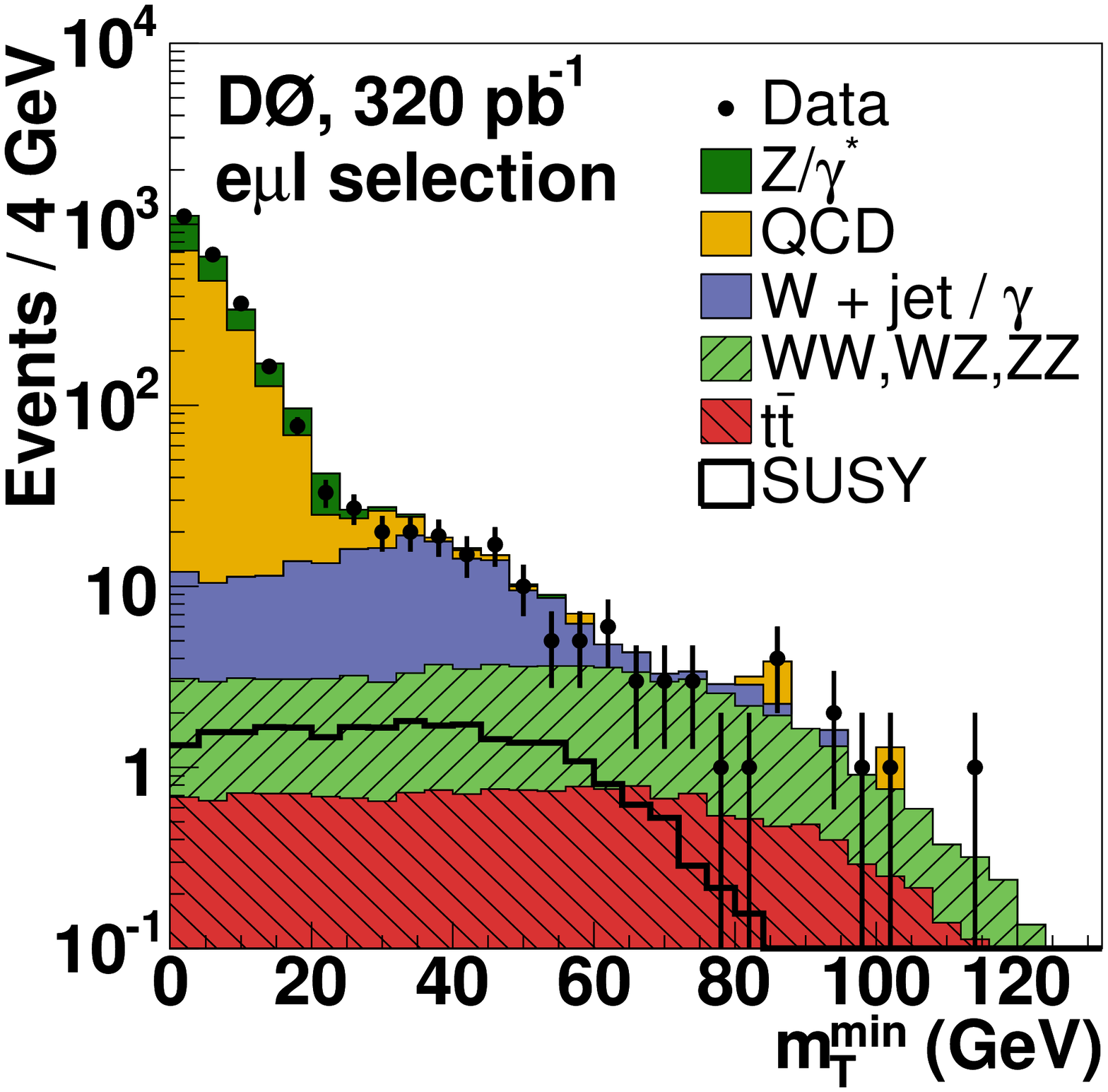} \hfill
\includegraphics[width=0.48\textwidth]{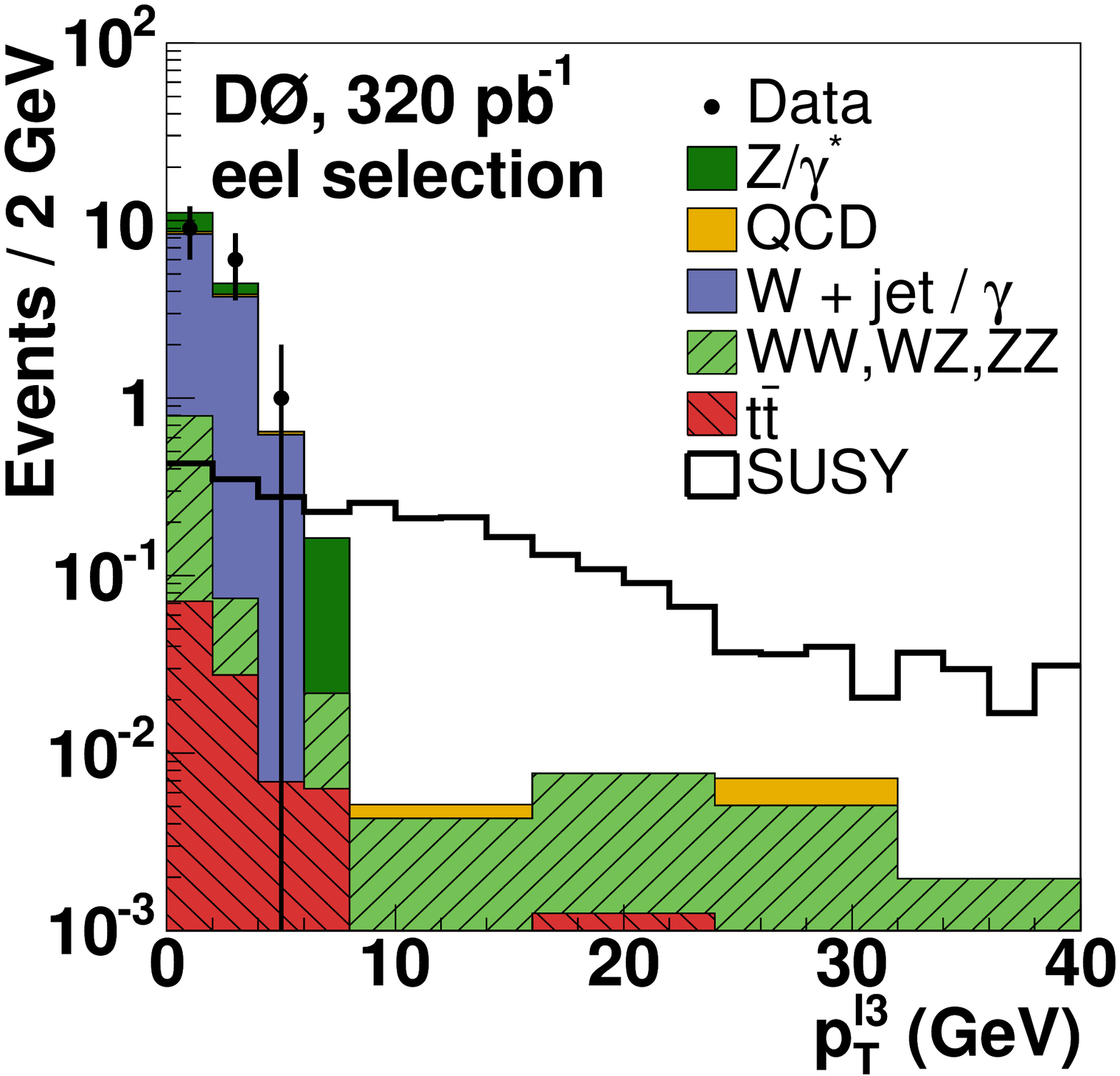}
\end{minipage}  \hfill
\begin{minipage}{0.32\textwidth}
\begin{turn}{270}
\includegraphics[height=\textwidth,width=5.5cm]{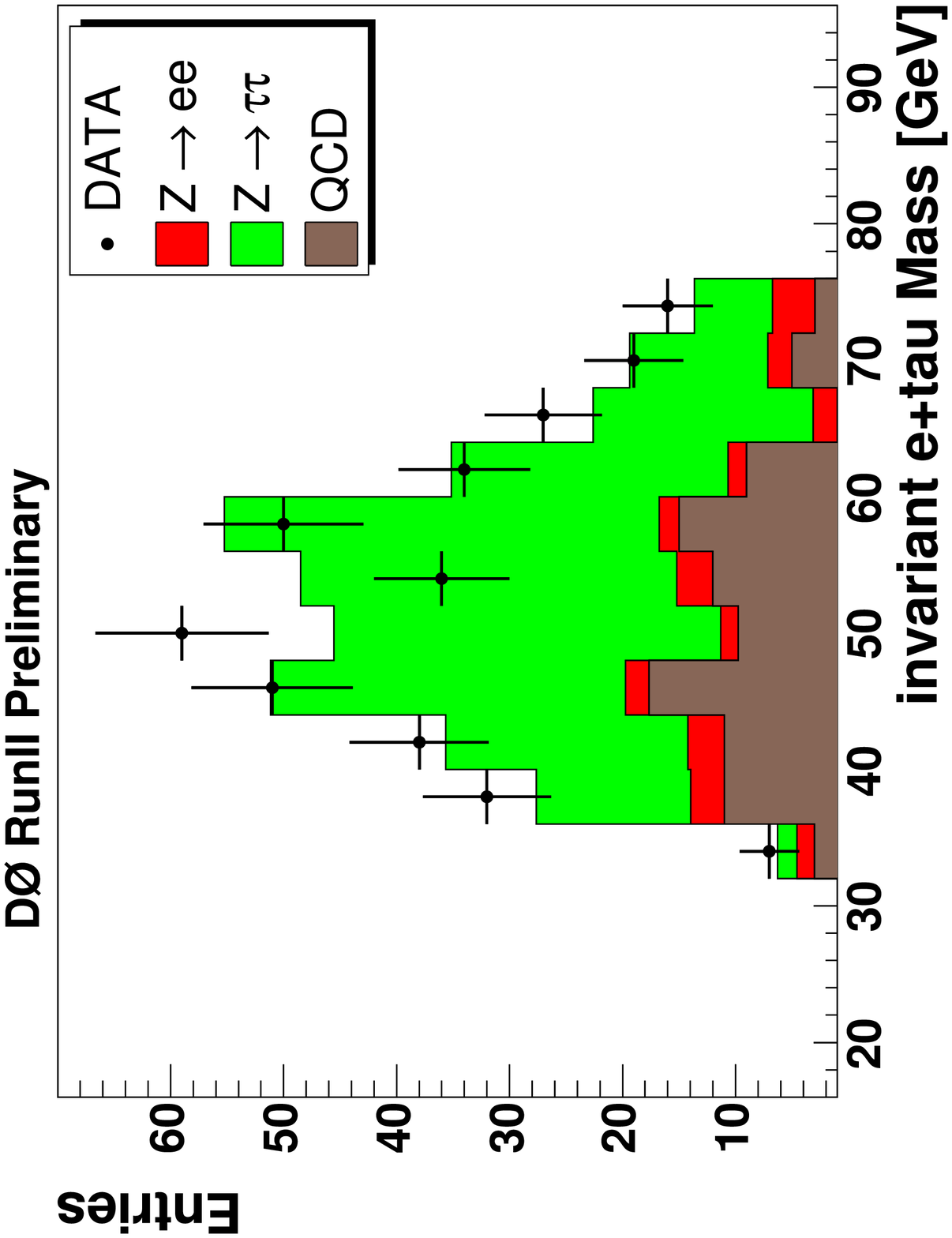}
\end{turn}
\end{minipage}
\vspace*{-0.4cm}
\caption{\label{fig:d0-tri-l-2}
D\O. Left: minimum transverse mass $m_{\rm T}^{\rm min}$.
Center: transverse momentum of the third track $p_{\rm T}^{\rm \ell 3}$ for data, 
background and simulated signal.
Right: invariant mass of $\rm e\tau$.
}
\vspace*{-0.2cm}
\end{figure}

\begin{figure}
\includegraphics[width=0.32\textwidth,height=6cm]{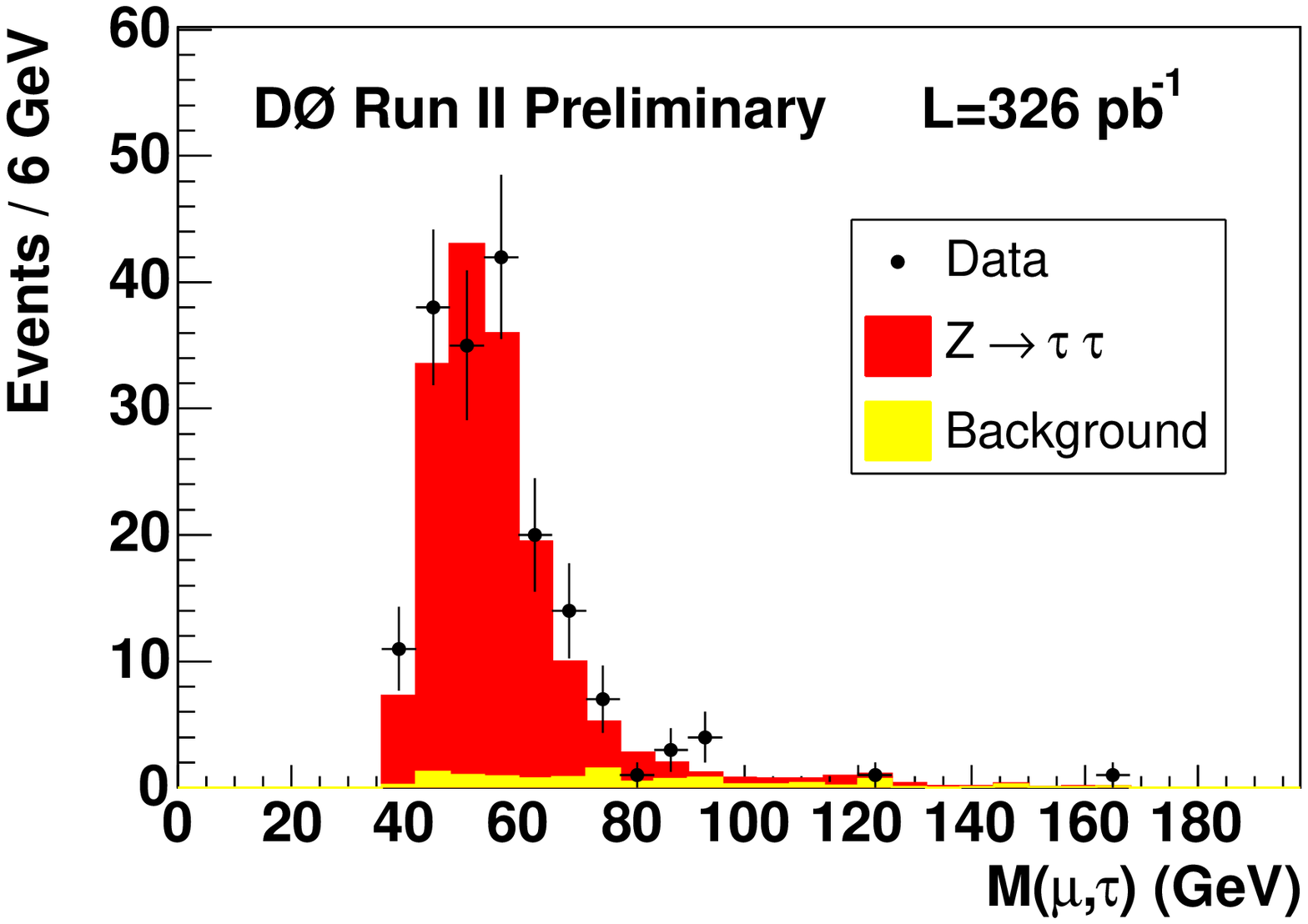} \hfill
\includegraphics[width=0.32\textwidth,height=5.5cm]{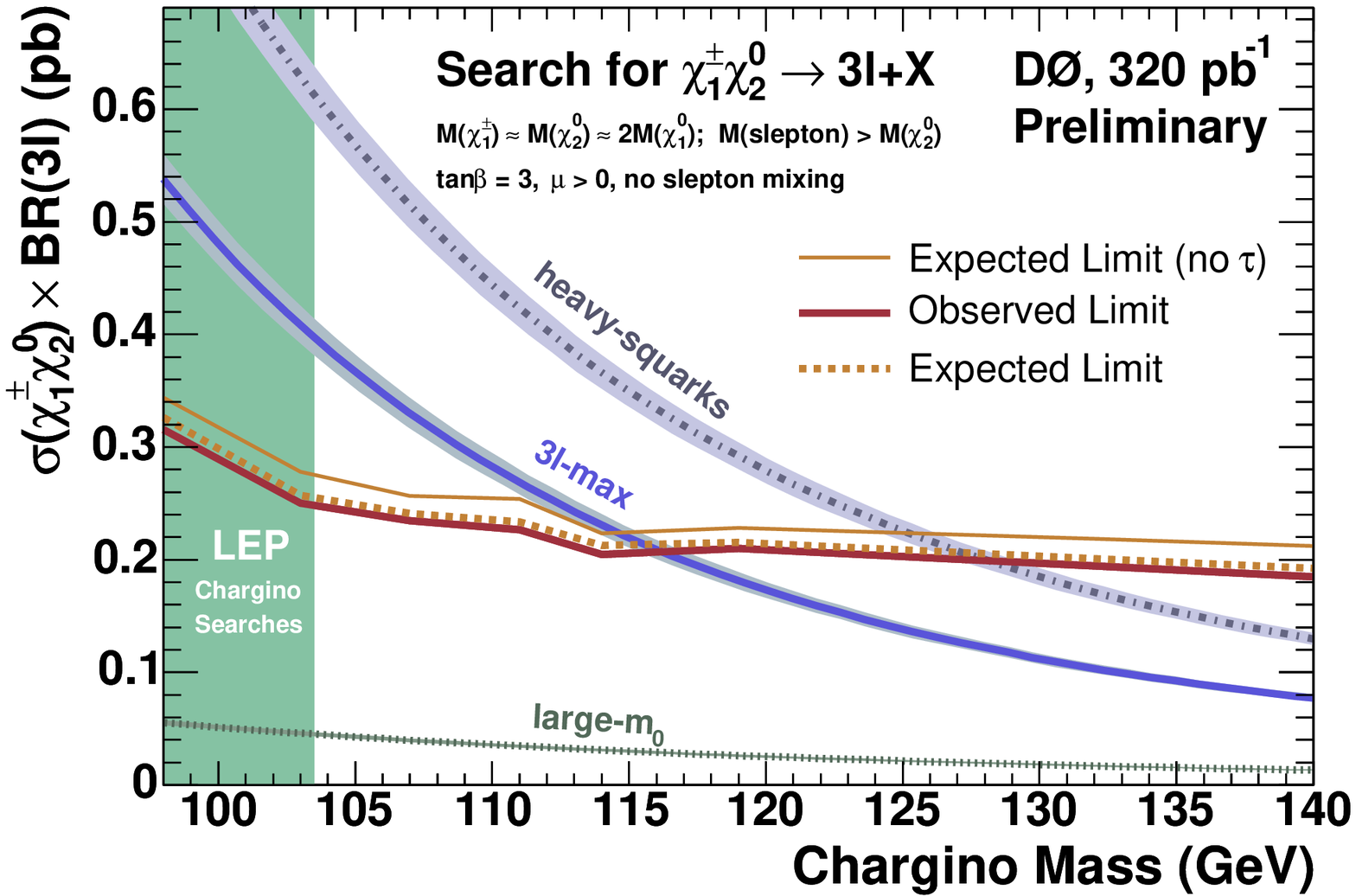}\hfill
\includegraphics[width=0.32\textwidth,height=5.5cm]{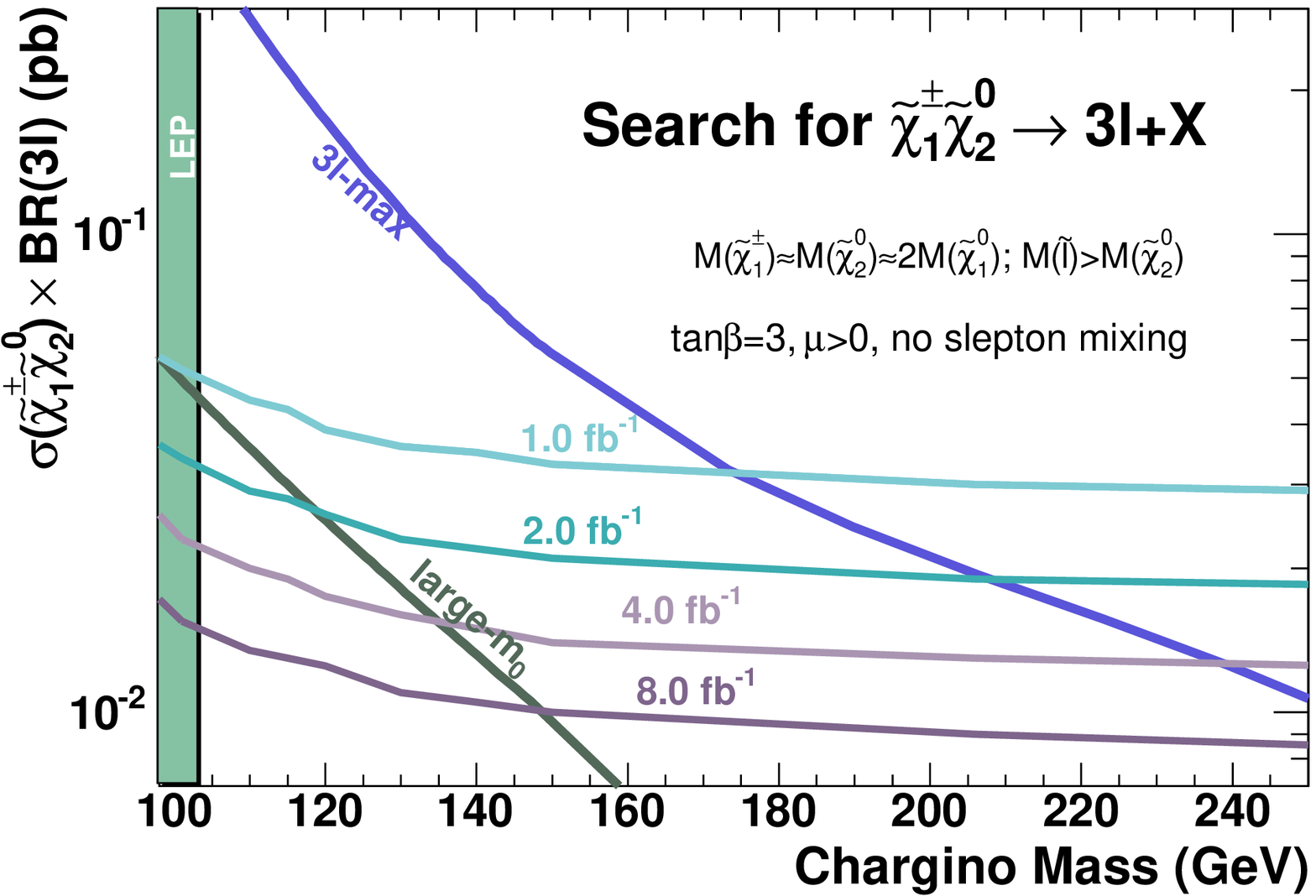}
\vspace*{-0.4cm}
\caption{
D\O. Left: invariant mass of $\rm \mu\tau$.  
Center:  limits from the tri-lepton searches.
Right: sensitivity expectations from the tri-lepton searches with larger luminosities.
}
\label{fig:d0-tri-l-3}
\vspace*{-0.3cm}
\end{figure}

The sparticle production with a tri-lepton final state is illustrated in 
Fig.~\ref{fig:d0-tri-l-1}. Figures~\ref{fig:d0-tri-l-1},~\ref{fig:d0-tri-l-2}, and~\ref{fig:d0-tri-l-3}
(from~\cite{d0-tri}) show the good agreement between data and simulated background for various 
final states with a tri-lepton signatures.
There is no indication of a signal and resulting limits are given as a function of the chargino mass.
With increasing luminosities 1, 2, 4, 8~\fb\ and for a scenario with maximal leptonic branching ratio, 
the expectations for the chargino mass reach are 170, 210, 235 and 265~GeV,
as shown in Fig.~\ref{fig:d0-tri-l-3} (from~\cite{expectations}).

\subsection{Scalar Quarks and Gluinos}

\begin{figure}[htbp]
%\vspace*{-0.2cm}
\includegraphics[width=0.48\textwidth]{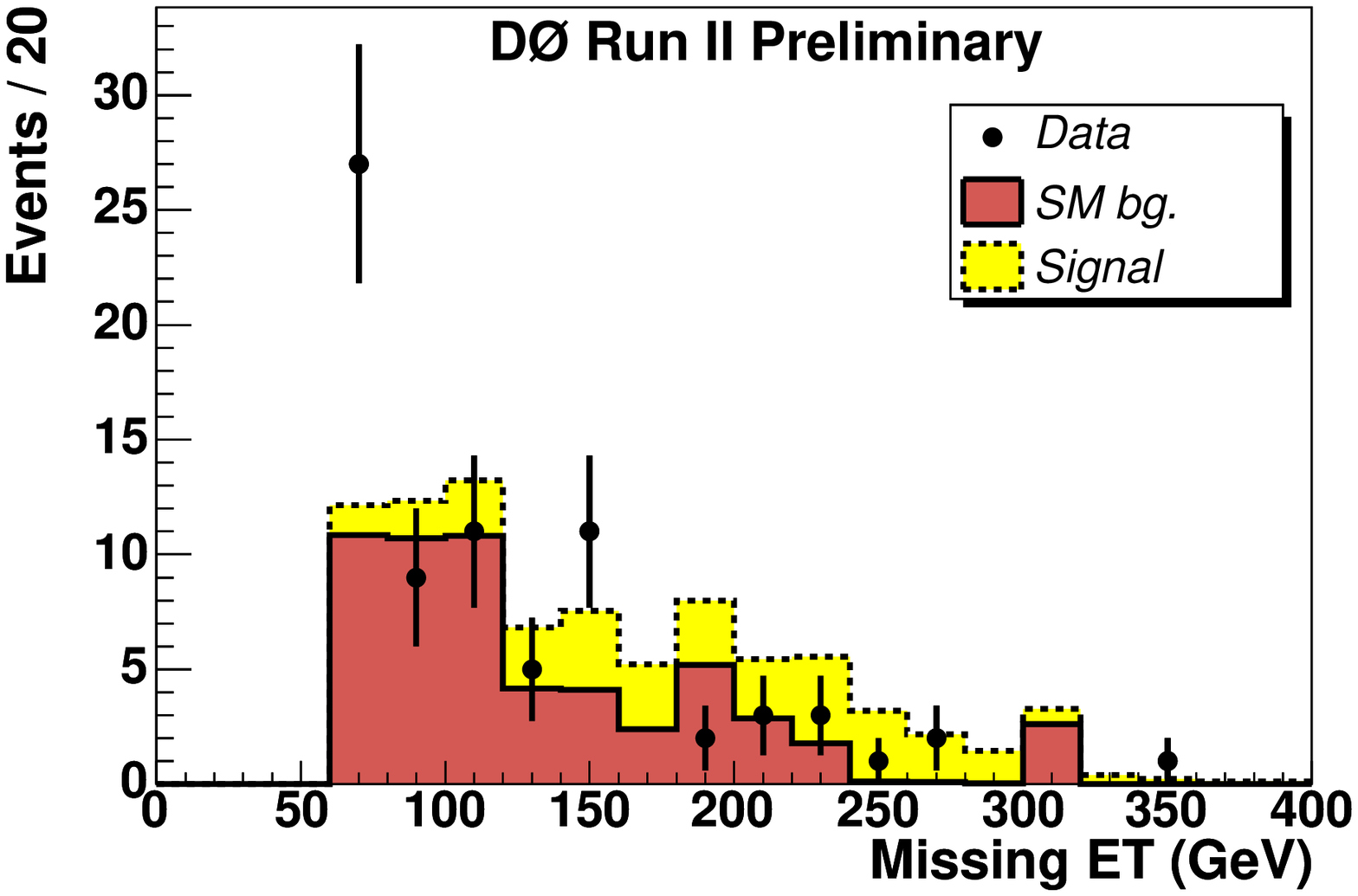} \hfill
\includegraphics[width=0.48\textwidth]{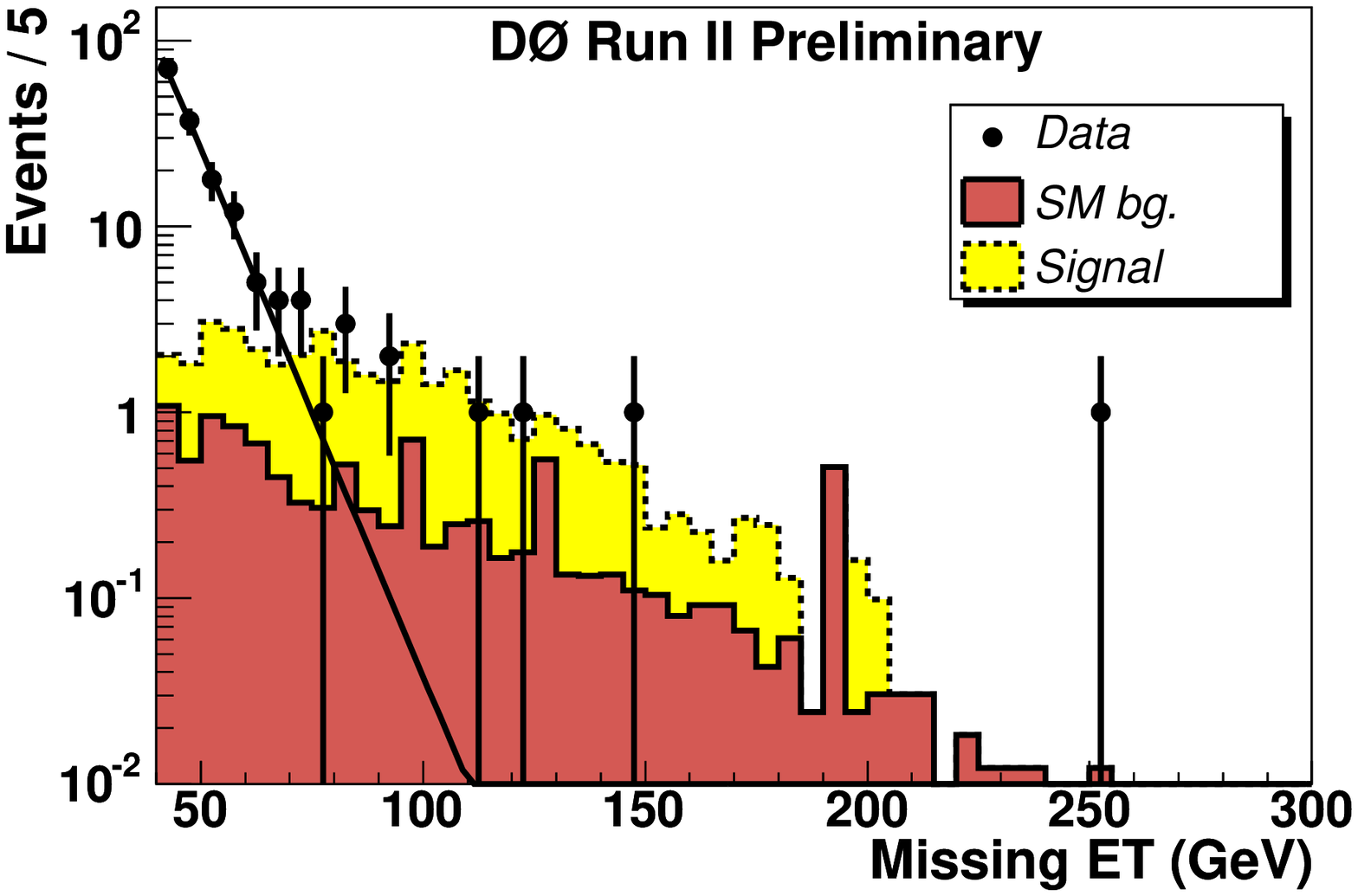} 
\vspace*{-0.2cm}
\caption{
D\O. Left: gluino heavier than squark, squark pair-production is expected, leading to acoplanar di-jets.
The excess at low missing $E_T$ is due to QCD background.    
Right: squark heavier than gluino, gluino pair-production is expected, leading to four and more jets.
} \label{fig:d0-sq-gl}
\vspace*{-0.2cm}
\end{figure}

\begin{figure}[htbp]
%\vspace*{-0.9cm}
\includegraphics[width=0.48\textwidth]{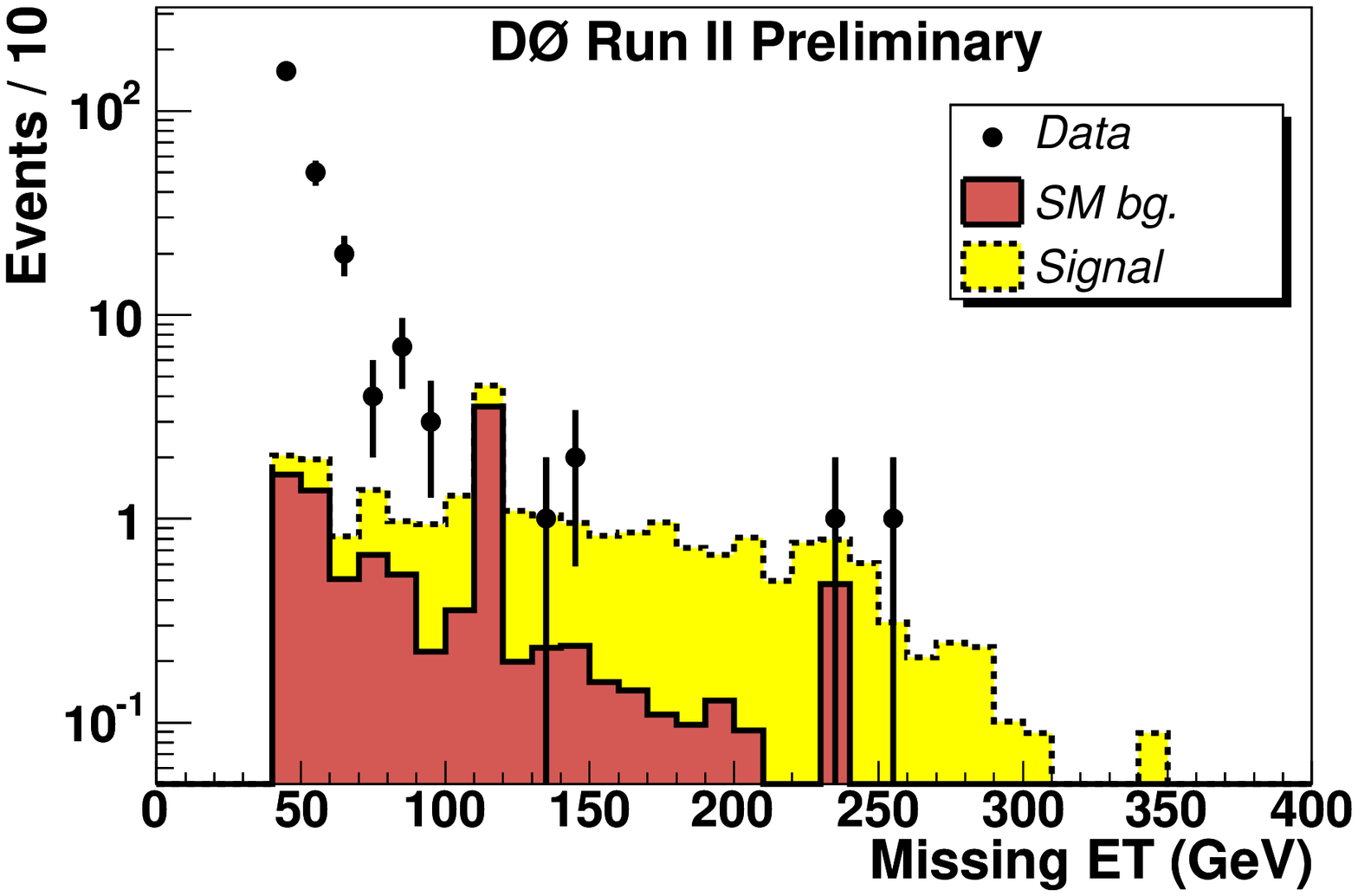} \hfill
\includegraphics[width=0.4\textwidth]{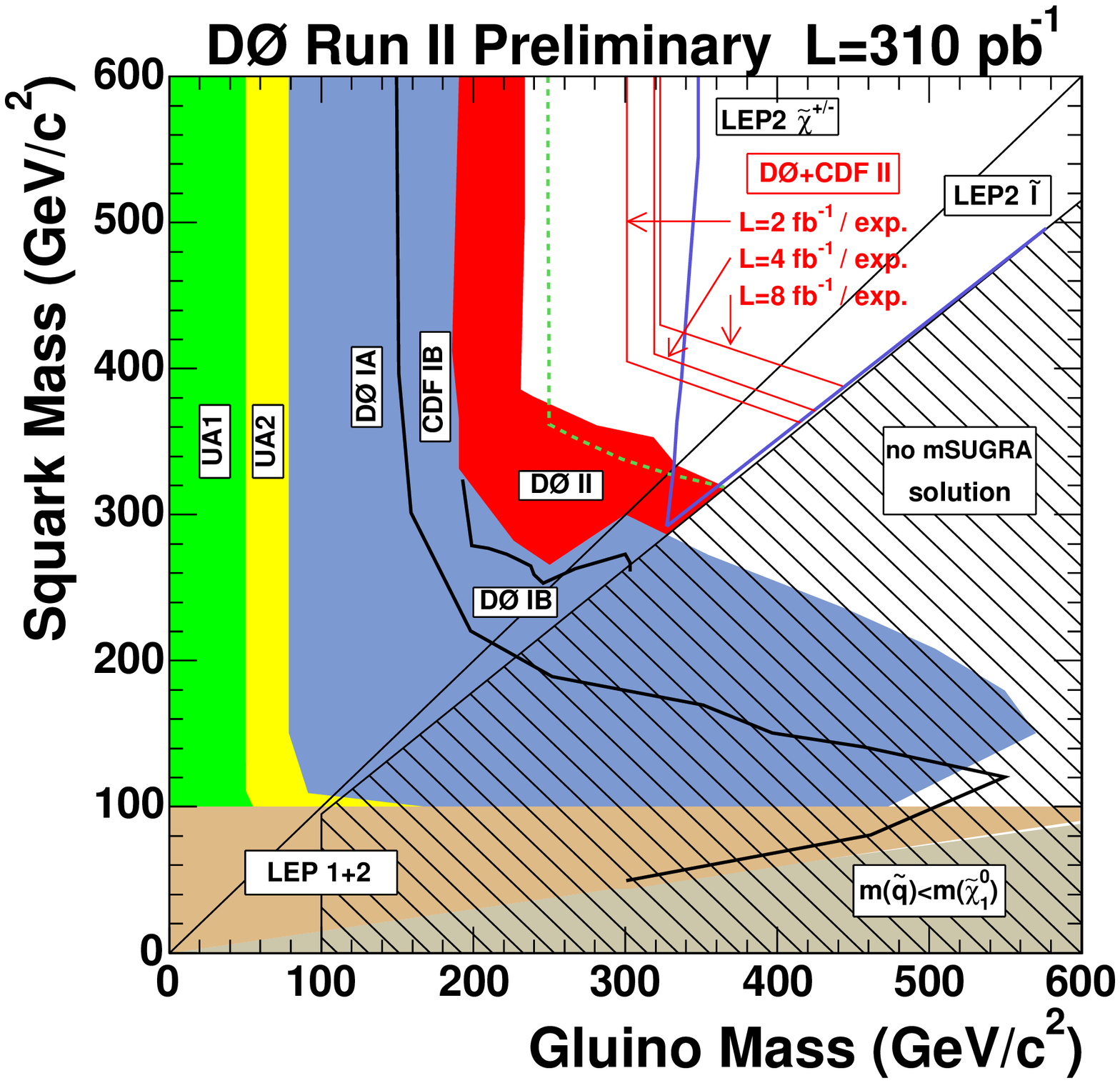}
\vspace*{-0.2cm}
\caption{
D\O. Left: gluino and squark masses are about equal, squark-gluino production is expected, leading typically to 
three jets.
The excess at low missing $E_T$ is due to QCD background.    
Right: squark-gluino mass limits. With the current luminosity, only a 
small region beyond  the LEP limits near the diagonal is excluded. The sensitivity reach with larger
luminosities is also shown.
} \label{fig:d0-sq-gl-limit}
%\vspace*{-0.7cm}
\end{figure}

Depending on the masses of squarks and gluinos, either $\rm \tilde q \bar{\tilde q}$, $\rm\tilde g \tilde g$, 
or $\rm\tilde q \tilde g$ could be produced. For gluinos heavier than squarks, squark pair-production dominates,
leading to a signature of two acoplanar jets and missing energy from the escaping neutralino. If the
squarks are heavier than gluinos, gluino pair-production is expected, leading to four, and more, jets. 
If both masses are about equal, squark-gluino production is expected.
Figures~\ref{fig:d0-sq-gl} and~\ref{fig:d0-sq-gl-limit} (from~\cite{d0-sq-gl}) 
show the missing $E_T$ distributions for expected signal and background in the 
three cases, and also the resulting mass limits and future sensitivity expectations~\cite{expectations}.

\subsection{Scalar Tops and Scalar Bottoms}

\begin{figure}[htbp]
%\vspace*{-0.4cm}
\includegraphics[width=0.32\textwidth,height=5cm]{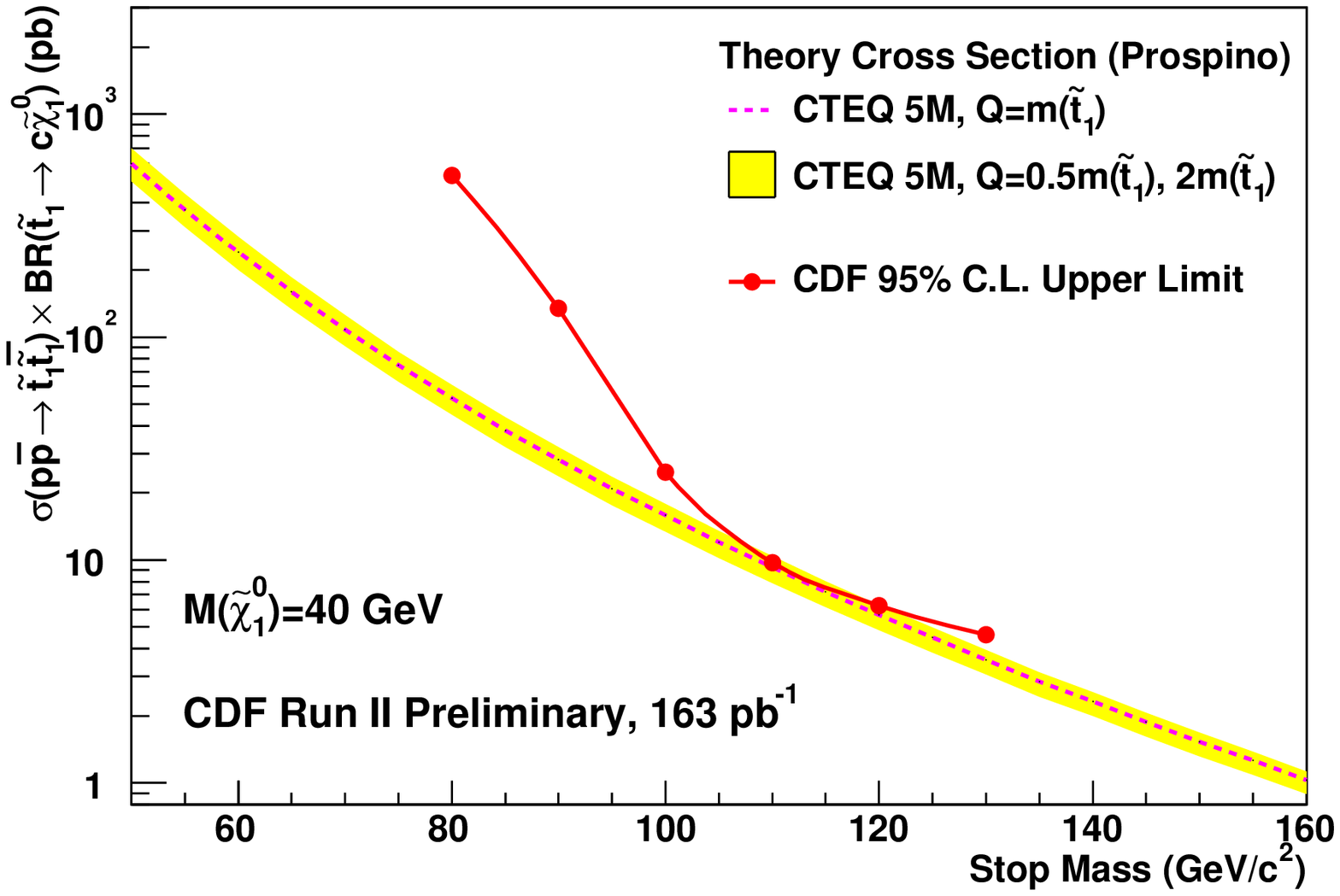} \hfill
\includegraphics[width=0.32\textwidth,height=5cm]{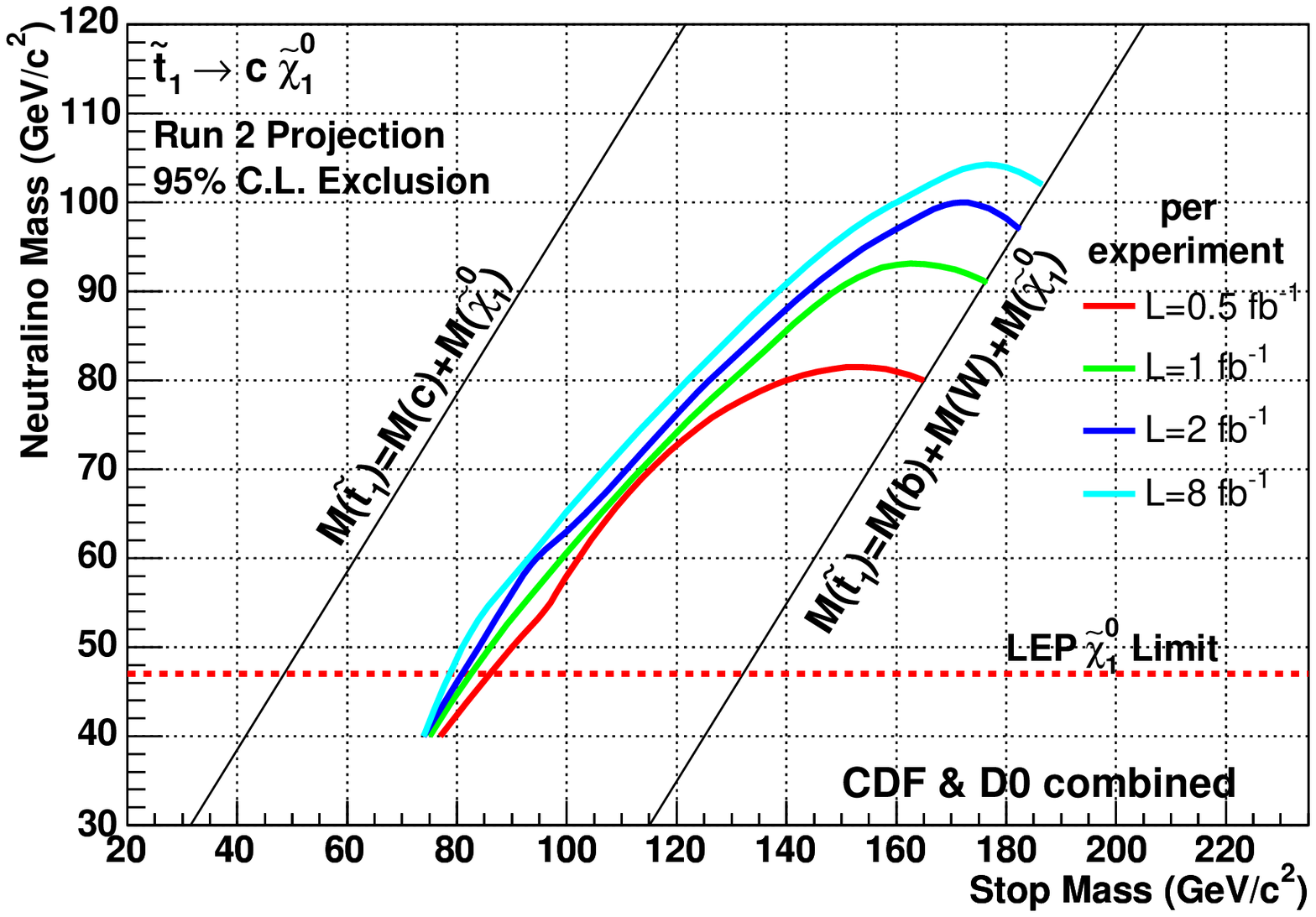}\hfill
\includegraphics[width=0.32\textwidth,height=5cm]{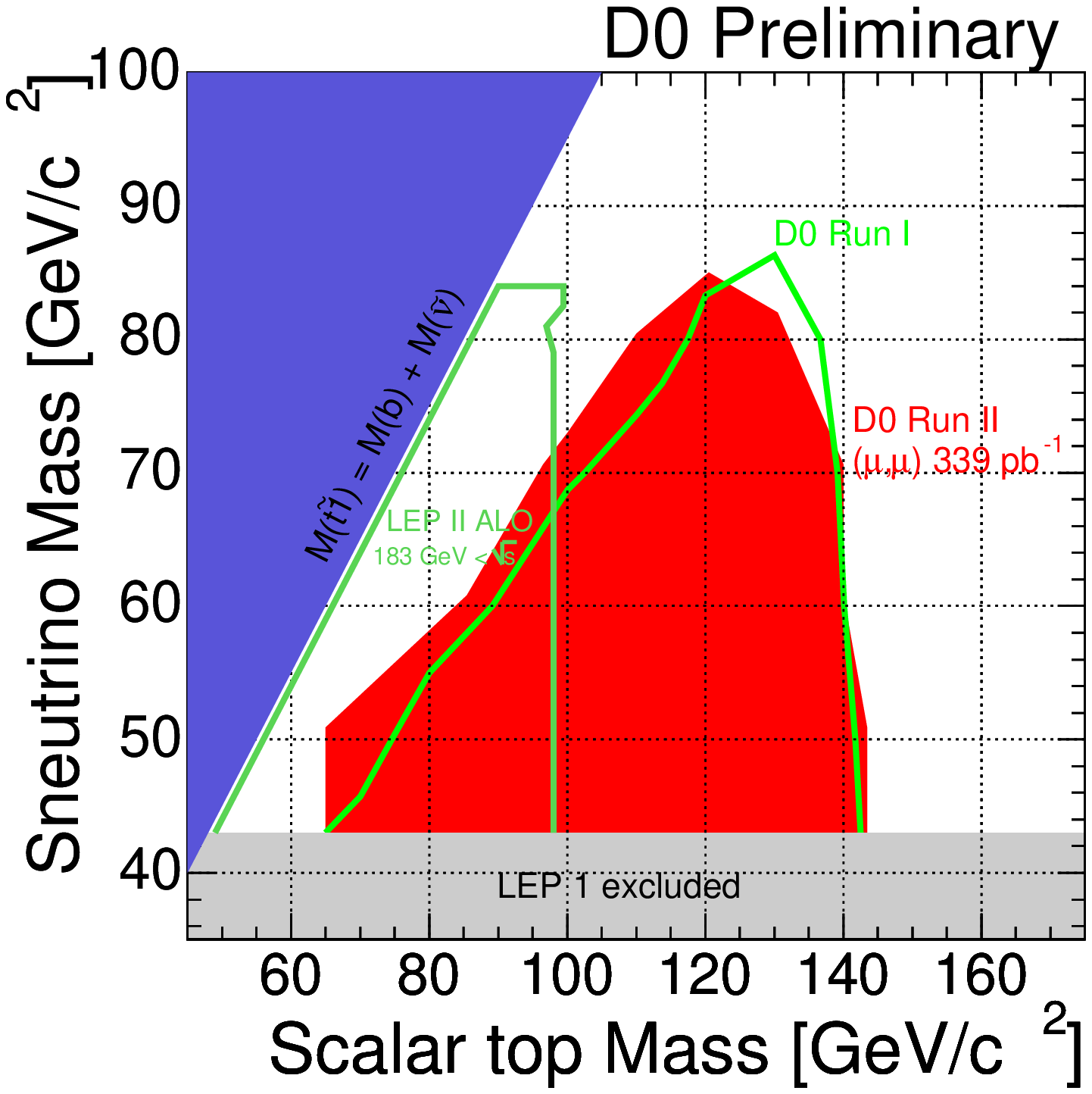}
\vspace*{-0.2cm}
\caption{Left: CDF scalar top cross section limit.
         Center: CDF and D\O\ scalar top sensitivity reach.
         Right: D\O\ $\rm \tilde t_1 \bar{\tilde t}_1\to b \bar b\mu^+\mu^-\tilde \nu\bar{\tilde \nu}$ limits.
         The dark region in the center is excluded at 95\%CL.
}
\label{fig:cdf-stops}
%\vspace*{-0.6cm}
\end{figure}

The pair-production of light scalar top quarks is characterized by two c-quark jets and missing energy if the decay
into $\rm b\chi^\pm$ is kinematically suppressed.
Currently the experimental cross section sensitivity approaches the expected signal
cross section. More luminosity is required to reach sensitivity for the scalar top mass of up to about
160 to 180~GeV, as shown in Fig.~\ref{fig:cdf-stops} (from~\cite{cdf-stops}).
A minimum stop-neutralino mass difference of about at least 35~GeV is required,
unlike LEP~\cite{lep-susy} or the ILC~\cite{ILC-stops}, where also small mass differences are covered.
In addition, the search for the reaction 
$\rm \tilde t_1 \bar{\tilde t}_1\to b \bar b\mu^+\mu^-\tilde \nu\bar{\tilde \nu}$ 
has been performed and resulting
limits are shown also in Fig.~\ref{fig:cdf-stops} (from~\cite{d0-bbmmnn}).

\begin{figure}[htbp]
\includegraphics[width=0.48\textwidth,height=6cm]{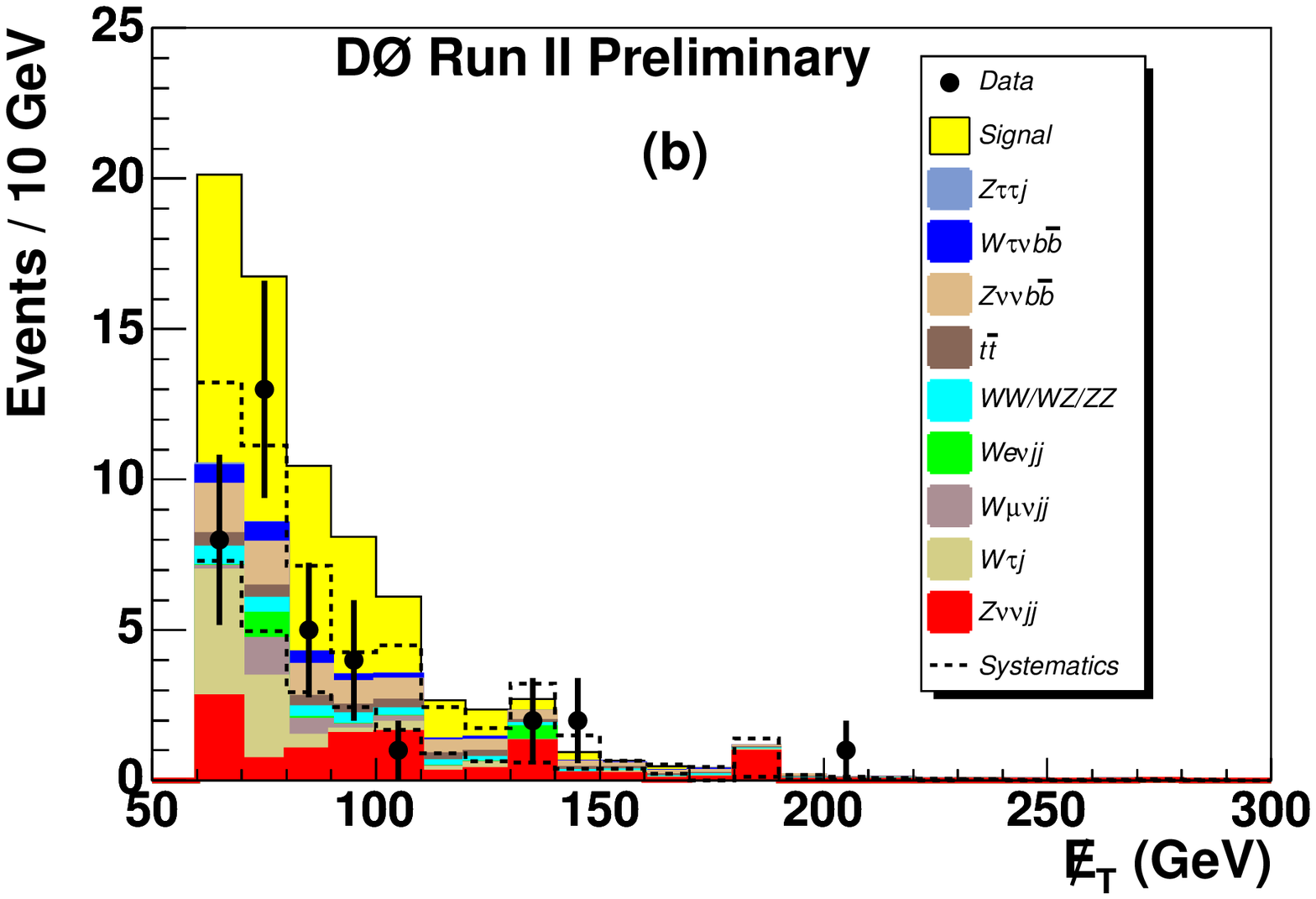} \hfill
\includegraphics[width=0.48\textwidth,height=6cm]{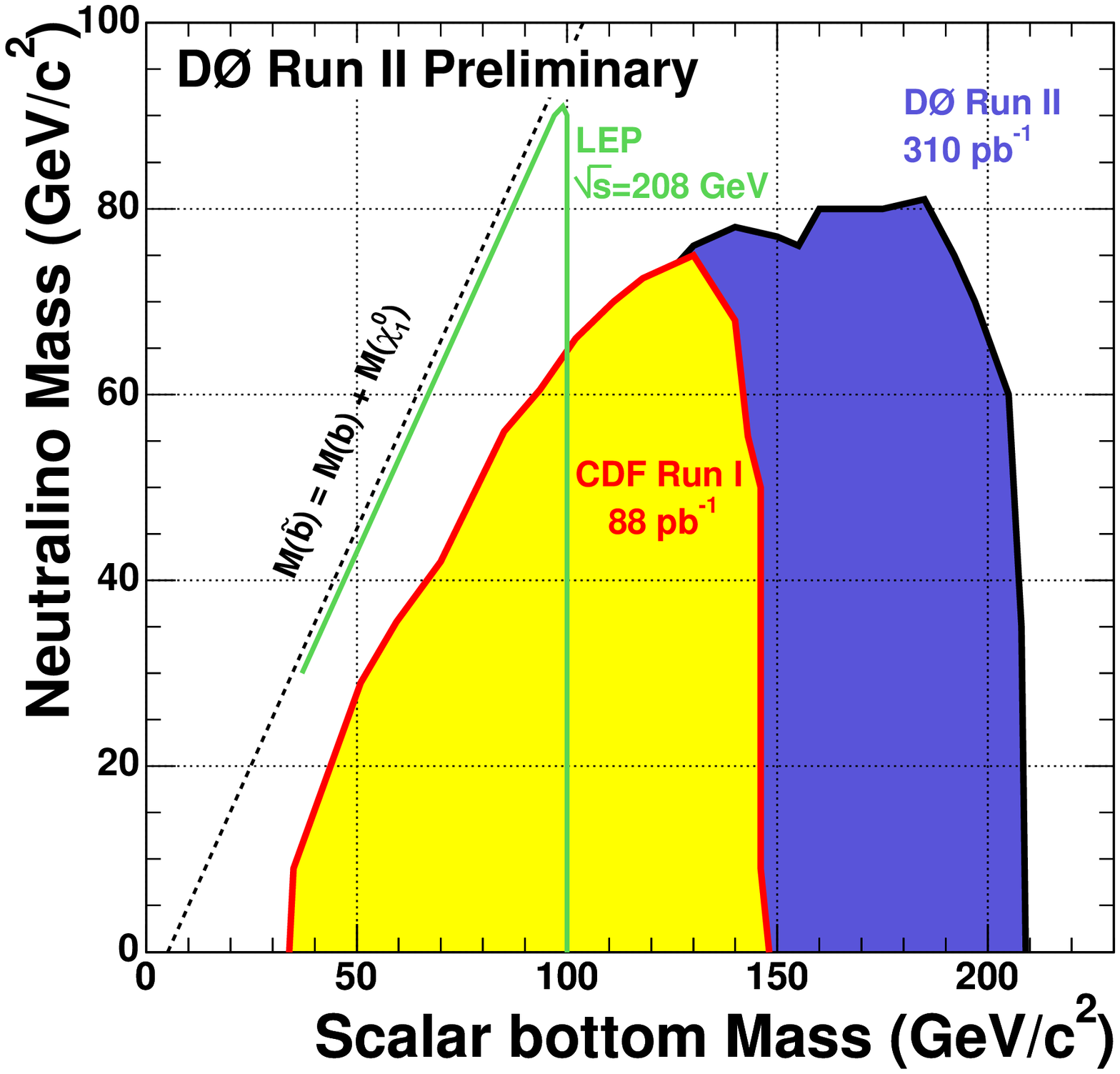}
\vspace*{-0.2cm}
\caption{D\O.
Left: sbottom missing $E_T$ distribution.
Right: sbottom-neutralino mass limits.
}
\label{fig:d0-gl-sb}
%\vspace*{-0.7cm}
\end{figure}

Two b-jets and missing energy from the neutralinos are expected in the final state of sbottom
pair-production. The missing $E_T$ and mass limits in the sbottom-neutralino plane are shown in 
Fig.~\ref{fig:d0-gl-sb} (from~\cite{d0-gl-sb}).

\subsection{Charged Massive Particles }

\begin{figure}[htbp]
\includegraphics[width=0.32\textwidth,height=5cm]{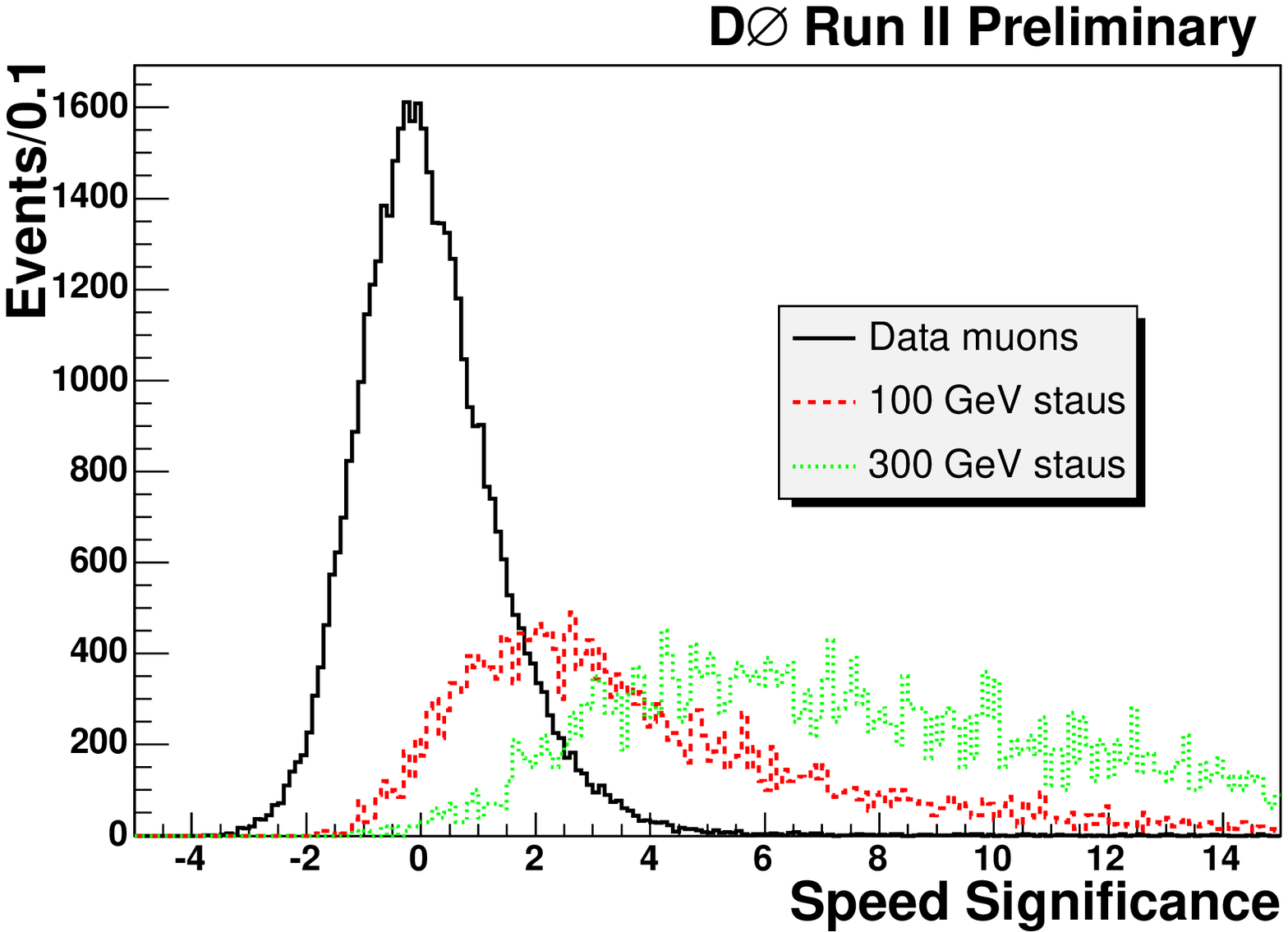}
\includegraphics[width=0.32\textwidth,height=5cm]{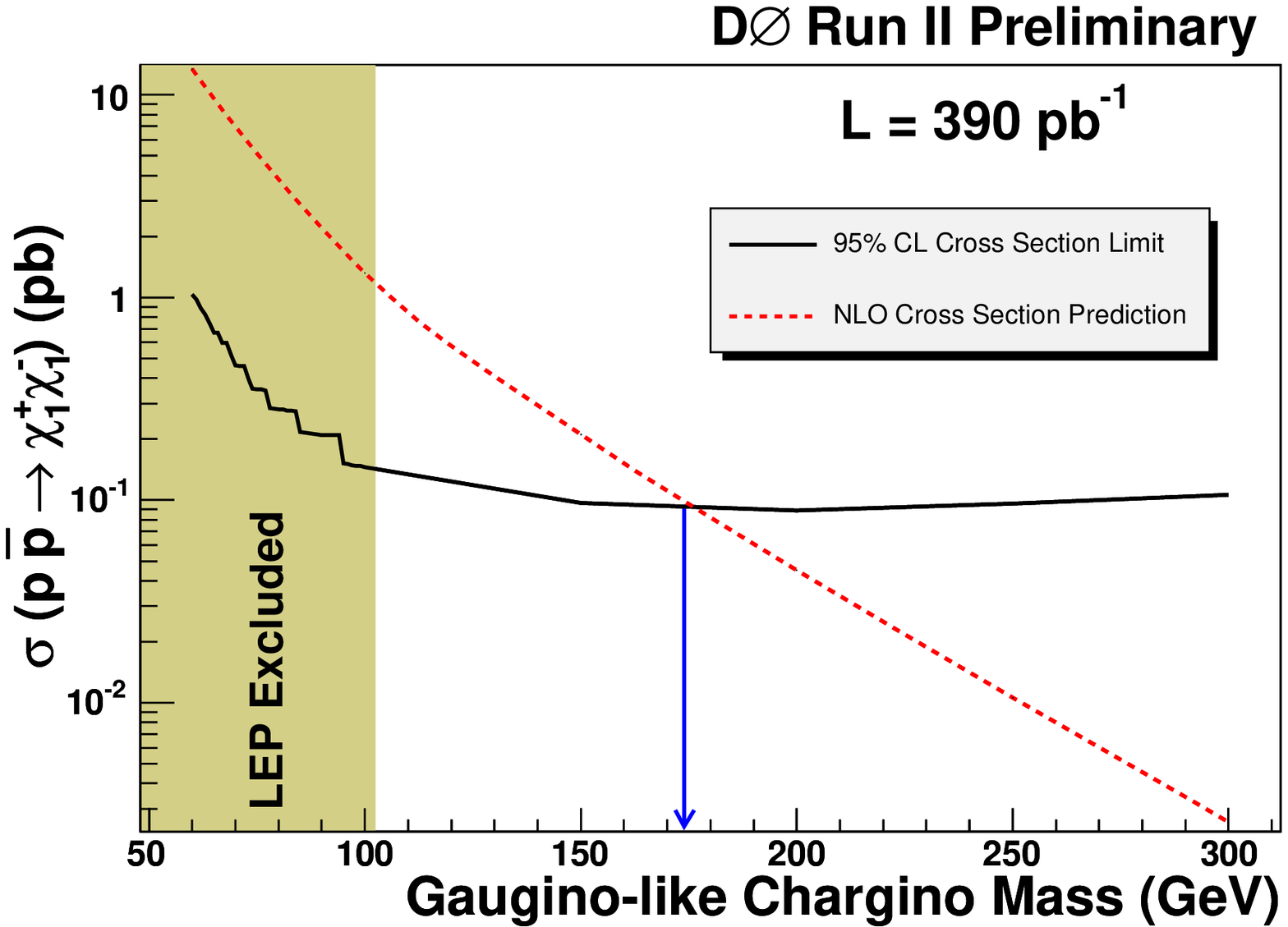}  \hfill
\includegraphics[width=0.32\textwidth,height=5cm]{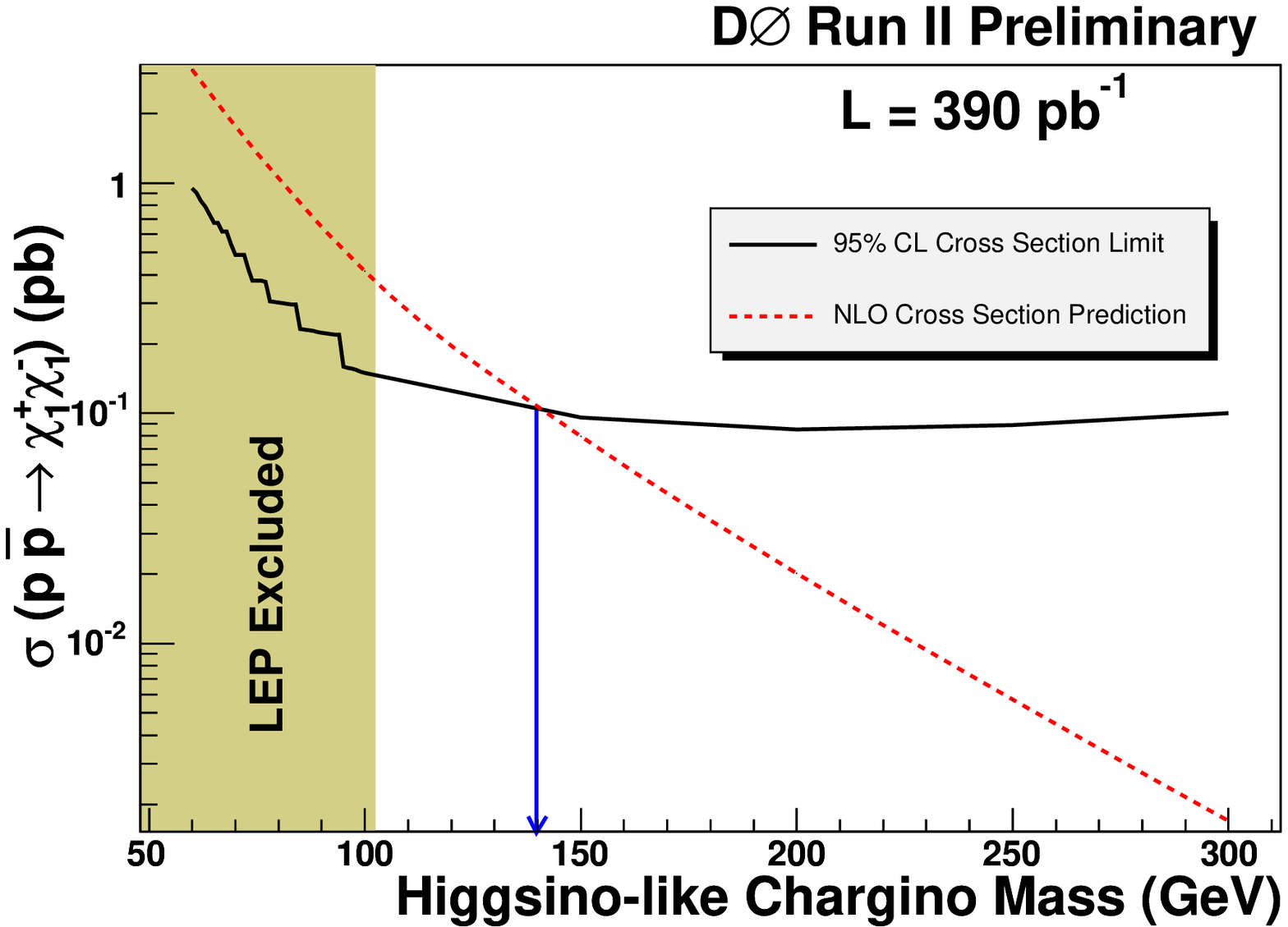}
\vspace*{-0.2cm}
\caption{
D\O. Left: speed significance of charged massive particle.
Center: AMSB gaugino interpretation.
Right: MSSM higgsino interpretation.
}
\label{fig:d0-amsb-limit}
\vspace*{-1cm}
\end{figure}

The production of stable scalar tau leptons would lead to a signature in the detector
like a pair of muons, but with invariant mass and speed inconsistent with the production
of muon pairs. The expected speed of these charged massive particles is expected to be 
significantly different compared to muons as shown in Fig.~\ref{fig:d0-amsb-limit} (from~\cite{d0-champ}).
An interpretation of the results as limits on chargino masses are given in the nearly mass-degenerate 
neutralino-chargino scenario, which occurs naturally 
in the Anomaly Mediated Symmetry Breaking (AMSB) model. Also shown is an example in the MSSM when the chargino
is Higgsino-like.

\clearpage
\section{LHC Prospects}

The LHC will operate at 14~TeV, which is about seven times the center-of-mass energy of the 
Tevatron. This important increase extends the reach to discover new particles. 
The first collisions at the LHC are scheduled for fall 2007, 
and the physics data run in 2008. 

\subsection{SM Higgs Boson}

\subsubsection{Production and Decay}

\begin{figure}[hbp]
\begin{minipage}{0.48\textwidth}
\begin{turn}{270}
\includegraphics[height=\textwidth,width=6cm]{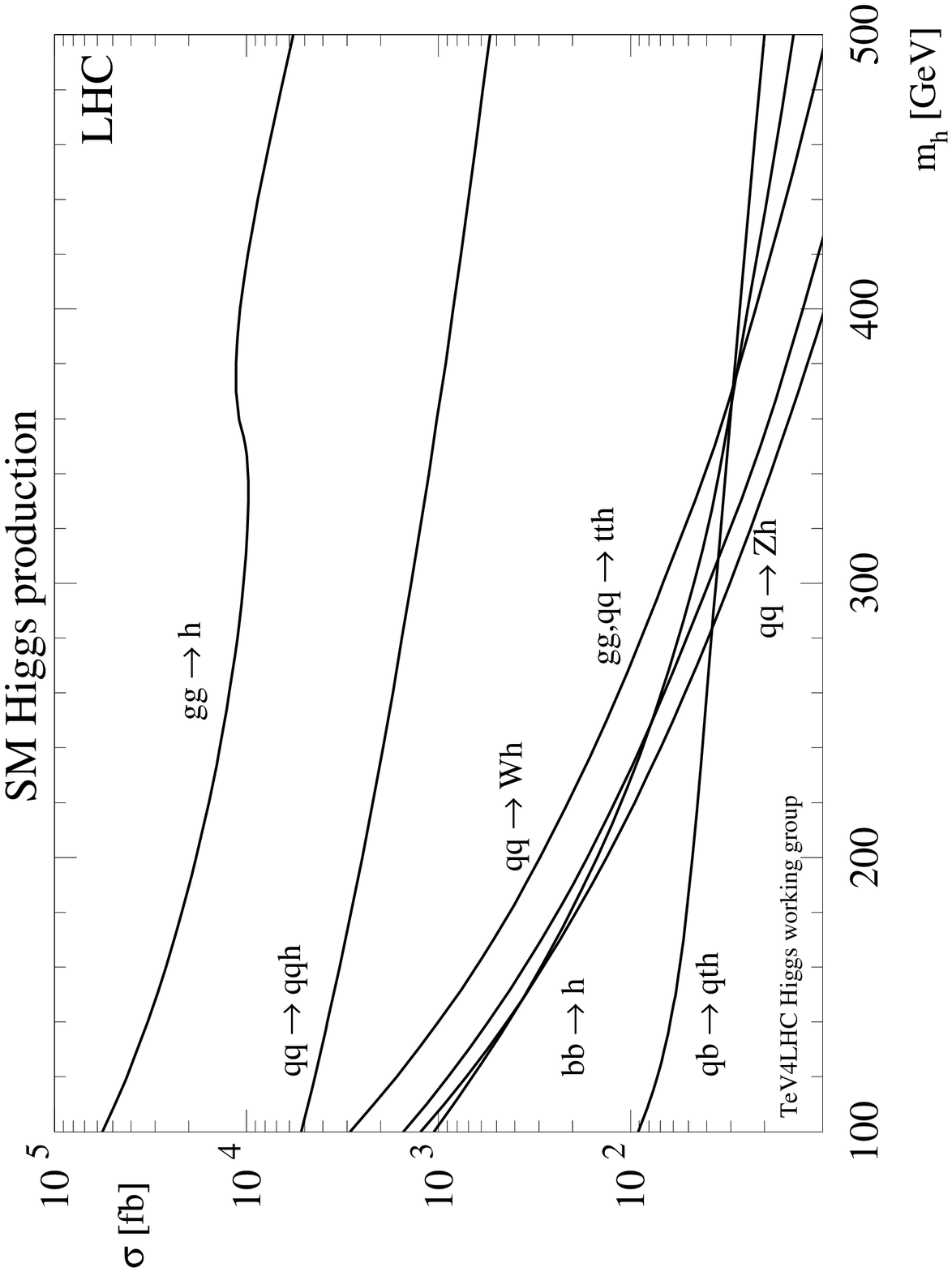}
\end{turn}
\end{minipage} \hfill
\begin{minipage}{0.48\textwidth}
\begin{turn}{270}
\includegraphics[height=\textwidth,width=6cm]{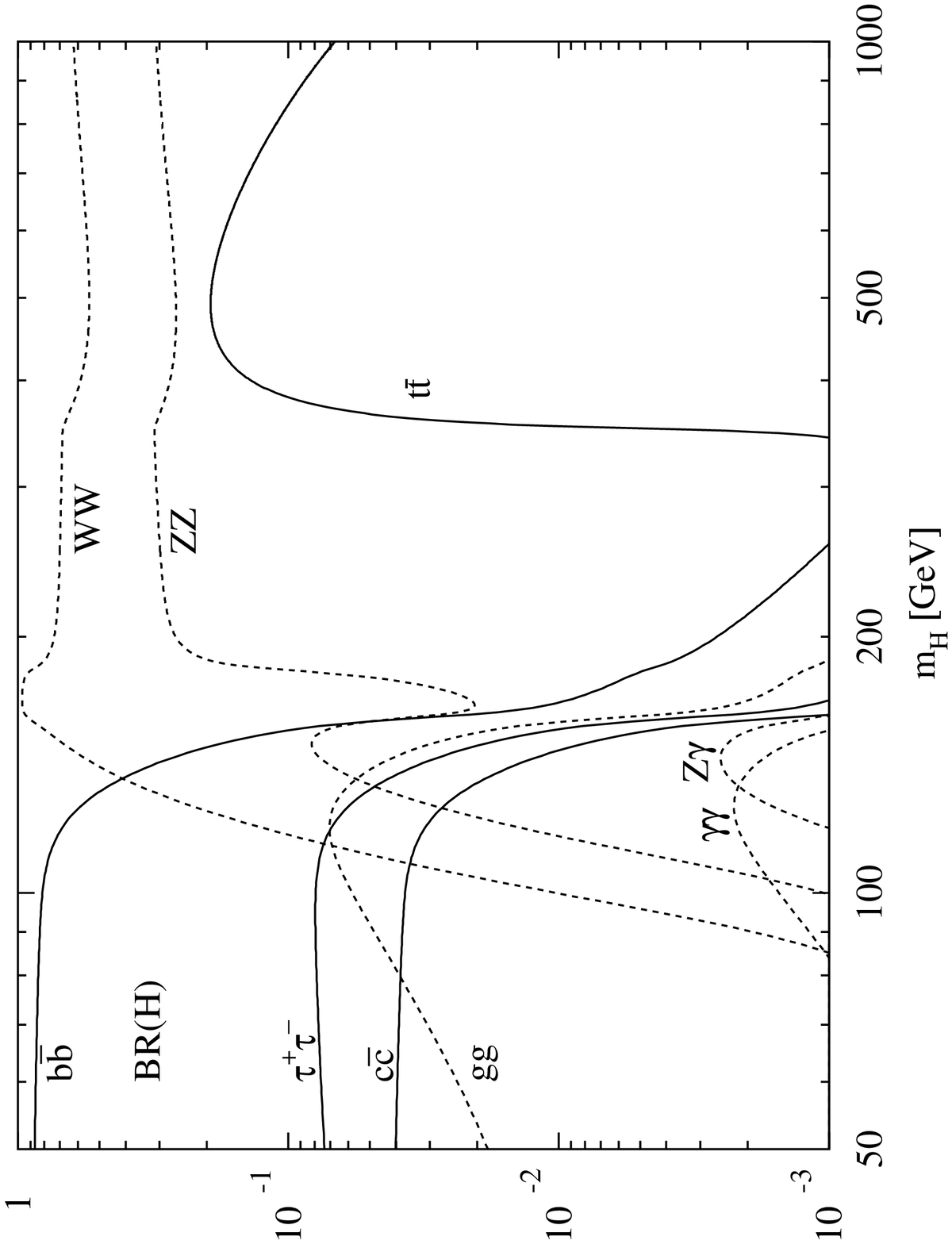}
\end{turn}
\end{minipage}
\vspace*{-0.2cm}
\caption{
Left: expected SM Higgs boson production cross sections at the LHC (14~TeV).
Right: expected Higgs boson decay branching ratios for a SM Higgs boson masses up to 1~TeV. 
}
\label{fig:lhc-xsec}
\end{figure}

The search of the 
SM Higgs boson will consist of a variety of search channels, defined by the production and 
decay properties of the Higgs boson. 
The expected production cross sections and branching 
ratios are summarized in Fig.~\ref{fig:lhc-xsec} (from~\cite{xsec} and~\cite{br}). 
The expected production cross sections for a 120 GeV Higgs boson are for the  Tevatron (LHC): 
1.01~\pb (42.1~\pb) gluon fusion,
0.071~\pb (4.41~\pb) vector boson fusion (VBF),
0.153~\pb (1.65~\pb) associated W production, and 0.093~\pb (0.89~\pb) 
associated Z production~\cite{xsec}. Thus, for this Higgs boson mass the expected rates are typically 
larger by a factor 10 to 40 at the LHC compared to the Tevatron.

\subsubsection{Sensitivity Reach}

\begin{figure}[thp]
%\vspace*{-0.5cm}
\includegraphics[width=0.48\textwidth,height=5.8cm]{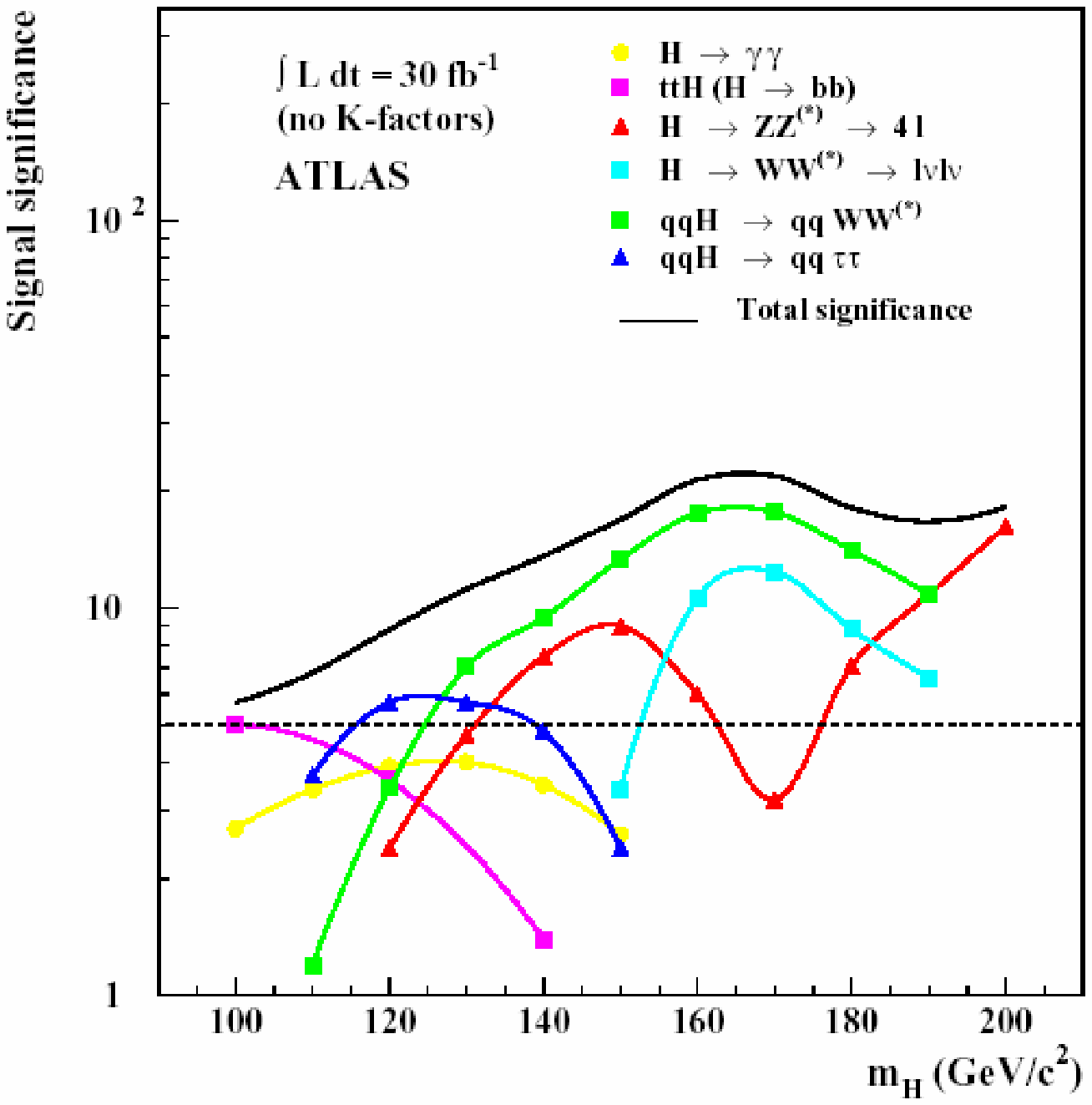} \hfill
\includegraphics[width=0.36\textwidth,height=5.5cm]{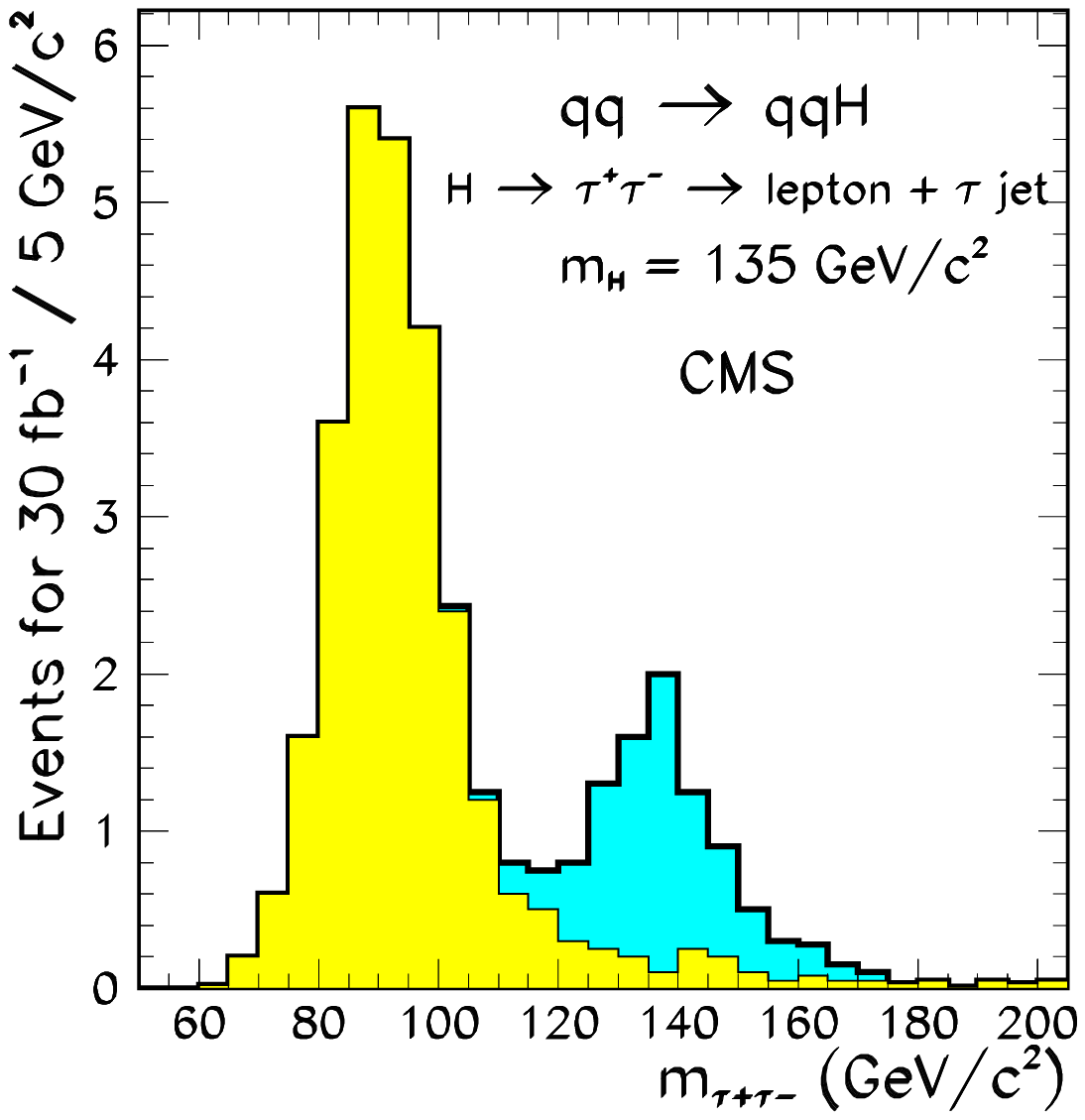}
\vspace*{-0.2cm}
\caption{
Left: ATLAS signal significance as a function of the SM Higgs boson mass.
Right: CMS $\rm qqH$ ($\rm H\to \tau^+\tau^-$) reconstructed invariant $\tau^+\tau^-$ mass for simulated 
signal and background.  
}
\label{fig:atlas-cms-potential}
\end{figure}

For an integrated luminosity of 30~\fb, which is expected to be collected in 3 years of 
data-taking (2008-2010), the highest sensitivity for a SM Higgs boson with mass around 
120~GeV SM Higgs boson is expected from the VBF process qqH ($\rm H\to \tau^+\tau^-$).
The signal significances are summarized in 
Fig.~\ref{fig:atlas-cms-potential} (from~\cite{atlas-sm}).
The strong potential to separate a 135 GeV signal from the background, which arises mainly from
$\rm Z\to \tau^+\tau^-$, is shown in Fig.~\ref{fig:atlas-cms-potential} (from~\cite{cms-tautau}).
The LHC will be able to cover the full mass range of interest with 10-30 fb$^{-1}$ and will
give a definitive experimental answer on the existence of the Higgs boson mechanism.

\subsubsection{Parameter Determination}

\begin{figure}[h!]
%\vspace*{-0.25cm}
\includegraphics[width=0.32\textwidth,height=5.3cm]{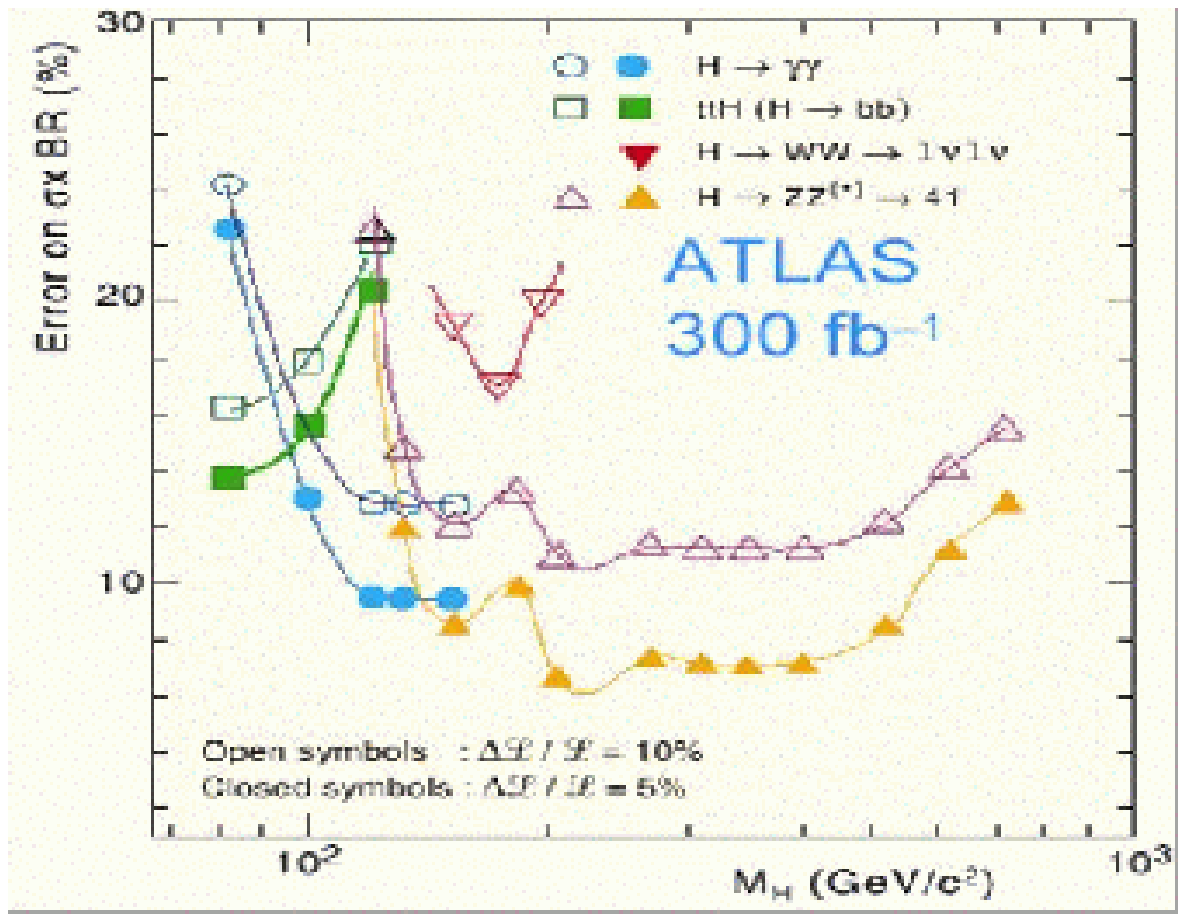}\hfill
\includegraphics[width=0.32\textwidth,height=5.3cm]{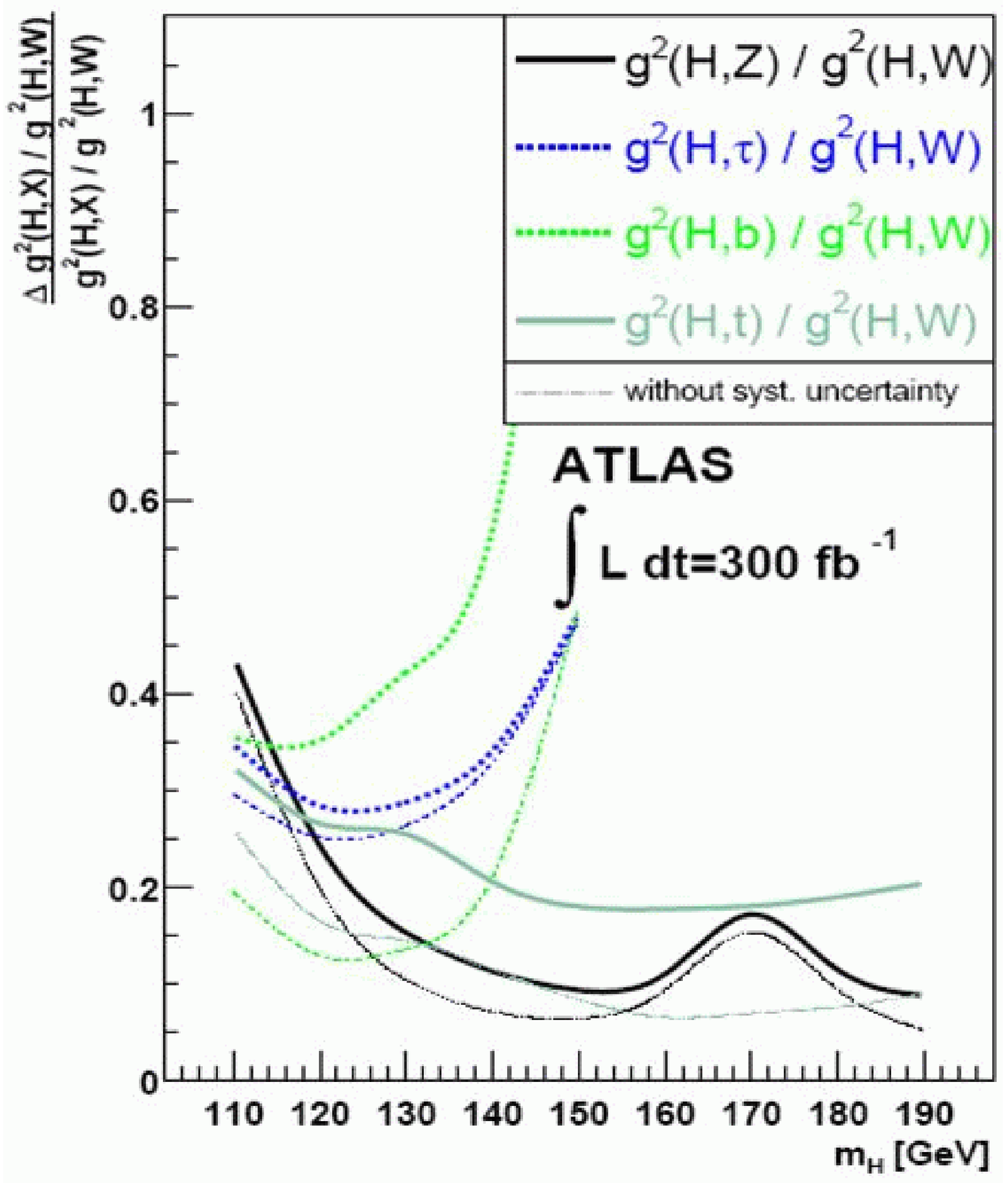}
\includegraphics[width=0.32\textwidth,height=5.3cm]{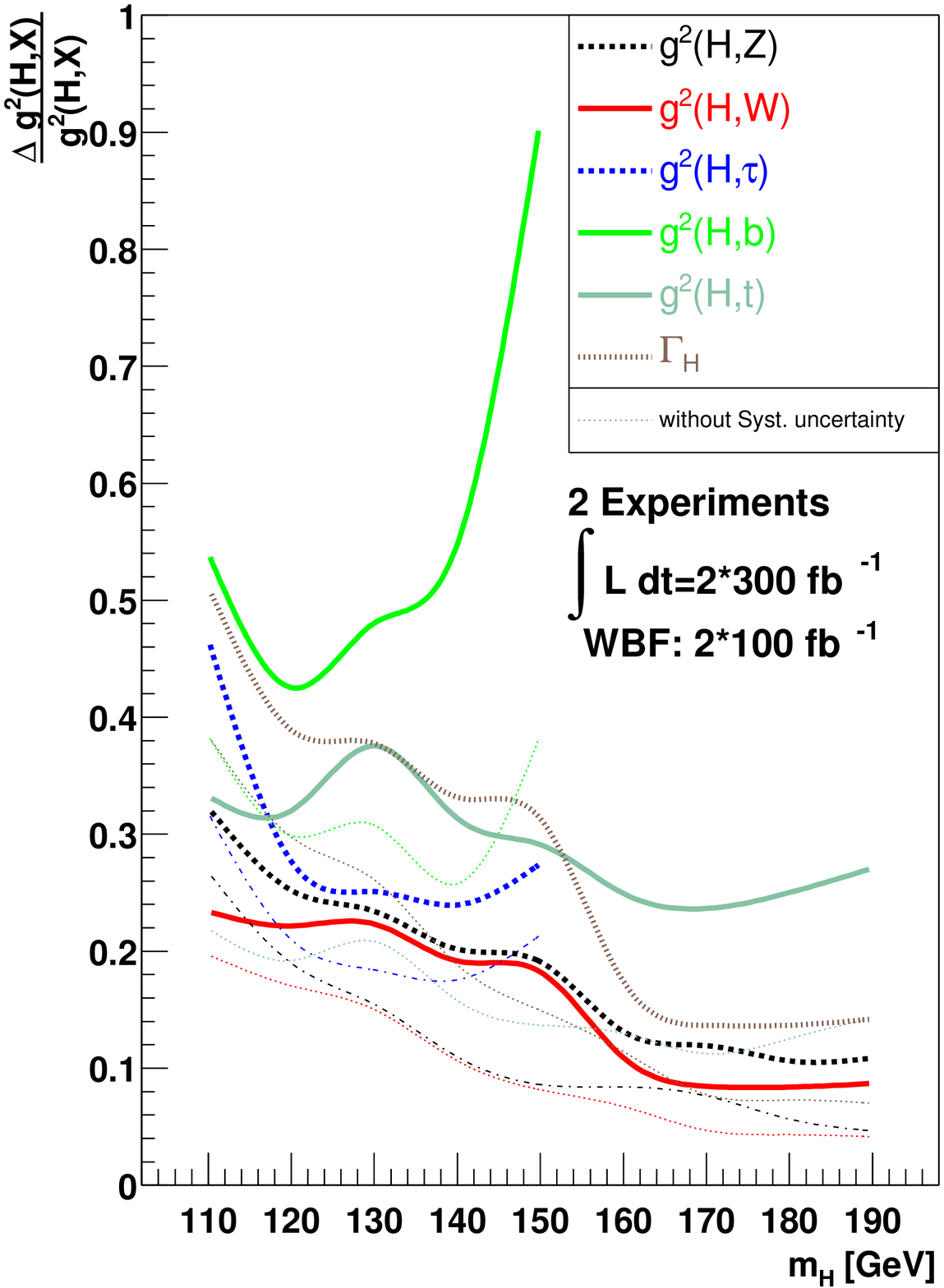}
\vspace*{-0.2cm}
\caption{
ATLAS. Left: expected precision on SM Higgs boson branching ratios times production cross sections.
Center: expected precision on SM Higgs boson coupling ratios.
Right: expected precision on SM Higgs branching ratios.
}
\label{fig:xsec-br-precision}
\end{figure}

\begin{figure}[hp]
\vspace*{-0.2cm}
\includegraphics[width=0.48\textwidth,height=5.5cm]{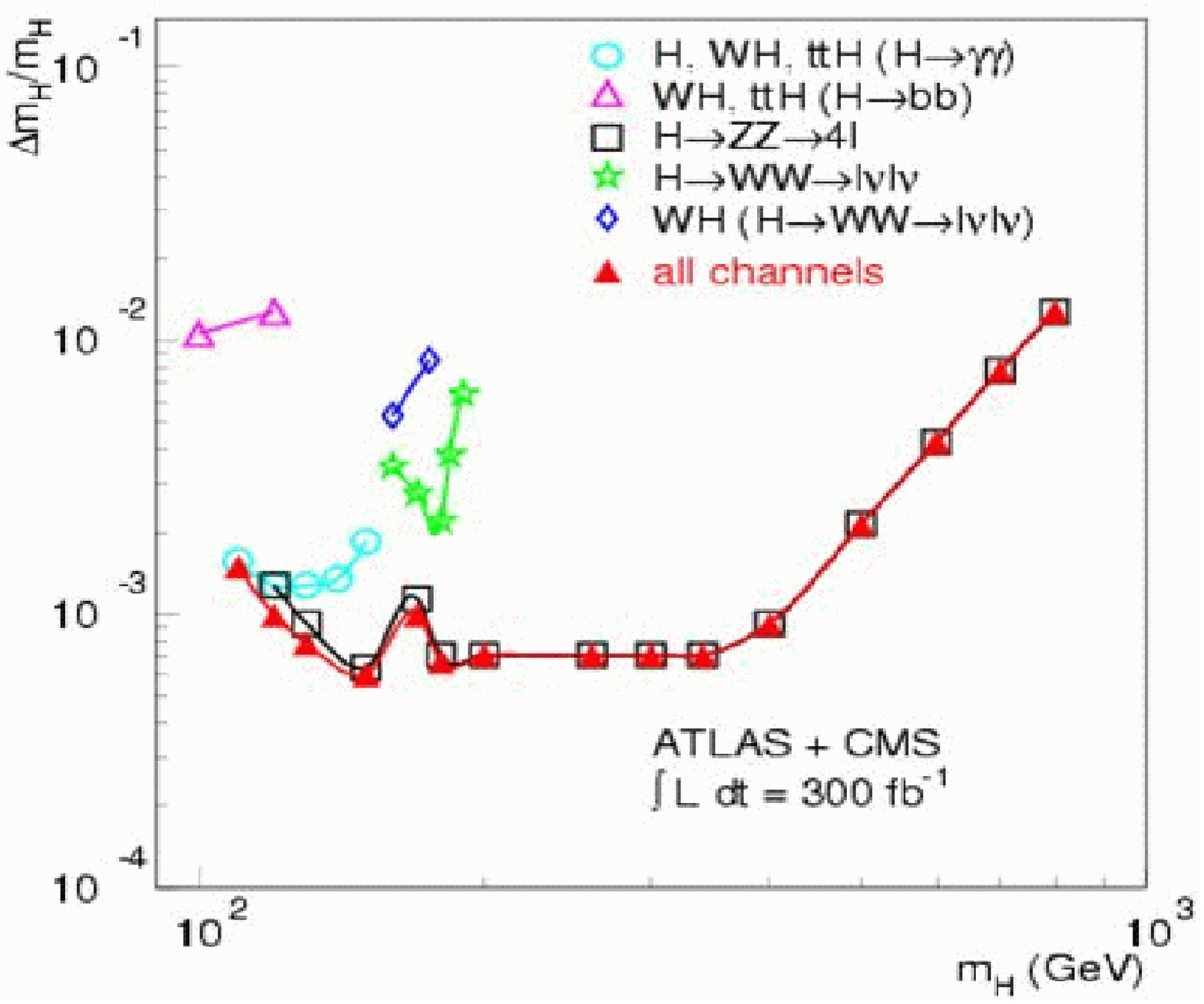}\hfill
\includegraphics[width=0.48\textwidth,height=5.5cm,bbllx=50pt,bblly=185pt,bburx=504pt,bbury=650pt,clip=]{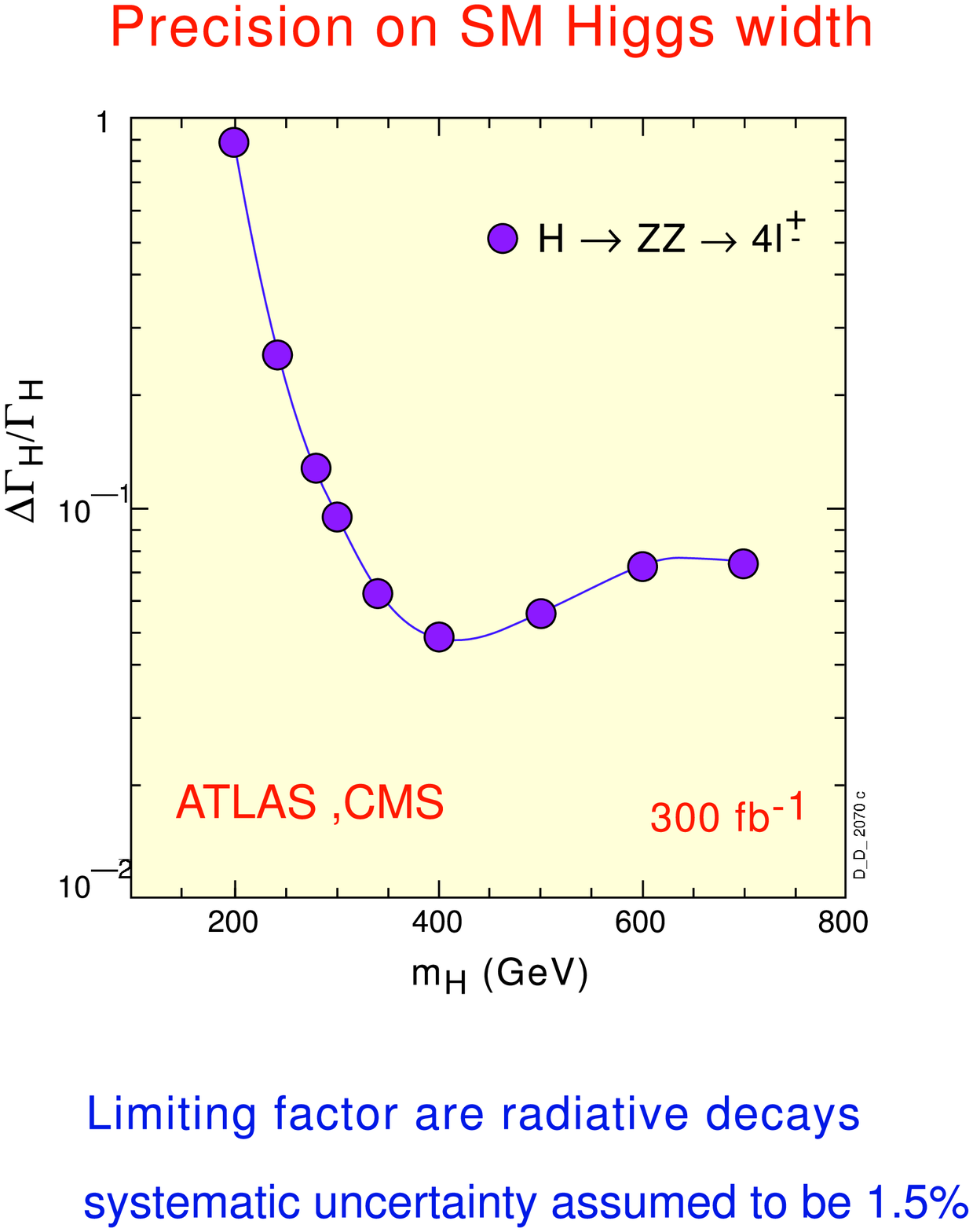}
\vspace*{-0.5cm}
\caption{
ATLAS and CMS expected precision. Left: on the SM Higgs boson mass.
Right: on the SM Higgs boson width.
}
\label{fig:sm-mass-width}
\vspace*{-0.5cm}
\end{figure}

First measurements of the SM Higgs boson branching ratios times production cross sections are
expected already with a luminosity of 10-30~\fb.
The precision depends on the Higgs boson mass, and on the luminosity uncertainty,
as shown in Fig.~\ref{fig:xsec-br-precision} (from~\cite{xsec-br-precision}) for 300~\fb.
The figure shows also the precision on the SM Higgs boson coupling ratios.
With an upper bound on the Higgs--gauge-boson couplings the absolute couplings rather than coupling 
ratios can be extracted as shown in Fig.~\ref{fig:xsec-br-precision} (from~\cite{lhc-couplings}).
The expected precision on the measurement of the SM Higgs boson mass and its width is given in 
Fig.~\ref{fig:sm-mass-width} (from~\cite{mass-width}).

\begin{figure}[hp]
\vspace*{-0.4cm}
\includegraphics[width=0.48\textwidth,height=5.5cm]{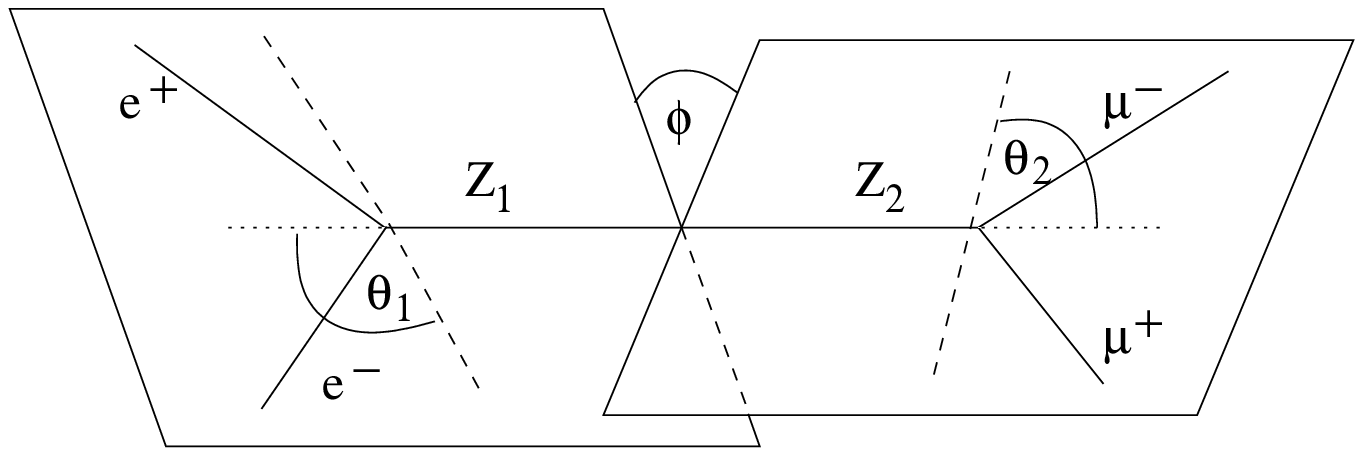}\hfill
\includegraphics[width=0.48\textwidth,height=5.5cm]{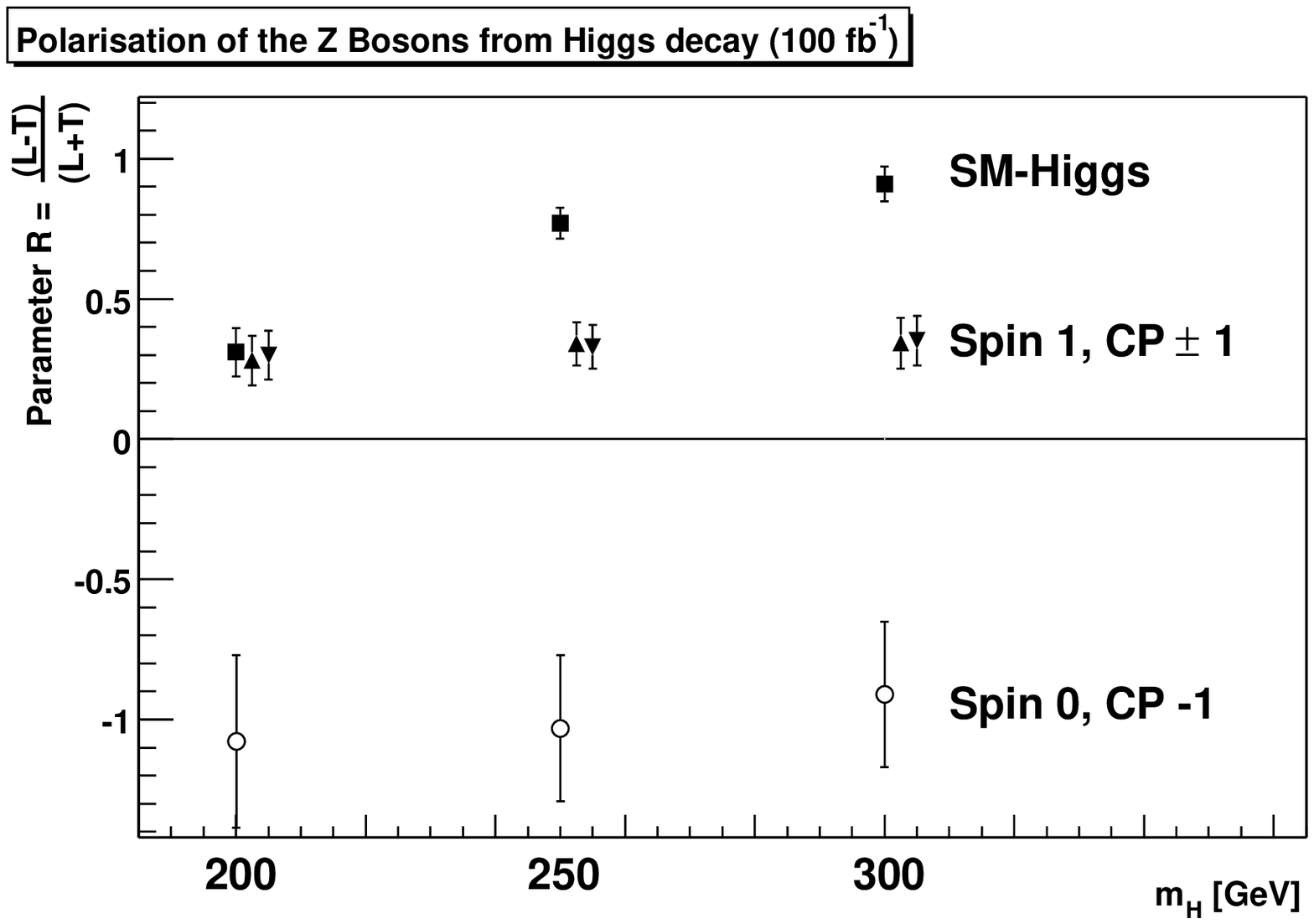}
\vspace*{-0.2cm}
\caption{Left: angle $\phi$ between the two planes defined by the leptons 
 from the decay of the two Z  bosons in the  rest frame of the Higgs.
 The dashed lines represent the direction of motion of the 
 leptons in the rest frame of the Z boson from which they originate. Angles $\theta_1$ and $\theta_2$ 
 are between the negatively charged leptons and the direction of motion of the 
 corresponding Z in the Higgs boson rest frame.
Right: parameter $R$, related to the decay angles, for different Higgs boson masses. Spin and CP values can
be well separated within the expected uncertainties.
}
\label{fig:atlas-spin}
\vspace*{-0.7cm}
\end{figure}

The spin and CP-value of the Higgs boson can be determined from studying the 
$\rm gg \rightarrow H \rightarrow ZZ \rightarrow 4\ell$ process, where the $ZZ$ production is kinematically allowed. 
Figure~\ref{fig:atlas-spin} (from~\cite{atlas-spin})
illustrates the sensitive variables and shows the clear separation between spin and CP values.
A future Linear Collider will improve both the mass and width determination~\cite{ilc-mass-width} and 
has in addition the potential to measure the Higgs boson self-coupling which is directly related to the shape of the 
Higgs boson potential.

%\clearpage
\subsection{Beyond the SM}

\subsubsection{Invisibly Decaying Higgs Bosons}

\begin{figure}[htbp]
\vspace*{-0.5cm}
\includegraphics[width=0.34\textwidth]{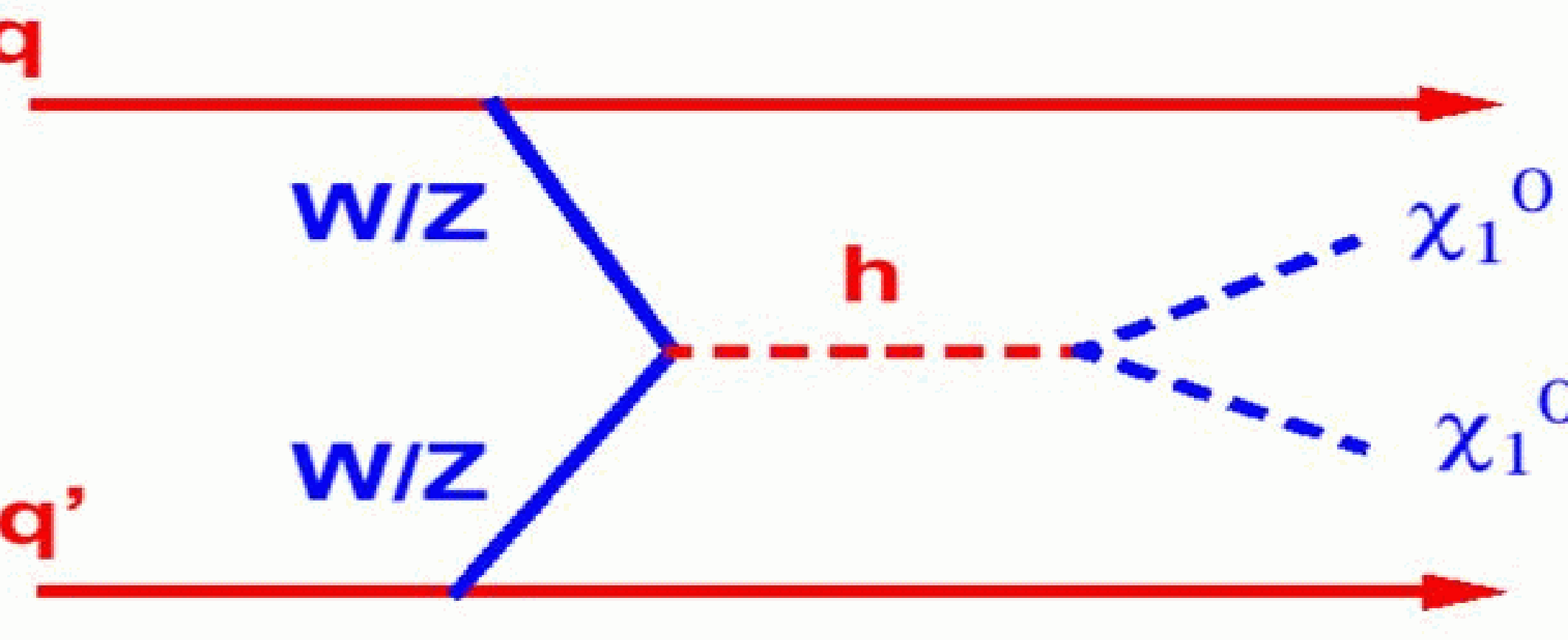}\hfill
\includegraphics[width=0.48\textwidth]{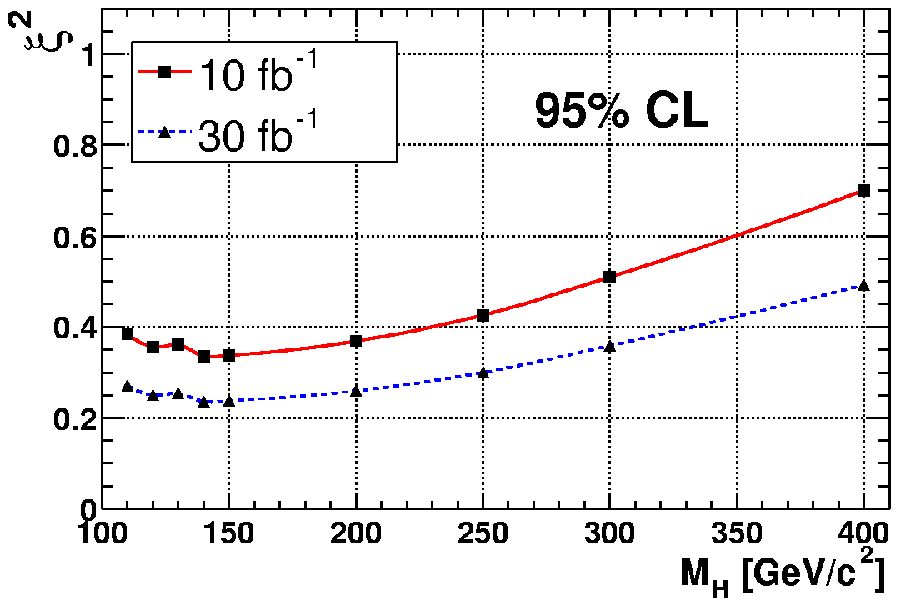}
\vspace*{-0.5cm}
\caption{ATLAS invisibly decaying Higgs bosons. Left: production and decay graph. Right: sensitivity reach.
}
\label{fig:inv}
\vspace*{-0.5cm}
\end{figure}

In models beyond the SM, the Higgs boson could decay invisibly, for example into a pair of
neutralinos, as illustrated in Fig.~\ref{fig:inv} (from~\cite{lhc-inv}). The parameter $\xi$
is defined as $g({\rm hZZ}) / g({\rm H_{\rm SM}ZZ})$. The sensitivity at the LHC extends the 
LEP results~\cite{lep05}.

\subsubsection{$\rm H\to \tau^+\tau^-$ and $\tan\beta$ Determination}

\begin{figure}[htbp]
\vspace*{-0.4cm}
\includegraphics[width=0.33\textwidth,height=6cm]{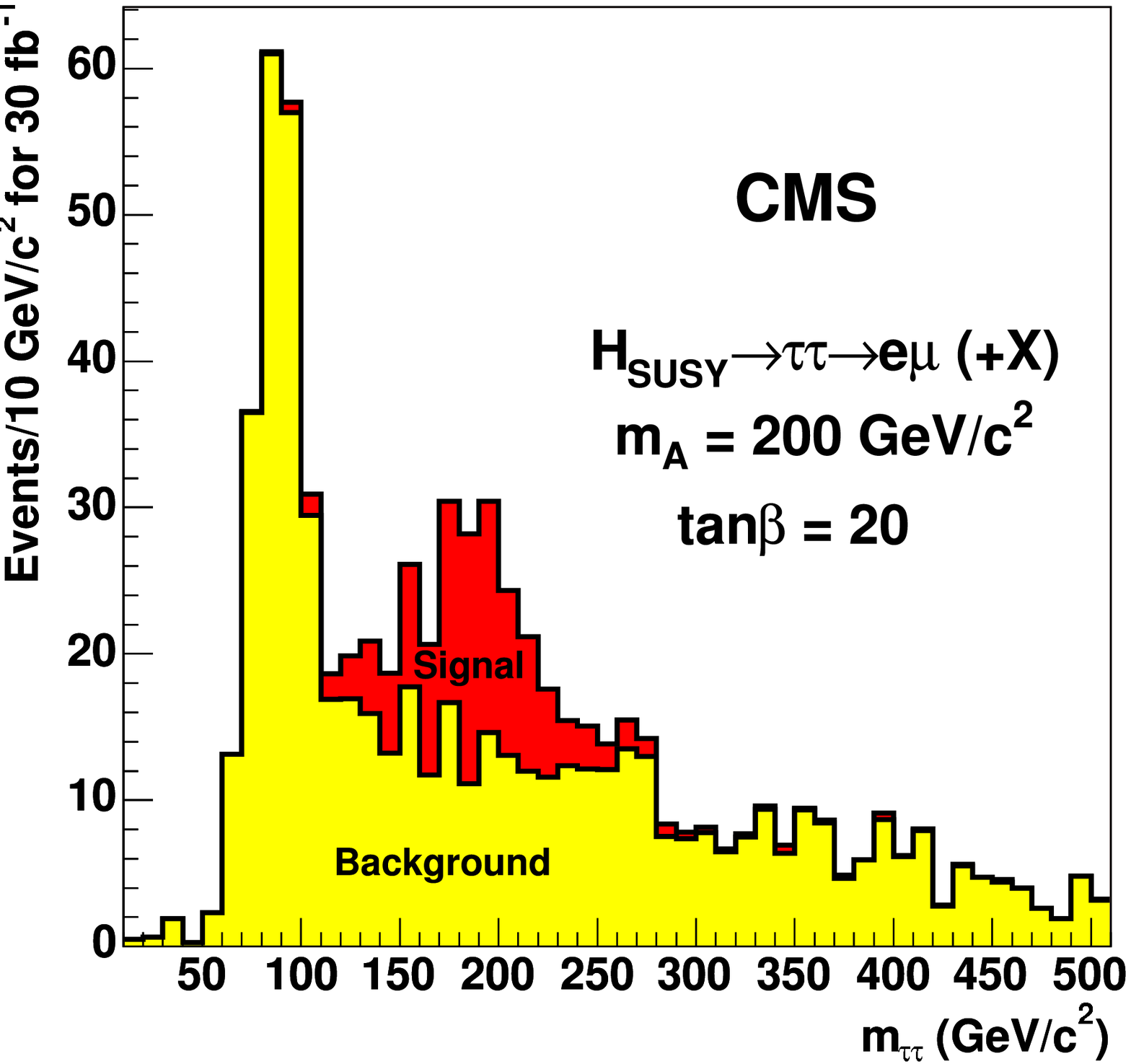}\hfill
\includegraphics[width=0.33\textwidth,height=5.5cm]{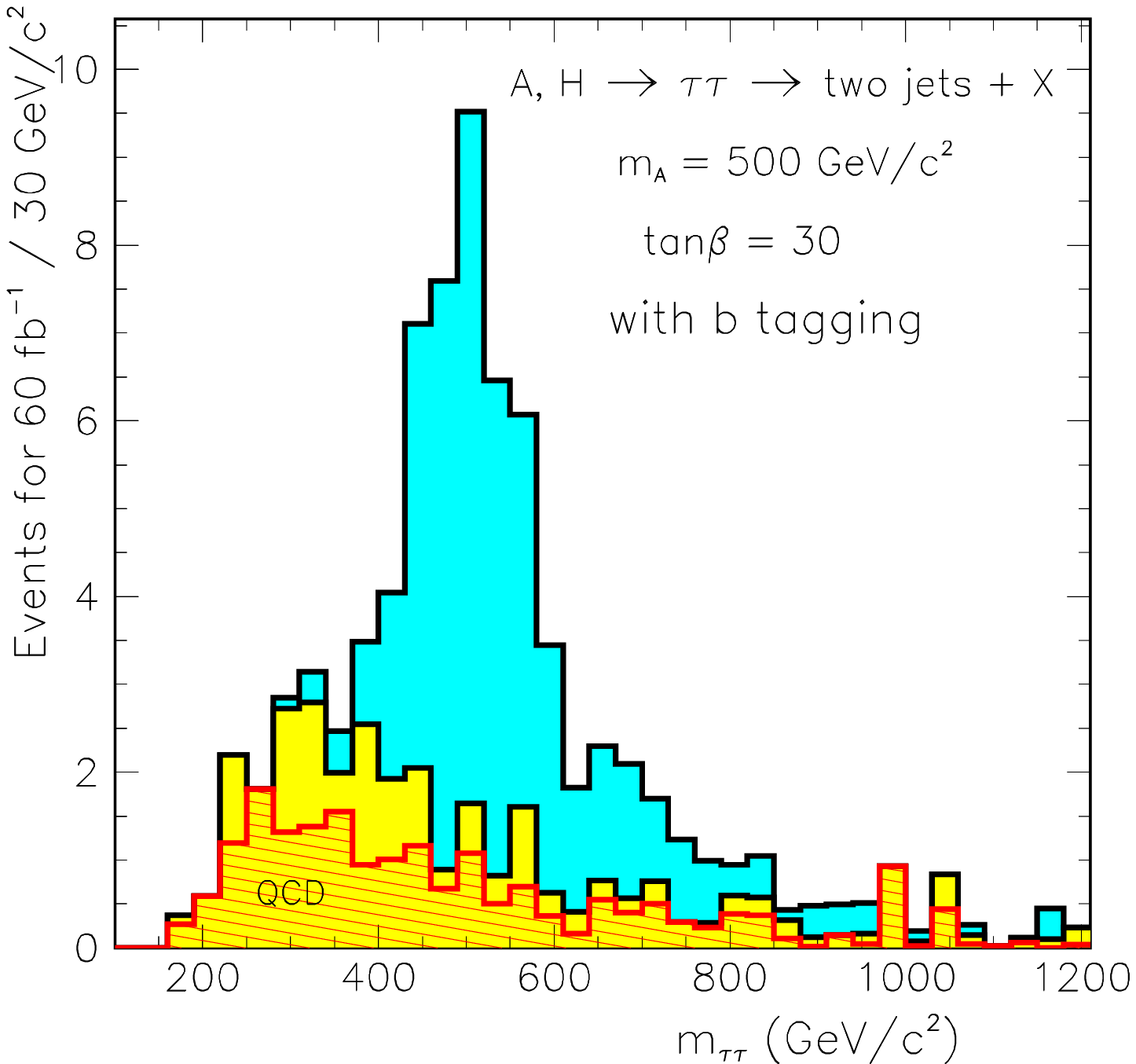}\hfill
\includegraphics[width=0.33\textwidth,height=6cm]{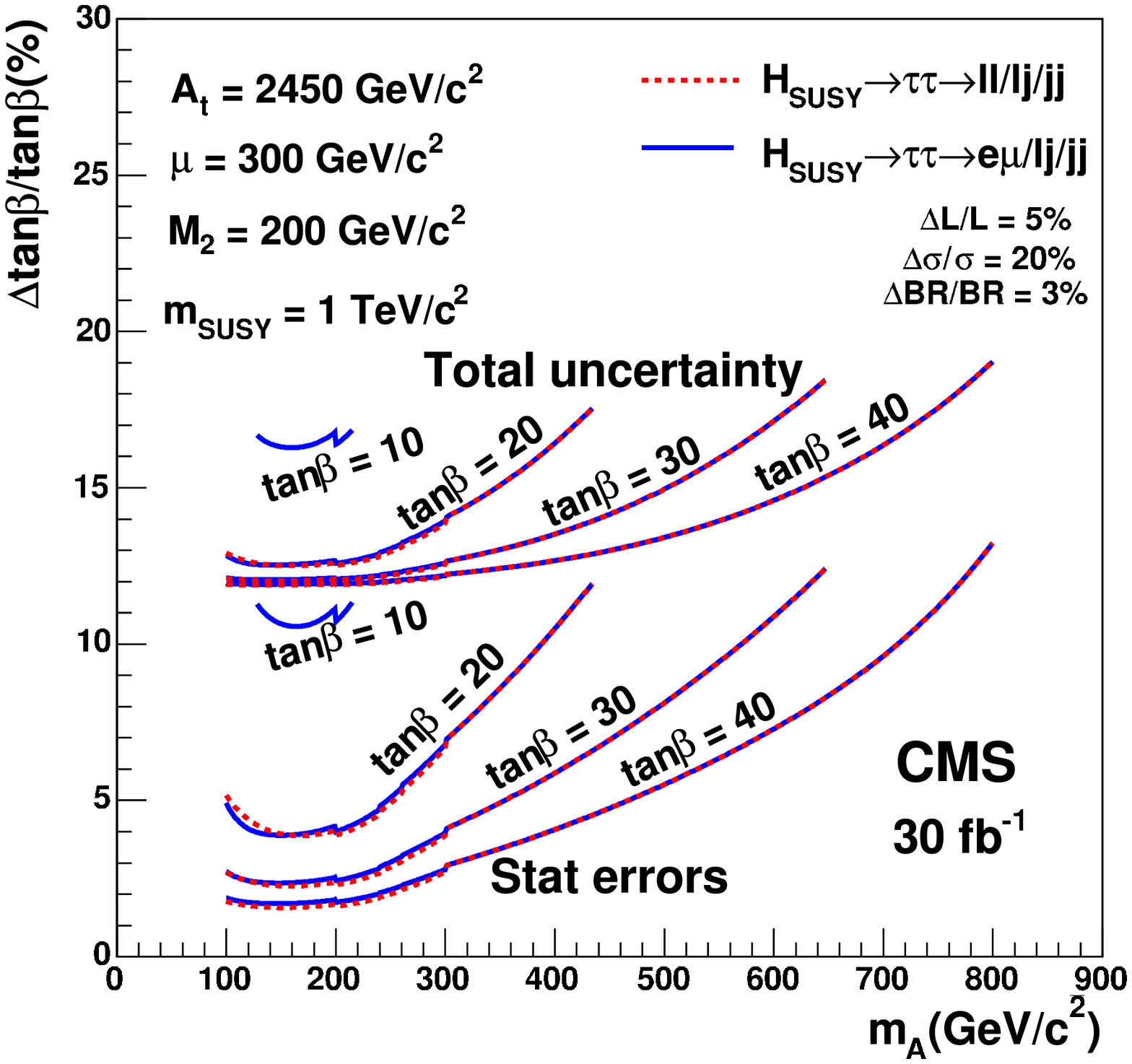}
\vspace*{-0.4cm}
\caption{
CMS. Left: $\rm H\to \tau^+\tau^- \to e\mu$ (+X) expected signal and background.
Center: $\rm H\to \tau^+\tau^- \to jets$ (+X) expected signal and background. 
Right: expected sensitivity on $\tan\beta$ including statistical and systematic uncertainties
(upper lines).
}
\label{fig:cms-tau}
\vspace*{-1cm}
\end{figure}

In the general 2-doublet Higgs model, the decay mode $\rm H\to \tau^+\tau^-$ is expected to be enhanced for large
values of $\tan\beta$. A clear signal over the background will be obtained as shown 
in Fig.~\ref{fig:cms-tau} (from~\cite{cms-tau}). The measurement of the expected rate will
allow the determination of $\tan\beta$ with the expected precision as given in Fig.~\ref{fig:cms-tau}. 
The precise determination of other SUSY parameters is an important requirement.
Further measurable dependencies on $\tan\beta$ of Higgs boson parameters have been discussed 
in the Linear Collider context~\cite{ilc-tgb}.

\begin{figure}[htbp]
\vspace*{0.1cm}
\begin{center}
\includegraphics[width=0.7\textwidth,height=8cm]{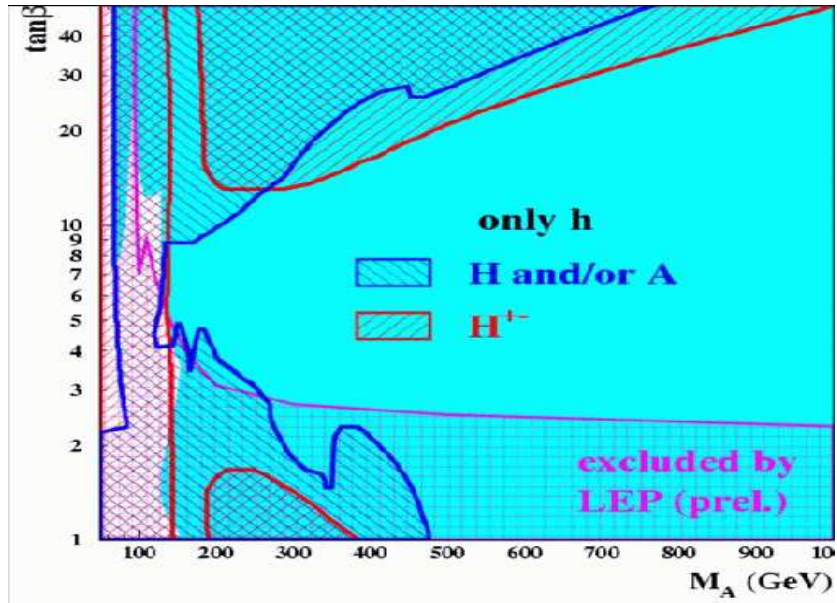}
\end{center}
\vspace*{-0.7cm}
\caption{
Expected parameter region of detection in the $m_{\rm A}$ versus $\tan\beta$ plane of the MSSM
for one or several Higgs bosons.
} \label{fig:lhc-mssm}
\vspace*{-0.2cm}
\end{figure}

In the MSSM, the LHC will cover the entire $(m_{\rm A},\tan\beta)$ parameter space and one or more Higgs bosons will
be detected (Fig.~\ref{fig:lhc-mssm} from~\cite{lhc-mssm}).
For large $\tan\beta$ several Higgs bosons are expected to be detected, while for 
medium $\tan\beta$ values and $m_{\rm A}$ above about 200~GeV, only one SM-like Higgs boson is expected to
be detected. In this region the determination of the underlying model is particularly challenging.
Further measurements with higher precision to determine the underlying process 
will be possible at a future Linear Collider (for an overview see for example Ref.~\cite{ICHEP02Higgs}).

\subsection{Supersymmetric Particles}

\begin{figure}[hcbp]
\includegraphics[width=\textwidth,height=9.5cm,bbllx=36pt,bblly=36pt,bburx=870pt,bbury=778pt,clip=]{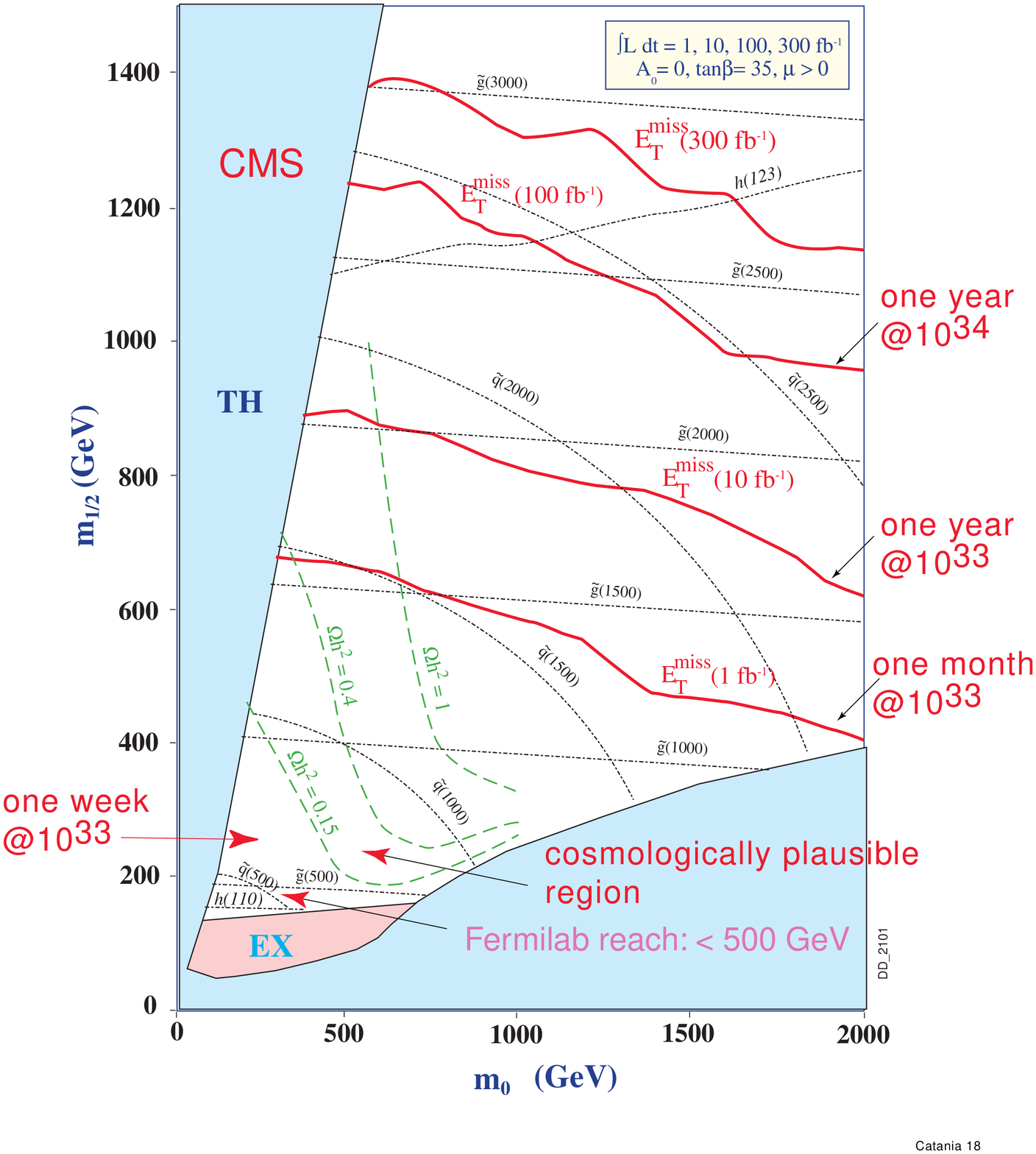}
\vspace*{-0.8cm}
\caption{Squark-qluino $5\sigma$ discovery reach with a jets plus missing energy signal in the mSUGRA model.
         In the forbidden region on the left (small $m_0$), the neutralino is not the LSP and in the lower region
         (small $m_{1/2}$) there is no electro-weak symmetry breaking.
         The cosmologically plausible region is in particular near the forbidden region on the left 
         (co-annihilation) and near the forbidden region near the bottom (focus-point). These regions could be
         covered within about the first year of data-taking. With a similar luminosity squarks of up to 2500~GeV
         could be discovered.   
} \label{fig:LHC-mSUGRA}
\vspace*{-0.6cm}
\end{figure}

In searches for Supersymmetric particles, the LHC has the potential for a discovery within
a short period of data-taking. An example in the mSUGRA model is shown in
Fig.~\ref{fig:LHC-mSUGRA} (from~\cite{lhc-msugra}). A variety of Supersymmetric reactions
will be in reach of the LHC. While a signature from Supersymmetry at the LHC cannot escape detection,
measuring the underlying structure is a very challenging task.
Further precision measurements will be possible at a Linear Collider, as reviewed for example in Ref.~\cite{susy03lc}.
It will be particular difficult for the LHC to determine the relevant scalar top parameters is the cosmologically
interesting region of stop-neutralino co-annihilation~\cite{borjanovic-EPS-susy},
where a future Linear Collider can perform precision measurements~\cite{ILC-stops}.

\clearpage
\section{Conclusions}
Much has been learned from the searches for Higgs bosons and other new particles at LEP.
The Tevatron Run-II searches are well under way and already have set several limits exceeding the previous LEP limits.
For the SM Higgs boson, searches for gluon fusion with WW decays, associated production
WH with $\rm \bb$ and WW decays, and $\rm ZH\to \nu\nu\bb$ decays have been performed.
Beyond the Standard Model, the searches at
the Tevatron for bbA, H$^{++}$, $\rm h\rightarrow\gamma\gamma,~and~\tau^+\tau^-$ have also much more 
potential for a discovery with larger luminosity.
In Supersymmetry, searches for neutralinos and charginos in GMSB and mSUGRA models, as well as
searches for squark and gluino, scalar tops, scalar bottoms, and scalar taus already have set
new limits. In the near future, the LHC will extend the discovery reach in the Higgs sector,
and has also the potential to measure the masses, branching ratios and $\tan\beta$.
Supersymmetric particles could be discovered in a short time of operation
after detector calibration and a good understanding of the SM backgrounds. 
With the operation of the Tevatron at increasing data rates and the LHC ahead, it is an 
exciting time for new discoveries.
The close collaboration of phenomenologists and experimentalists is crucial to fully exploit the
potential of the current and future accelerators.

%\begin{acknowledgments}
\section*{Acknowledgments}
I like to thank the organizers of PHENO'05 and the organizers of the workshop 
`From the Tevatron to the LHC to the Linear Collider' at the Aspen Center for Physics'05 
for their invitation and the very kind hospitality.
I would also like to thank Gregorio Bernardi, Suyong Choi and Arnd Meyer from the D\O\ experiment,
Song Ming Wang and Tom Junk from the CDF experiment, Steve Mrenna from the Fermilab Theory and 
Computing Division, and Albert de Roeck from the LHC for comments on the manuscript.
%\end{acknowledgments}

\end{document}